\RequirePackage{fix-cm}
\documentclass[twocolumn]{svjour3}          
\smartqed  
\usepackage{diagbox}
\usepackage{graphicx}
\usepackage{enumitem}
\usepackage{hyperref}
\usepackage{pifont}
\usepackage{amsmath}
\usepackage{amssymb}
\usepackage{multirow}
\usepackage{subcaption}
\usepackage{url}
\usepackage[misc]{ifsym}

\usepackage[table,xcdraw]{xcolor}

\newcommand{\decrease}[1]{{\color[HTML]{00C42B}{#1}}}
\newcommand{\increase}[1]{{\color[HTML]{DF0404}{#1}}}

\begin{document}

\title{
How Good Are Multi-dimensional Learned Indices? An Experimental Survey
\thanks{*These three authors contributed equally to this work. \\
\dag Qiyu Liu is the corresponding author.}
}


\author{Qiyu Liu\textsuperscript{*,\dag}         \and
        Maocheng Li\textsuperscript{*}      \and
        Yuxiang Zeng\textsuperscript{*}     \and
        Yanyan Shen      \and
        Lei Chen
}

\authorrunning{Qiyu Liu et al.} 

\institute{
            Qiyu Liu \at
             Southwest University\\
              \email{qyliu.cs@gmail.com} 
           \and
           Maocheng Li \at
            Hong Kong University of Science and Technology\\
            \email{csmichael@cse.ust.hk}
            \and 
           Yuxiang Zeng \at
              Beihang University\\
              \email{yxzeng@buaa.edu.cn}
          \and
          Yanyan Shen \at
              Shanghai Jiao Tong University\\
              \email{shenyy@sjtu.edu.cn}
          \and
          Lei Chen \at
              Hong Kong University of Science and Technology \& Hong Kong University of Science and Technology (GZ)\\
              \email{leichen@ust.hk}
}

\date{Received: date / Accepted: date}

\maketitle

\begin{abstract}
Efficient indexing is fundamental for multi-dimensional data management and analytics. 
An emerging tendency is to directly learn the storage layout of multi-dimensional data by simple machine learning models, yielding the concept of \emph{Learned Index}. 
Compared with the conventional indices used for decades (e.g., $k$d-tree and R-tree variants), learned indices are empirically shown to be both space- and time-efficient on modern architectures. 
However, there lacks a comprehensive evaluation of existing multi-dimensional learned indices under a unified benchmark, which makes it difficult to decide the suitable index for specific data and queries and further prevents the deployment of learned indices in real application scenarios. 
In this paper, we present the first in-depth empirical study to answer the question of \emph{how good multi-dimensional learned indices are}. 
Six recently published indices are evaluated under a unified experimental configuration including index implementation, datasets, query workloads, and evaluation metrics. 
We thoroughly investigate the evaluation results and discuss the findings that may provide insights for future learned index design. 
\keywords{Learned Index \and Spatial Index \and Benchmark \and Experimental Study}
\end{abstract}

\section{Introduction}\label{sec:introduction}
Multi-dimensional data management and analytics play an important role in various domains such as business intelligence~\cite{chen2012business}, smart transportation~\cite{DBLP:journals/tits/ZhuYWNT19}, neural science~\cite{sejnowski2014putting}, climate studies~\cite{faghmous2014spatio}, etc. 
As the data volume grows at an exponential speed, conventional multi-dimensional indices like R-tree~\cite{DBLP:conf/sigmod/Guttman84} and its variants~\cite{DBLP:conf/sigmod/BeckmannKSS90,DBLP:conf/vldb/KamelF94,DBLP:journals/talg/ArgeBHY08,DBLP:conf/vldb/PapadiasTS03} are designed to speedup data access and query processing over big multi-dimensional databases. 

\begin{figure*}[t]
    \centering
    \includegraphics[width=.6\textwidth]{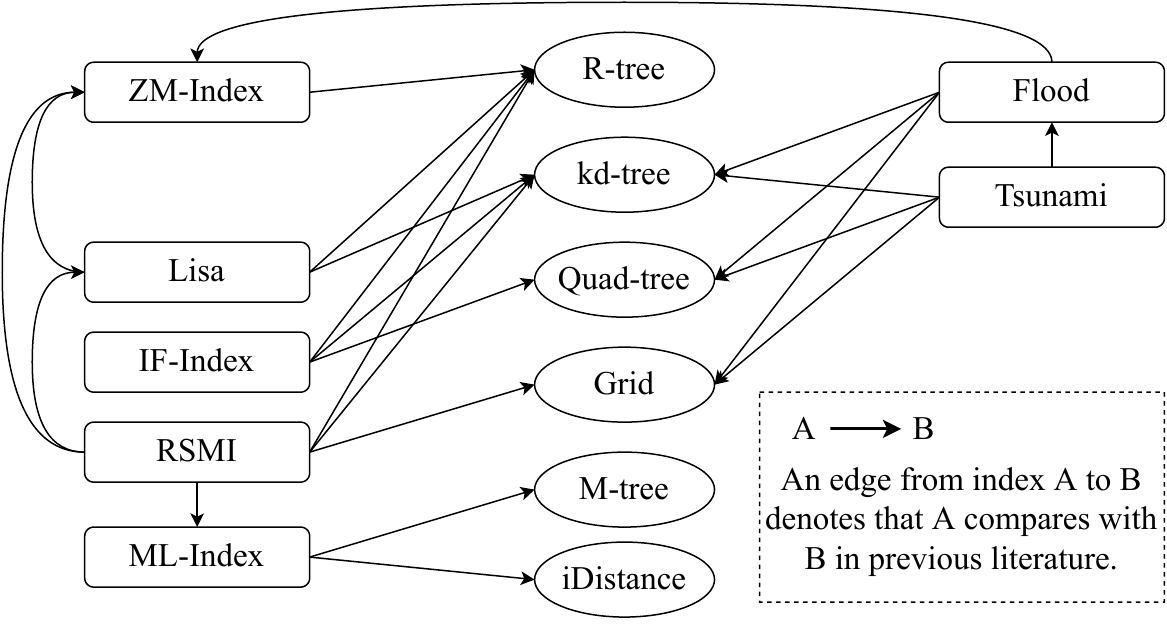}
    \caption{Illustration of limited comparison in previous studies.  }
    \label{fig:compare}
\end{figure*}

Although traditional index structures like B+-tree and R-tree have been studied and embedded into practical DBMS for decades (e.g., Oracle~\cite{DBLP:conf/sigmod/KanthRA02} and PostgreSQL~\cite{postgresql}), a recent proposal~\cite{DBLP:conf/sigmod/KraskaBCDP18} introduced a new index design paradigm called \emph{Learned Index} based on the observation that data indexing can be modeled as a machine learning problem where the input is a search key and the output is its corresponding location onto the storage. 
Supposing that a set of $N$ keys are sorted and stored as consecutive data pages, a B+-tree can be viewed as a mapping from key $x$ to its page ID. 
Thus, an error-bounded Cumulative Distribution Function (CDF) model is functionally equivalent to a B+-tree index. 
Compared with traditional indexes, learned index is supposed to be both space- and time-efficient as a trained model (e.g., piece-wise linear function) is usually compact and simple for inference. 


Inspired by the impressive results obtained from 1-dimensional learned index~\cite{DBLP:conf/sigmod/KraskaBCDP18,DBLP:journals/pvldb/MarcusKRSMK0K20,DBLP:journals/pvldb/FerraginaV20,DBLP:journals/pvldb/WuZCCWX21,DBLP:journals/pvldb/SunZL23}, learned indices on multi-dimensional data are intensively studied during the past years such as ZM-Index~\cite{DBLP:conf/mdm/WangFX019},  ML-Index~\cite{DBLP:conf/edbt/DavitkovaM020}, IF-Index~\cite{DBLP:conf/vldb/0001KH20}, RSMI~\cite{DBLP:journals/pvldb/QiLJK20}, LISA~\cite{DBLP:conf/sigmod/Li0ZY020}, Flood \cite{DBLP:conf/sigmod/NathanDAK20}, and Tsunami~\cite{DBLP:journals/pvldb/DingNAK20}. 
These works independently claim that they are empirically more performant than traditional spatial indices like R-tree or $k$d-tree. 
However, to the best of our knowledge, there lacks a comprehensive evaluation for published multi-dimensional learned indices under a unified experimental configuration, which obscures the impact and future direction of this prominent research field. 
The limitations of existing experiments and evaluations are summarized as follows. 

First, the newly proposed indices lack enough comparisons with previous studies. 
Figure~\ref{fig:compare} visualizes the comparison relationship of existing works where a node refers to an index and an edge $A\to B$ refers to that index $A$ compares with $B$ in previous literature. 
From Figure~\ref{fig:compare}, most of the previous works only compare with ZM-Index~\cite{DBLP:conf/mdm/WangFX019} that combines the space-filling curves and 1-dimensional learned index. 
However, the original ZM-Index implementation is less optimized (see Section~\ref{subsec:exp_range_time} for details), meaning that a weak baseline was most frequently compared. 
Besides, existing comprehensive benchmarks~\cite{DBLP:journals/pvldb/MarcusKRSMK0K20,DBLP:journals/pvldb/SunZL23} evaluate 1-D learned indices only, which cannot be reused for multi-dimensional cases. 




Second, the existing learned indices are not compared under a unified configuration including index implementation, datasets, query workloads, and evaluation metrics. 
For example, a class of multi-dimensional learned indices~\cite{DBLP:conf/mdm/WangFX019,DBLP:conf/cidr/KraskaABCKLMMN19,DBLP:conf/edbt/DavitkovaM020,DBLP:conf/sigmod/NathanDAK20,DBLP:journals/pvldb/DingNAK20} utilize 1-D learned index (e.g., RMI~\cite{DBLP:conf/sigmod/KraskaBCDP18}) as building blocks, but different implementations are used in different indices, leading to an unfair comparison and unconvincing results on their true performance.

Unlike the experiments in previous studies where only the ZM-Index~\cite{DBLP:conf/mdm/WangFX019} was compared, in this work, we \textbf{re-implement} and \textbf{optimize} six multi-dimensional learned indices to perform a comprehensive evaluation, which is a nearly complete coverage to the best of our knowledge. 
To make the comparison fair and the evaluation results more convincing, we standardize the experiment configurations, including the index implementation, datasets, and evaluation query workloads. 
This also benefits future research as newly proposed multi-dimensional learned indices can be easily evaluated on our benchmark. 
In addition, though IO-efficiency and performance in distributed DBMS are also important, we focus on the \textbf{in-memory} and \textbf{single-machine}\footnote{Although there exist multi-dimensional learned indices such as LISA~\cite{DBLP:conf/sigmod/Li0ZY020} and RSMI~\cite{DBLP:journals/pvldb/QiLJK20} claiming that they are disk-based indices, unfortunately, they do not adopt any disk-based optimization objectives (e.g., page access or IO cost). 
In these works, the index structures are loaded in the memory, but data pages are stored on the disk, which is intrinsically the same as other memory-based learned indices. } performance of indices over both synthetic and real query workloads, which is similar to a recent 1-D learned index benchmark SOSD~\cite{DBLP:journals/pvldb/MarcusKRSMK0K20}. 
Though losing generality to some extent, our empirical study can provide insightful results for in-memory analytical applications, which are becoming increasingly prominent~\cite{DBLP:conf/gis/YuWS15,DBLP:conf/sigmod/XieL0LZG16,DBLP:journals/pvldb/TangYMOA16}.

In summary, our experimental study makes the following contributions.
\begin{itemize}[leftmargin=*]
    \item To deliver a fair comparison among existing multi-dimensional learned indices, we unify the index implementation, model training process, query workloads, and evaluation pipeline. 
    Our benchmark implementation is fully open-sourced\footnote{\url{https://github.com/qyliu-hkust/learnedbench}} to benefit future studies in this prominent direction. 
    \item To the best of our knowledge, our work is the first \textbf{comprehensive} and \textbf{in-depth} evaluation for multi-dimensional learned indices under a uniform experimental environment. 
    \item By thoroughly analyzing the evaluation results, we give the answer to the vital question concerned by practical DBMS users, i.e., ``\textit{How good are multi-dimensional learned indices?}'' In addition, based on the key experimental findings, we also identify the potential research opportunities for future studies. 
\end{itemize}



The rest of this paper is structured as follows. 
We review the background of learned index and formulate the multi-dimensional data model and corresponding queries in Section~\ref{sec:preliminaries}. 
We establish a taxonomy and provide an overview of the existing multi-dimensional learned indices in Section~\ref{sec:index}. 
Section~\ref{sec:exp_setup} introduces the index implementation details and experimental setups. 
Section~\ref{sec:exp_results} presents the evaluation results and in-depth discussions. 
Finally, we conclude the paper and discuss future directions in Section~\ref{sec:conclusion}. 

\section{Preliminaries and Background}\label{sec:preliminaries}
In this section, we overview the backgrounds of multi-dimensional data indexing and learned index structures. 

\subsection{Multi-dimensional Data Indexing}
As adopted by most of the existing works, we consider a collection of $N$ points from $d$-dimensional Euclidean space, and we focus on the range query and $k$ nearest neighbor query ($k$NN) defined as follows. 

\begin{definition}[Point]
    A point $o\in\mathbb{R}^d$ is a vector in $d$-dimensional Euclidean space. Let $o_i$ denote the $i$-th coordinate of $o$ and $\mathcal{O}$ denote the collection of $N$ points. 
\end{definition}

\begin{definition}[Range Query]\label{def:range_query}
    Given a collection of points $\mathcal{O}$, a range query takes a $d$-dimensional hyper-rectangle $R$ as input and returns all the points lying in $R$, i.e., $range(R)=\{o|o\in R, o\in\mathcal{O}\}$.
\end{definition}



\begin{definition}[$k$-Nearest Neighbor Query]
    A $k$-nearest neighbor query $kNN(q, k)$ retrieves $k$ objects from $\mathcal{O}$ whose distances to $q$ are ranked in ascending order. Formally, for $\forall o\in kNN(q, k)$, $\nexists o'\in \mathcal{O}\setminus kNN(q,k)$ such that $||o',q||_2\leq ||o,q||_2$.
\end{definition}



Indexing and corresponding query processing over multi-dimensional data have been studied and applied in commercial DBMS for decades. 
For low-dimensional data, conventional indices include R-tree~\cite{DBLP:conf/sigmod/Guttman84} and its variants like R*-tree~\cite{DBLP:conf/sigmod/BeckmannKSS90}, STR-tree~\cite{DBLP:conf/icde/LeuteneggerEL97} and Hilbert R-tree~\cite{DBLP:conf/vldb/KamelF94}, $k$d-tree~\cite{DBLP:journals/cacm/Bentley75}, Grid File~\cite{DBLP:journals/tods/NievergeltHS84}, etc. 
On the other hand, the data space is becoming inevitably sparse for high-dimensional space due to the curse of dimensionality, and query processing like $k$NN query based on the aforementioned indices will be no better than a na\"ive linear scan. 
To this end, pivot-based methods are generally adopted to index high-dimensional data, e.g., iDistance~\cite{DBLP:journals/tods/JagadishOTYZ05}, vantage-point (VP) tree~\cite{yianilos1993data}, MVP-tree~\cite{bozkaya1999indexing}, etc. 
A recent proposal~\cite{chen2017pivot} surveyed and evaluated the family of pivot-based indices on high-dimensional query processing. 

In this work, we focus on evaluating indices over \textit{multi-dimensional points} instead of complicated geometries like lines, rectangles, or polygons. 
This is because all of our evaluated learned indices do not support these non-point geometries. 
Moreover, it is non-trivial to make such extensions by following the extended routines \cite{DBLP:books/daglib/0014978} of conventional indexes (e.g., R-trees) due to two reasons. 
First, existing learned indexes have assumed that the object is a spatial point
and utilized techniques (e.g., space-filling curve) that heavily rely on this assumption.
Second, these complicated geometries involve more complicated spatial predicates, such as \texttt{intersect} and \texttt{touch}~\cite{boost-geometry}, 
which are currently not supported by existing learned indices.

\subsection{Learned Data Indexing}
\noindent\underline{\textbf{1-Dimensional Learned Index.} }
A 1-D learned index is intrinsically an error-bounded CDF model (scaled by the data size $N$). 
In the seminal work~\cite{DBLP:conf/sigmod/KraskaBCDP18}, Kraska et~al.~proposed the first learned index RMI that is empirically shown to be Pareto optimal compared with a B+-tree index. 
Following RMI, Kip et~al.~proposed a simple learned index called RadixSpline~\cite{DBLP:conf/sigmod/KipfMRSKK020} that requires only a single pass of data to construct. 
Furthermore, PGM-Index~\cite{DBLP:journals/pvldb/FerraginaV20} adopted the optimal piece-wise linear approximation~\cite{o1981line} as the underlying CDF model, leading to strong theoretical results on the space and time complexity. 
To handle dynamic operations like key insertion and deletion, Ding et~al.~proposed ALEX~\cite{DBLP:conf/sigmod/DingMYWDLZCGKLK20} by using a gapped array in their structure to handle record updates; LIPP~\cite{DBLP:journals/pvldb/WuZCCWX21} further improved the update efficiency by reducing the last-mile search error in leaf nodes. 
More discussions and comparisons about 1-D learned indices can be found in a benchmark paper SOSD~\cite{DBLP:journals/pvldb/MarcusKRSMK0K20}. 
Besides data indexing, there are emerging attempts of embedding learned models into conventional data structure and algorithm design, e.g., learned Bloom filters~\cite{DBLP:conf/sigmod/KraskaBCDP18,DBLP:conf/nips/Mitzenmacher18,DBLP:journals/pvldb/LiuZSC20}, learned sorting~\cite{DBLP:conf/sigmod/KristoVCMK20}, learned data compression~\cite{DBLP:conf/alenex/BoffaFV21,liu2022hap}, etc.

\begin{table*}
    \centering
    \footnotesize
    \caption{Overview of the existing multi-dimensional learned index structures.}
    \label{tab:index_summary}
    \begin{tabular}{|c|c|c|c|c|c|c|c|}
    \hline
    Index & {ZM-Index}  & {ML-Index} & {LISA} & {IF-Index} & {RSMI}  & {Flood} & {Tsunami} \\\hline
    Reference & \cite{DBLP:conf/mdm/WangFX019} &  \cite{DBLP:conf/edbt/DavitkovaM020} & \cite{DBLP:conf/sigmod/Li0ZY020} &  \cite{DBLP:conf/vldb/0001KH20} & \cite{DBLP:journals/pvldb/QiLJK20}  & \cite{DBLP:conf/sigmod/NathanDAK20} & \cite{DBLP:journals/pvldb/DingNAK20} \\\hline
    Type & projection & projection & projection & augmentation & augmentation & grid & grid \\\hline
    Data Ordering & Z-curve & \begin{tabular}[c]{@{}c@{}}projection\\function\end{tabular} & \begin{tabular}[c]{@{}c@{}}projection\\function\end{tabular} & \begin{tabular}[c]{@{}c@{}}selected\\dimension\end{tabular} & Z-curve & \begin{tabular}[c]{@{}c@{}}selected\\dimension\end{tabular} & \begin{tabular}[c]{@{}c@{}}selected\\dimension\end{tabular}\\\hline
    Data Layout & order-based & order-based & order-based & \begin{tabular}[c]{@{}c@{}}space\\partition\end{tabular} & \begin{tabular}[c]{@{}c@{}}space\\partition\end{tabular} & grid & grid \\\hline
    Model & RMI & RMI & \begin{tabular}[c]{@{}c@{}}piece-wise\\ linear\end{tabular} & \begin{tabular}[c]{@{}c@{}}linear\\interpolation\end{tabular}  & MLP & RMI & RMI \\\hline
    Model Training & algorithmic & algorithmic & numpy & algorithmic & PyTorch & algorithmic & algorithmic \\\hline
    Updatable & \ding{53} & \ding{53} & $\checkmark$ & $\checkmark$ & $\checkmark$ & \ding{53} & \ding{53}\\\hline
    Support Dim. & $\geq2$ & $\geq2$ & $\geq2$ & $\geq2$ & 2 & $\geq2$ & $\geq2$ \\\hline
    Range Query & $\checkmark$ & $\checkmark$ & $\checkmark$ & $\checkmark$ & $\checkmark$ & $\checkmark$ & $\checkmark$\\\hline
    $k$NN Query & \ding{53} & $\checkmark$ & $\checkmark$ & \ding{53} & $\checkmark$ & \ding{53} & \ding{53} \\\hline
    \end{tabular}
\end{table*}

\noindent\underline{\textbf{Multi-Dimensional Learned Index.}}
Similar to how an R-tree is a multi-dimensional analog to a B+-tree index, it is natural to extend the 1-dimensional learned index to multi-dimensional datasets by directly learning the mapping from multi-dimensional keys to their storage location. 
Designing multi-dimensional learned indices has rapidly become a promising research direction in the past years. 
Typical works in this area include  ZM-Index~\cite{DBLP:conf/mdm/WangFX019},  ML-Index~\cite{DBLP:conf/edbt/DavitkovaM020}, IF-Index~\cite{DBLP:conf/vldb/0001KH20}, RSMI~\cite{DBLP:journals/pvldb/QiLJK20}, Lisa~\cite{DBLP:conf/sigmod/Li0ZY020}, Flood~\cite{DBLP:conf/sigmod/NathanDAK20}, and Tsunami~\cite{DBLP:journals/pvldb/DingNAK20}. 
In addition, a recent system SageDB~\cite{DBLP:conf/cidr/KraskaABCKLMMN19} also incorporates a learned grid index, which can be viewed as a simplified version of Flood~\cite{DBLP:conf/sigmod/NathanDAK20}. 
The details of these learned indices will be discussed in Section~\ref{sec:index}. 
As we will reveal in this work later, existing multi-dimensional learned indices mainly focus on read-only and single-machine workloads.
In addition, the query support of these learned indices is still preliminary (majorly range query and $k$NN query), remaining a tremendous number of research opportunities (see discussions in Section~\ref{sec:exp_results} and Section~\ref{sec:conclusion}). 


\section{Multi-dimensional Learned Index}\label{sec:index}
In this section, we first investigate the data layouts adopted by different indices in Section~\ref{subsec:data_layout}. 
Then, we provide an index taxonomy in Section~\ref{subsec:taxonomy} and overview indices in each category in Section~\ref{subsec:projection_index}--\ref{subsec:grid_index}. 
Table~\ref{tab:index_summary} summarizes the major technical features of existing works.

\subsection{Data Layout}\label{subsec:data_layout}
We first discuss the multi-dimensional data layout, based on which we establish our taxonomy for existing multi-dimensional learned index structures. 

Data layout specifies how an index organizes the data points onto the storage (i.e., disk or memory), which plays an important role in multi-dimensional index design. 
As shown in Table~\ref{tab:index_summary}, we identify that the existing learned indices usually employ one of the three types of data layouts: \ding{182} order-based layout, \ding{183} space partition-based layout, and \ding{184} grid-based layout.

\noindent\underline{\textbf{Order-based Layout.}} 
Multi-dimensional data points are organized on storage consistent with a pre-defined sorting order. 
Different from the 1-D case, there is no intrinsic sorting order for multi-dimensional data, and existing works usually select a sorting dimension or employ the space-filling curves (e.g., Z-order curve~\cite{DBLP:journals/tods/QiTCZ20} or Hilbert curve~\cite{DBLP:conf/vldb/KamelF94}) to order data. 

\noindent\underline{\textbf{Partition-based Layout.}} 
The space partition-based layout recursively divides the data space under some strategy (e.g., the middle-point strategy in $k$d-tree~\cite{DBLP:journals/cacm/Bentley75}) until a partition threshold is reached. 
In this case, data points within the same partition are grouped together and sequentially materialized to the storage. 

\noindent\underline{\textbf{Grid-based Layout.}} 
The grid-based layout can be regarded as a special case of the partition-based layout. 
The data space is partitioned into grid cells, and points falling into the same grid cell are placed in storage consecutively.


\subsection{Taxonomy}\label{subsec:taxonomy}
Based on different data layouts and how learned CDF models are integrated, we classify the existing multi-dimensional learned index studies into three categories: \ding{182} projection-based index, \ding{183} augmentation-based index, and \ding{184} grid-based index. 

\noindent\underline{\textbf{Projection-based Index}} adopts a projection function to map $k$-dimensional keys to 1-D values, and then train a 1-D learned index (e.g., RMI~\cite{DBLP:conf/sigmod/KraskaBCDP18} or PGM-Index~\cite{DBLP:journals/pvldb/FerraginaV20}) over the set of mapped values. 
To preserve the spatial locality, existing studies usually employ space-filling curves like Z-order or Hilbert curves as the choice for projection functions. 
Such an idea is not new and has been commonly adopted in conventional spatial index design (e.g., UB-tree~\cite{DBLP:conf/vldb/RamsakMFZEB00} and Hilbert R-tree~\cite{DBLP:conf/vldb/KamelF94}). 
The projection-based indices include SageDB~\cite{DBLP:conf/cidr/KraskaABCKLMMN19}, LISA~\cite{DBLP:conf/sigmod/Li0ZY020}, ZM-Index~\cite{DBLP:conf/mdm/WangFX019}, and ML-Index~\cite{DBLP:conf/edbt/DavitkovaM020}.

\noindent\underline{\textbf{Augmentation-based Index}} extends the traditional index structures that are based on recursive space partitioning but augment the ordinary node search with model-based search. 
These indices employ similar space partition and node split strategies from the existing multi-dimensional indices like R-tree and $k$d-tree, thus inheriting their high generality and wide applications. 
The augmentation-based indices include IF-Index \cite{DBLP:conf/vldb/0001KH20} and RSMI~\cite{DBLP:journals/pvldb/QiLJK20}. 

\noindent\underline{\textbf{Grid-based Index}} employs grids as the data layout. 
Different from the ordinary grid indices, the learned grid index does not physically store grid cells; instead, learned CDF functions over a set of selected dimensions are trained to locate the correct grid cell. 
Suppose that the $j$-th dimension in the grid is partitioned into $m$ buckets, a point $o$ will be placed in the $\lfloor CDF_j(o_j)\cdot m\rfloor$-th bucket. 
The grid-based indices include Flood~\cite{DBLP:conf/sigmod/NathanDAK20} and  Tsunami~\cite{DBLP:journals/pvldb/DingNAK20}.

Recent studies like Qd-tree~\cite{DBLP:conf/sigmod/YangCWGLMLKA20} and RLR-tree~\cite{DBLP:journals/corr/abs-2103-04541} also leverage machine learning techniques, especially deep reinforcement learning (DRL), to construct multi-dimensional data index. 
These DRL-based index design studies focus on finding a more reasonable data partition, whereas their indexing layer is still the traditional index like R-tree. The DRL techniques are used only for finding better data layout, not for locating data records on the storage. Thus, they are out of the scope of learned index and not evaluated in our work. 

\begin{figure}[t]
    \centering
    \includegraphics[width=0.45\textwidth]{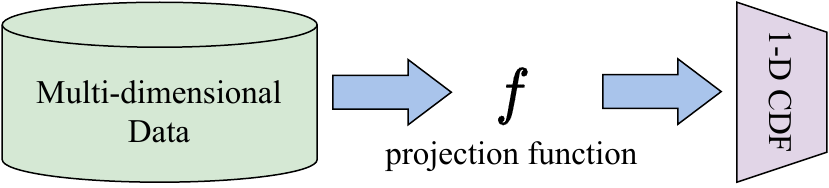}
    \caption{Workflow of projection-based indices.}
    \label{fig:projection}
\end{figure}

\subsection{Projection-based Index}\label{subsec:projection_index}
Figure~\ref{fig:projection} shows the basic workflow of projection-based indices. 
For a $d$-dimensional dataset $\mathcal{O}$, a projection function $f: \mathbb{R}^d\to\mathbb{R}$ is applied to convert $\mathcal{O}$ to a set of 1-D values $\mathcal{O}'$. 
Then, $\mathcal{O}'$ is sorted and stored in consecutive data pages, and a 1-D learned index (e.g., PGMIndex) is built to serve as the \emph{approximate} mapping to storage location (i.e., block ID or index in a dense sorted array). 
Similar to the 1-D learned indices, to ensure the correctness of point search, the projection function $f$ should be also a monotonic mapping, i.e., for two points $o$ and $o'$ where $o'$ dominates $o$ on each dimension, $f(o)\leq f(o')$. 
Then, any range search query can be transformed to 1-D interval search over the mapped values. 
The major difference of existing works in this class is the choice of projection function $f$. 

\noindent\underline{\textbf{ZM-Index}}~\cite{DBLP:conf/mdm/WangFX019} is the first multi-dimensional learned index where the Z-order curve is chosen as the projection function. 
To efficiently compute the Z-addresses, the data space should be partitioned into grids such that the bit interleaving technique can be used~\cite{DBLP:conf/vldb/RamsakMFZEB00}. 
To process a range query, the query box is first decomposed into intervals of Z-addresses using the same technique in the UB-tree~\cite{DBLP:conf/vldb/RamsakMFZEB00}, and then the trained CDF model is queried to efficiently find the corresponding storage location. 




\begin{figure}[t]
    \centering
    \includegraphics[width=0.45\textwidth]{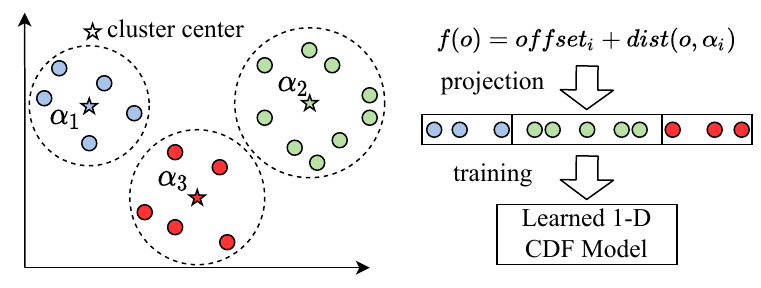}
        \caption{Illustration of the projection function of ML-Index~\cite{DBLP:conf/edbt/DavitkovaM020} where the $k$-means centers are used as reference points. }    
        \label{fig:mlindex_map}
\end{figure}

\noindent\underline{\textbf{ML-Index}}~\cite{DBLP:conf/edbt/DavitkovaM020} employs an improved iDistance function, which is usually used to index high dimensional data for efficient nearest neighbour searches~\cite{DBLP:journals/tods/JagadishOTYZ05}, to project the multi-dimensional data. 
As shown in Figure~\ref{fig:mlindex_map}, given a set of selected reference points (RP) $\alpha_1,\cdots,\alpha_m$ (e.g., obtained using the $k$-means algorithm), the input data $\mathcal{O}$ are partitioned into $m$ partitions based on the distance to each RP. 
For any point $o\in\mathcal{O}$, supposing that $\alpha_i$ is the closest RP to $o$, the ML-Index adopts the following projection function,
\begin{equation}\label{eq:mlindex_projection}
    \begin{aligned}
        f(o) =& offset_i + dist(o, \alpha_i), \\
        offset_i =& \sum_{j<i} \max_{o'\in O_j} dist(o', \alpha_j),
    \end{aligned}
\end{equation}
where $O_j=\{o|\alpha_j \text{ is the closest RP to } o, o\in\mathcal{O}\}$. 

Compared with the original iDistance method, i.e., $iDist(o)=i\cdot C + dist(o,\alpha_i)$ where $C$ is a large constant, 
Eq.~\eqref{eq:mlindex_projection} eliminates the overlap between different data partitions and reduces the gaps between consecutive partitions, making the learning of a CDF model on the mapped keys much easier. 
The query processing (range queries or $k$NN queries) on ML-Index is similar to the iDistance~\cite{DBLP:journals/tods/JagadishOTYZ05} method, where a B+-tree is constructed to maintain the iDistance values. 
Since the computation of the projection function in Eq.~\eqref{eq:mlindex_projection} only requires a valid distance metric $dist(\cdot,\cdot)$, the ML-Index is also available on data from general metric spaces (e.g., strings and graphs). 
However, as all the other learned indices do not support metric space indexing, we focus on the performance of ML-Index in Euclidean space and leave the extension to general metric space as an interesting future work.

\begin{figure}[t]
    \centering
    \includegraphics[width=0.45\textwidth]{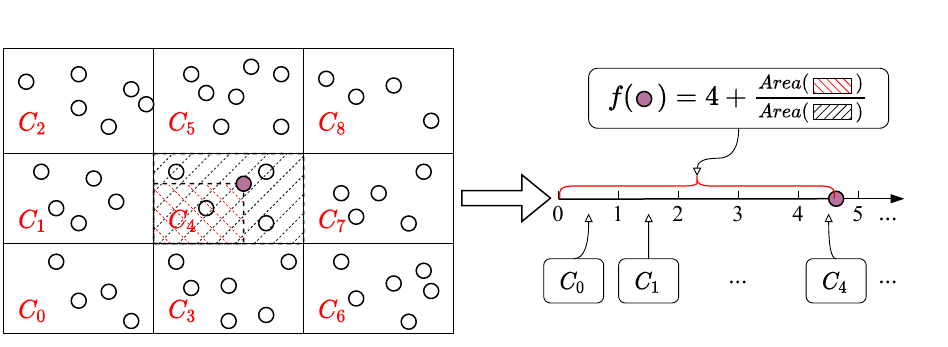}
        \caption{Illustration of the projection function of LISA~\cite{DBLP:conf/sigmod/Li0ZY020} based on a $3\times3$ grid partition. Note, the Lebesgue measure in 2-D space is the area of a rectangular region.}
        \label{fig:lisa_map}
\end{figure}

\noindent\underline{\textbf{LISA}}~\cite{DBLP:conf/sigmod/Li0ZY020} majorly attacks that the space-filling curve-based projection usually accesses data blocks that are irrelevant to the query rectangle. 
To solve this issue, LISA employs a grid-based projection function. 
For a $d$-dimensional dataset, data points are first partitioned into $T_1\times T_2\cdots\times T_d$ equal-depth grid cells, and each cell $C$ is associated with a unique ID $t$, i.e., $C_t=[\theta_l^{(1)},\theta_h^{(1)})\times\cdots\times[\theta_l^{(d)},\theta_h^{(d)})$ and $t = (((i_1\times T_2 + i_2)\times T_2 + i_3)\times\cdots)\times T_d + i_d$ where $\theta_l^{(j)},\theta_h^{(j)}$ are the grid cell boundaries on the $j$-th dimension and $i_j$ is the partition ID along the $j$-th dimension ($j\in 1,\cdots,d$). 
Then, for an arbitrary point $o$ falling into $C_t$, the projection function is defined as follows,
\begin{equation}\label{eq:lisa_projection}
    \begin{aligned}
        f(o) =& t + \frac{\lambda(H_t)}{\lambda(C_t)}, \\
        H_t =& [\theta_l^{(1)},o^{(1)})\times\cdots\times[\theta_l^{(d)},o^{(d)}),
    \end{aligned}
\end{equation}
where $\lambda(\cdot)$ is the Lebesgue measure (i.e., area in 2-D space). 
The projection function of LISA is illustrated in Figure~\ref{fig:lisa_map} where the areas of red and black dashed regions are the Lebesgue measures of $C_4$ and $H_4$ for the point in purple.
Intuitively, Eq.~\eqref{eq:lisa_projection} is also similar to the iDistance method~\cite{DBLP:journals/tods/JagadishOTYZ05} where grid cells can be viewed as reference points. 
Clearly, points falling into cell $C_t$ will be mapped to the same interval $[t,t+1)$, which preserves the spatial locality. 

After the grid construction, all the points are projected to 1-D space by using Eq.~\eqref{eq:lisa_projection} and partitioned to shards, and a model called \emph{shard prediction function} (SP), which should be a monotonic function, is trained to map each point to their shard ID. 
Finally, points belonging to the same shard are stored in data pages, and a local model (i.e., 1-D learned index) is trained to locate the correct data page. 

Note that, although a grid partition is required, instead of physically storing a grid index, LISA encodes all the grid information into a projection function (i.e., Eq.~\eqref{eq:lisa_projection}) and orders data based on the projected values. 
Thus, in Table~\ref{tab:index_summary}, we classify LISA into the projection-based index, instead of the grid-based index.


\begin{figure}
    \centering
    \includegraphics[width=0.45\textwidth]{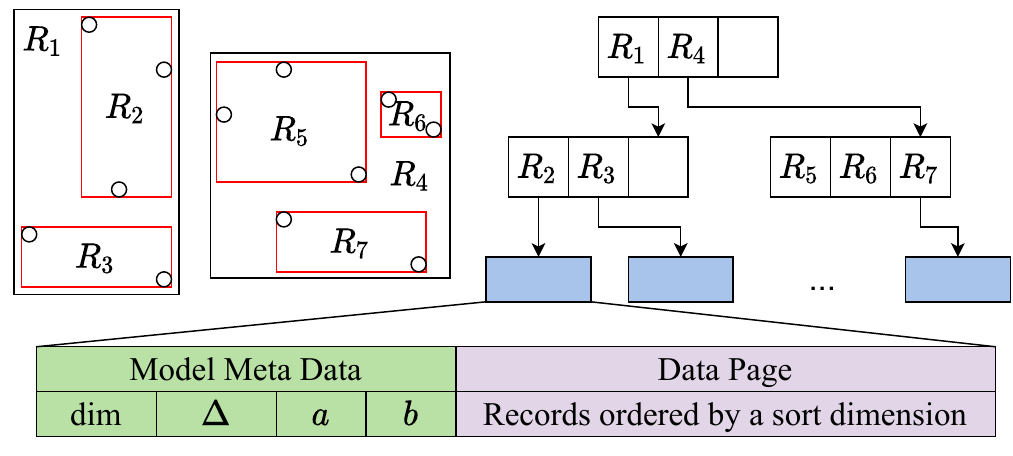}
    \caption{Illustration of IF-Index~\cite{DBLP:conf/vldb/0001KH20} where dim is the selected sorting dimension, $\Delta$ is the maximum prediction error, and $a$, $b$ are the slope and interception of the linear model.}
    \label{fig:handsoff}
\end{figure}

\begin{figure}
    \centering
    \includegraphics[width=0.47\textwidth]{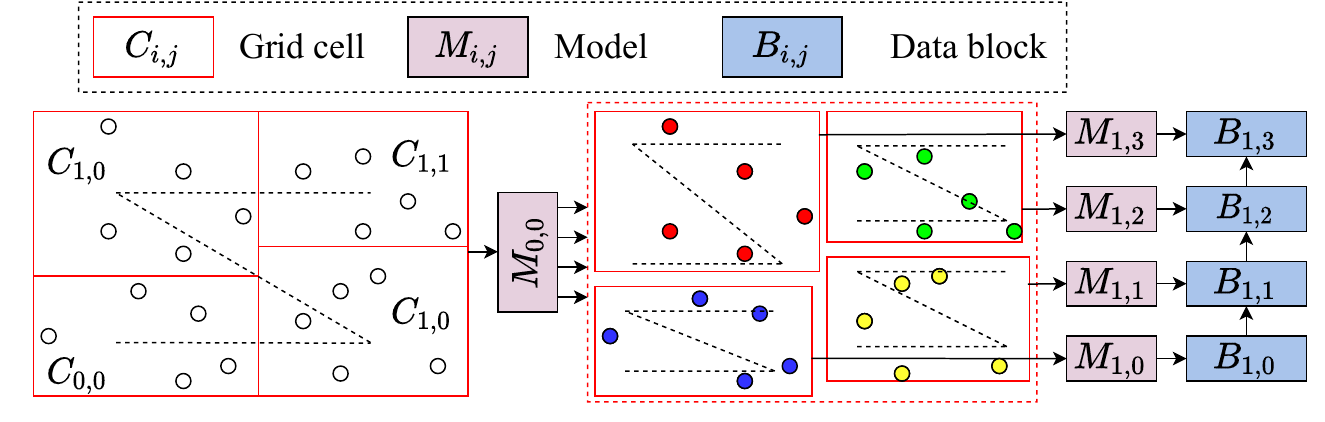}
        \caption{Illustration of RSMI~\cite{DBLP:journals/pvldb/QiLJK20} where the partition threshold $N'=5$. Note that each data block maintains a pointer to the next block for efficient scanning.}
        \label{fig:rsmi}   
\end{figure}

\subsection{Augmentation-based Index}\label{subsec:augmentation_index}
We then introduce the augmentation-based indices, where learned models are plugged into conventional index structures (e.g., R-tree or $k$d-tree) to accelerate the search efficiency.

\noindent\underline{\textbf{IF-Index}}~\cite{DBLP:conf/edbt/DavitkovaM020} replaces the leaf node search procedure in existing tree structures (e.g., R-tree) with 1-D learned index-based search. The structure of IF-Index based on the R-tree is illustrated in Figure~\ref{fig:handsoff}. 
In this structure, the non-leaf nodes are constructed the same way as an ordinary R-tree. 
However, in the leaf nodes, not only the corresponding data page is stored, but also the model metadata. 
The model metadata contains the dimension $dim$, which is used to order the points on the data page, as well as a linear interpolation model for predicting the location of the search key on the data page. 
In order to determine the best sorting dimension, IF-Index evaluates the interpolation cost for each dimension on every leaf node, resulting in a time complexity of $O(dN'\log N')$, where $d$ represents the number of dimensions and $N'$ denotes the number of points contained within the leaf node.

\noindent\underline{\textbf{RSMI}}~\cite{DBLP:journals/pvldb/QiLJK20} takes a further step by using the model-based search in both leaf and non-leaf nodes. 
As shown in Figure~\ref{fig:rsmi}, to construct the RSMI index, spatial points are first partitioned into $2^{\lfloor \log_4 N'/B \rfloor}\times 2^{\lfloor \log_4 N'/B \rfloor}$ equal-depth grid cells, where $B$ is the block size and $N'$ is the partition threshold parameter. 
In practice, $N'$ is typically assumed to be the number of points for which a learned function can achieve high accuracy. 
Then, grid cells are ordered using space-filling curves, and a learned model is trained to map each grid cell to its sorting order (e.g., $M_{0,0}$ in Figure~\ref{fig:rsmi}). 
As each grid cell can still contain many points (i.e., $\gg N'$), the above data partitioning and model training procedures are recursively invoked until each partition has at most $N'$ points. 
After completing the partitioning, the data points are packed into data blocks based on the order of rank space Z-order curve, and finally, leaf models are trained to predict the corresponding block ID for each point (e.g., $M_{1,0}\sim M_{1,3}$ in Figure~\ref{fig:rsmi}). 

Similar to a quad-tree, RSMI also employs grids to recursively partition the space, leading to a grid tree structure. 
The learned models in RSMI are used to perform faster searches in each level of the grid tree, instead of directly locating the grid cells. Thus, RSMI is classified as the augmentation-based index, instead of the grid-based index.

\begin{figure}[t]
    \centering
    \includegraphics[width=0.47\textwidth]{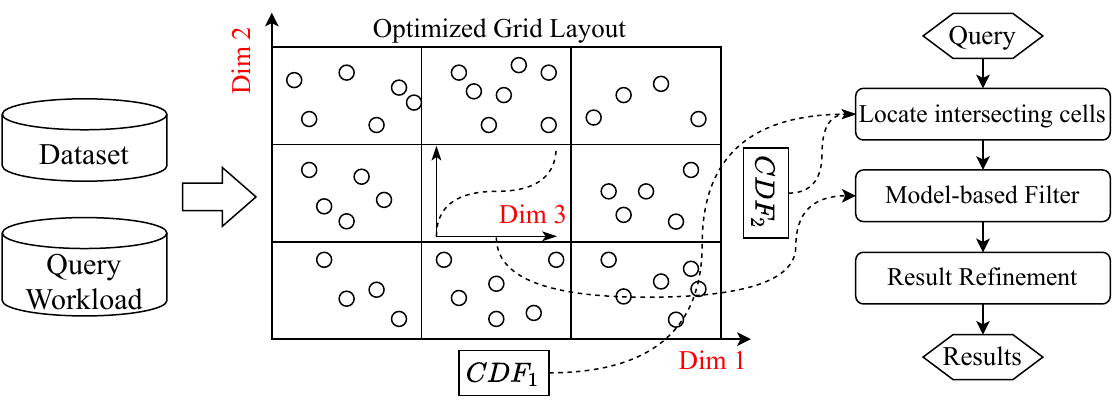}
        \caption{Illustration of the Flood~\cite{DBLP:conf/sigmod/NathanDAK20} index framework on a 3-dimensional dataset, where (1) Dim1 and Dim2 are used to generate a grid layout, and (2) Dim3 is used to sort the data within each grid cell.}
        \label{fig:flood}
\end{figure}

\subsection{Grid-based Index}\label{subsec:grid_index}
The grid-based indices Flood~\cite{DBLP:conf/sigmod/NathanDAK20} and Tsunami~\cite{DBLP:journals/pvldb/DingNAK20} target on learning compact multi-dimensional grids to efficiently process orthogonal range predicates. 

\noindent\underline{\textbf{Flood}}~\cite{DBLP:conf/sigmod/NathanDAK20} adopts the common grid index as the data layout. 
To construct Flood on a $d$-dimensional dataset, a sorting dimension is first chosen for ordering the data within each grid cell, and the remaining $d-1$ dimensions are adopted to overlay a grid. 
Assume, w.l.o.g., that the $d$-th dimension is the sorting dimension. 
Different from an ordinary grid index, Flood uses learned CDF models (i.e., 1-D learned index) to construct grid partitions. 
Specifically, a Flood grid is typically a multi-dimensional array $G$ of cells, and the corresponding cell of an arbitrary point $o$ in $G$ is 
\begin{multline}\label{eq:flood_cell}
    G(o) = (\left\lfloor cdf_1(o_1)\cdot K_1 \right\rfloor,\left\lfloor cdf_2(o_1)\cdot K_2 \right\rfloor, \cdots, \\
    \left\lfloor cdf_{d-1}(o_{d-1})\cdot K_{d-1}\right\rfloor ),
\end{multline}
where $cdf_i(\cdot)$ is the CDF trained on the $i$-th dimensional values of the dataset, and $K_i$ is the partition number for the $i$-th dimension. 
To enable faster refinement of a range filter, for each bucket, an auxiliary CDF model is also trained using the sorting dimension. 

As shown in Figure~\ref{fig:flood}, Flood establishes a cost model and uses a historical query workload to select the sorting dimension and tune the hyper-parameters (e.g., partition number $K_i$). 
To process a range query, cells intersecting the query hyper-rectangle are first retrieved by Eq.~\eqref{eq:flood_cell}, and then their intra-cell CDF models are queried to apply efficient filtering based on the range predicates.



\begin{figure}[t]
    \centering
    \includegraphics[width=0.47\textwidth]{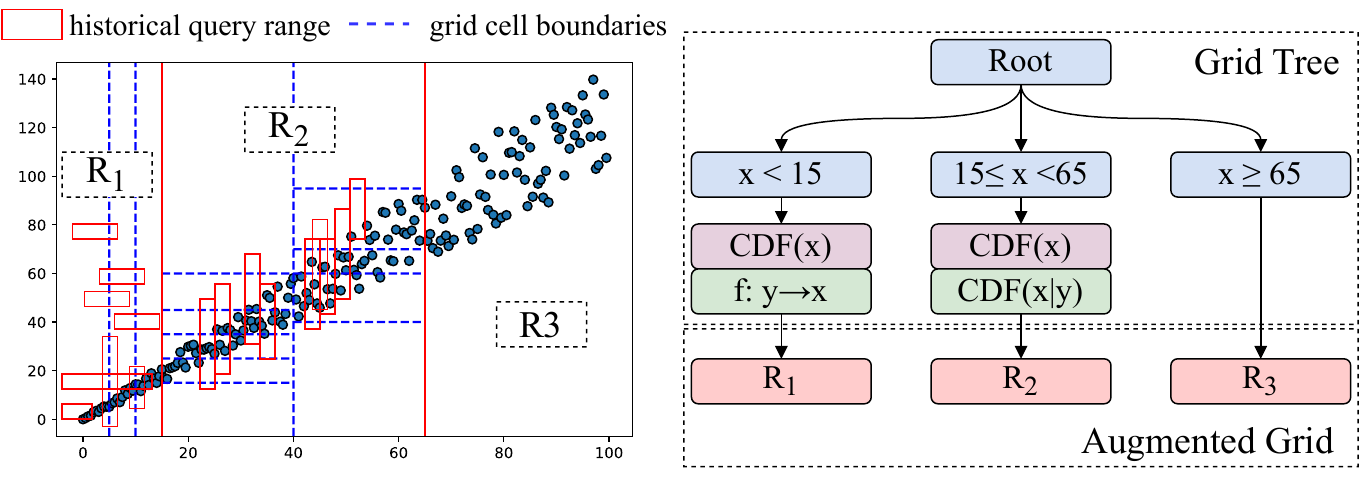}
        \caption{Illustration of a Tsunami~\cite{DBLP:journals/pvldb/DingNAK20} index built on a 2-dimensional correlated dataset and tuned by skewed query workloads. The red solid lines refer to the partition boundaries of the grid tree, and the blue dashed lines refer to the bucket boundaries of augmented grids.}
        \label{fig:tsunami}
\end{figure}

\noindent\underline{\textbf{Tsunami}}~\cite{DBLP:journals/pvldb/DingNAK20} further improves Flood's performance on correlated data and skewed query workloads. 
As shown in Figure~\ref{fig:tsunami}, Tsunami's structure consists of two parts: a grid tree that adapts to skewed query workloads and an augmented grid index that is optimized to capture data correlations. 
Similar to a $k$d-tree, the grid tree is also a space partitioning tree based on a selected subset of dimensions, which divides the whole space into several disjoint regions such that the query skewness of a historical workload is reduced within each region (e.g., $R_1$, $R_2$, $R_3$ in Figure~\ref{fig:tsunami}). 
Then, an augmented grid is constructed for each region. 

Different from Flood, where a grid cell is determined by independent CDF models on each dimension, Tsunami captures correlation patterns using both functional dependency ($f: y\to x$ in $R_1$) and conditional CDF model ($CDF(x|y)$ in $R_2$). 
To determine the granularity of each augmented grid, historical workloads are considered such that frequently queried regions are intensively partitioned (e.g., region $R_2$) while less queried regions are mildly partitioned (e.g., region $R_1$ and $R_3$). 
The range query processing on Tsunami is similar to that of Flood, i.e., locating the intersected regions and grids and then refining the results. 

Both Flood and Tsunami set up a periodical mechanism to monitor the workload shifts. 
Once the workload distribution significantly differs from the one used to construct the index, the whole grid layout and learned models will be re-tuned using the latest collected workload characteristics.

\subsection{Discussion}\label{subsec:discussion_index}
In this section, we discuss several omitted details related to the design choices of multi-dimensional learned indices, including \ding{182} index update, \ding{183} model selection, and \ding{184} model training. 

\noindent\underline{\textbf{Index Update.}} Most of the multi-dimensional learned indices except RSMI~\cite{DBLP:journals/pvldb/QiLJK20} and LISA~\cite{DBLP:conf/sigmod/Li0ZY020} do not support dynamic operations like insertion and deletion. 
The major bottleneck comes from the hardness of updating outdated models. 
An interesting finding is that both RSMI and LISA adopt a model-based data layout to handle dynamic updates. 
To insert (or delete) a record, RSMI and LISA simply queries the underlying learned models to obtain the ID of block to be inserted into and finalize the insertion if the block is not full. 
In this case, the block ID predicted by the model is always regarded to be \textit{correct} due to the model-based layout. 

Such a strategy to handle dynamic operations is similar to that of existing updatable 1-D learned indices (e.g., ALEX~\cite{DBLP:conf/sigmod/DingMYWDLZCGKLK20} and LIPP~\cite{DBLP:journals/pvldb/WuZCCWX21}), where new keys are inserted to a sorted array with gaps based on the model prediction. 
However, existing indices merely consider the query performance decay problem when the data distribution significantly shifts, and in the worst case, re-constructing the index is inevitable. 

\noindent\underline{\textbf{Model Selection.}} A learned index can be conceptually regarded as a combination of ``data layout + learned model''. 
Though playing an important role, in most cases, the underlying learned models in existing works can be safely replaced by another one as long as it is error-bounded and monotonic to ensure the correctness of query processing results. 

\begin{figure}[t]
    \centering
    \includegraphics[width=0.4\textwidth]{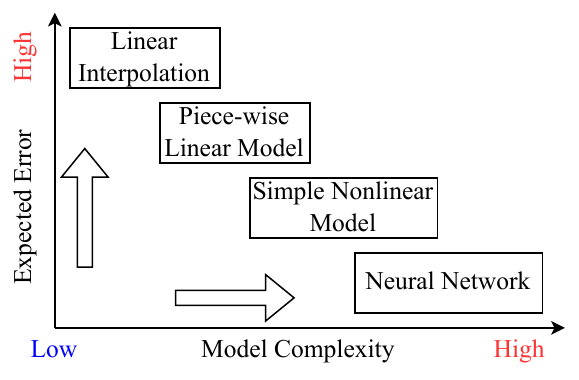}
    \caption{Characteristics of different model choices.}
    \label{fig:model_compare}
\end{figure}

Figure~\ref{fig:model_compare} roughly depicts the popular learned model choices used for index design. 
Clearly, simple models like linear interpolation or piece-wise linear approximation (PLA) are efficient to learn and infer with a sacrifice of limited model capacity, leading to potentially higher error. 
On the other hand, complex models like neural networks can achieve potentially better accuracy while suffering from extra training and inference overheads caused by deep learning runtimes. 
This increases the freedom of index design and enables a series of trade-offs. 
For example, a sophisticated model is usually costly to learn and store; however, it can effectively filter out unnecessary points to be examined when processing a range query. 

\noindent\underline{\textbf{Model Training and Inference.}} Besides model selection, effective and efficient model training and inference also play an important role in the learned index framework. 
As shown in Table~\ref{tab:index_summary}, existing works employ either self-designed algorithms (i.e., algorithmic approach) or utilize mature libraries (usually based on Python) like Pytorch~\cite{pytorch}, Tensorflow~\cite{Tensorflow}, or numpy~\cite{Numpy}. 
Although these external libraries show much higher flexibility in terms of model design choices, they usually require extra runtime overheads (e.g., \texttt{libtorch} for Pytorch) and suffer from longer training time. 
On the other hand, well-designed algorithmic approaches are usually more efficient for training and inference with a sacrifice of model design flexibility, e.g., the error-bounded piece-wise linear approximation~\cite{o1981line} used in PGM-Index~\cite{DBLP:journals/pvldb/FerraginaV20} and the top-down training strategy used in RMI~\cite{DBLP:conf/sigmod/KraskaBCDP18}. 
From our evaluation results (Section~\ref{sec:exp_results}), compared with indices internally using PGM-Index, the Pytorch-based solution is not significantly better in terms of query processing efficiency, but takes $\sim 9000\times$ longer time to train the model.

\section{Experiment Setup}\label{sec:exp_setup}
This section introduces the index implementation details and experiment setups. 
We implement the whole benchmark in C++ and compile it using GCC 9.5 where the optimization level is set to O3. 
All experiments are performed on a Ubuntu Linux machine with Intel(R) Core(TM) i7-10700K CPU and 32 GB memory. 
We disable the CPU's Turbo Boost feature and lock the CPU frequency to 4.67 GHz in BIOS.

\subsection{Index Implementation Details}\label{subsec:impl_details}
We first present the implementation details of all the compared multi-dimensional indices (learned and non-learned). 
Table~\ref{tab:baselines} summarizes all 13 compared methods where 7 indices are conventional indices and 6 indices are learned indices. 

\noindent\underline{\ding{182} \textbf{\texttt{ZMI}}}: the ZM-Index~\cite{DBLP:conf/mdm/WangFX019} that combines the Z-order curve and 1-D learned index. 
To compute the Z-order curve values, for a dataset of $N$ $d$-dimensional points, we uniformly partition each dimension into $N^{1/d}$ buckets, implying an equal-width grid layout of $N$ buckets. 
A fine-grained grid can improve the pruning power of unnecessary data access but increase the space and time efficiency of index construction and query processing. 
We try different grid partition resolutions and find that a $N^{1/d}\times\cdots\times N^{1/d}$ uniform grid can robustly achieve the best performance on different datasets.

\noindent\underline{\ding{183} \textbf{\texttt{MLI}}}: the ML-Index~\cite{DBLP:conf/edbt/DavitkovaM020} that combines the improved iDistance function (Eq.~\eqref{eq:mlindex_projection}) and 1-D learned index. 
To appropriately set the reference points, Lloyd's $k$-means algorithm is invoked with the $k$-means++ initialization strategy~\cite{DBLP:conf/soda/ArthurV07}. 
According to~\cite{DBLP:journals/tods/JagadishOTYZ05}, for pivot-based indices, the query efficiency improvement is minor when the number of reference points (i.e., $P$) is larger than 25. 
In our implementation, the partition number $P$ is set to 20 for datasets of size $<20$M and 40 for datasets of size $>20$M.

\begin{table}[t]
    \centering
    \caption{Summary of the compared indices.}\label{tab:baselines}
    \begin{tabular}{|c|c|c|c|c|}
    \hline
        Index & Type & Range & $k$NN & Updatable \\\hline
        \texttt{ZMI} & learned  & $\checkmark$ & $\checkmark$ & \ding{53} \\
        \texttt{MLI} & learned & $\checkmark$ & $\checkmark$ & {\ding{53}} \\
        \texttt{RSMI} & learned & $\checkmark$ & $\checkmark$ & {$\checkmark$} \\
        \texttt{IFI} & learned & $\checkmark$ &  \ding{53} & {\checkmark} \\
        \texttt{LISA} & learned & $\checkmark$ & $\checkmark$ & {$\checkmark$} \\
        \texttt{Flood} & learned & $\checkmark$ & \ding{53} & {\ding{53}} \\ \hline
        \texttt{STRtree} & tree & $\checkmark$ & $\checkmark$ & {$\checkmark$}\\
        \texttt{R*tree} & tree & $\checkmark$ & $\checkmark$ & {$\checkmark$}\\
        \texttt{kdtree} & tree & {$\checkmark$} & $\checkmark$ & {$\checkmark$}\\
        \texttt{qdtree} & tree & $\checkmark$ & \ding{53} & {$\checkmark$}\\
        \texttt{ANN} & tree & \ding{53} & $\checkmark$ & {$\checkmark$}\\
        \texttt{UG} & grid & $\checkmark$ & \ding{53} & {$\checkmark$}\\
        \texttt{EDG}& grid & $\checkmark$ & \ding{53} & {$\checkmark$}\\\hline
    \end{tabular}
\end{table}

\noindent\underline{\ding{184} \textbf{\texttt{LISA}}}: the LISA index~\cite{DBLP:conf/sigmod/Li0ZY020} that employs the grid-based projection function (Eq.~\eqref{eq:lisa_projection}). 
The original implementation of \texttt{LISA} is in Python and highly depends on Numpy. 
As all the learned indices are based on C++, we re-implement and optimize the LISA index structure at our best. 
To compute Eq.~\eqref{eq:lisa_projection}, we construct an equal-depth grid where each grid cell contains roughly $B=2000$ points. 
By such setting, the range of Eq.~\eqref{eq:lisa_projection} is in $[0, N/B + 1]$. 

\noindent\underline{\ding{185} \textbf{\texttt{IFI}}}: the IF-Index~\cite{DBLP:conf/vldb/0001KH20} (a.k.a., Hands-off) based on R-tree. 
The original IF-Index employs linear interpolation to estimate the key location, whose prediction error is large, especially for non-uniform data. 
To reduce the error and thus improve the query efficiency, we train linear models using the least square method, yielding a 2\%--15\% performance improvement with a sacrifice of 10\%--20\% more training time. 
Besides, through our practice, we find that it is unnecessary to select different sorting dimensions for each leaf node, as we do not know the distribution of query rectangles when building the index. 
The capacities of leaf and non-leaf nodes are set to 1000 and 64, which can achieve the best trade-off between index size and performance on different datasets. 


\noindent\underline{\ding{186} \textbf{\texttt{RSMI}}}: the recursive spatial model index~\cite{DBLP:journals/pvldb/QiLJK20}. 
We select their original implementation and follow the same index tuning strategies as discussed in their paper (e.g., selection of the partition threshold parameter $N'$). 
Note that, \texttt{RSMI} only supports 2-D datasets as it mainly targets spatial applications. 
The model training and inference of \texttt{RSMI} is based on Pytorch~\cite{pytorch}. 
Thus, to make it fair when comparing with other indices, we choose the CPU-only version of Pytorch but enable multi-threading to accelerate the model training (otherwise it fails to terminate in 5 hours for a 20M dataset). 

\noindent\underline{\ding{187} \textbf{\texttt{Flood}}}: the learned grid index~\cite{DBLP:conf/sigmod/NathanDAK20}. As \texttt{Flood} is not open-sourced at the current stage, we implement \texttt{Flood} at our best. 
The original \texttt{Flood} index requires a query workload to tune its hyper-parameters (e.g., selection of the sorting dimension). 
However, since other indices are not optimized using query histories, we do not implement this part to make the comparison fair. 
Similarly, \texttt{Tsunami}~\cite{DBLP:journals/pvldb/DingNAK20} is a fully workload-driven index and thus is not compared in this benchmark due to the same reason. 
To avoid the exponentially growing number of grid cells w.r.t.~data dimension, the partition number on each dimension except the sorting dimension is set to $(N/B)^{1/(d-1)}$ where $B=2000$ is the (rough) number of points in each cell.

Except \texttt{RSMI} and \texttt{IFI} that employ deep learning and linear regression model, we choose the PGMIndex~\cite{DBLP:journals/pvldb/FerraginaV20} as the default underlying 1-D learned index of the multi-dimensional learned indices (i.e., \texttt{ZMI}, \texttt{MLI}, \texttt{LISA}, and \texttt{Flood}) to unify their implementations. 
Note that, the default error parameter $\epsilon$ for PGMIndex is set to 64, which is an empirically robust and optimized value across different datasets and configurations (see Section~\ref{subsec:exp_error} for more details). 

However, according to the results of a recent benchmark on 1-D learned indices~\cite{DBLP:journals/pvldb/MarcusKRSMK0K20}, a well-optimized recursive model index (RMI) can slightly outperform the PGMIndex. 
The reasons that we do not choose RMI~\cite{DBLP:conf/sigmod/KraskaBCDP18} as the underlying 1-D learned index in our implementation are threefold. 
\begin{itemize}[leftmargin=*]
    \item The state-of-the-art RMI implementation~\cite{DBLP:journals/pvldb/MarcusKRSMK0K20} is intrinsically an ``index compiler'' that takes a dataset as input and generates \emph{static} index data and header files separately, which is less flexible to be embedded into other indices. 
    \item The parameter tuning of PGMIndex is much easier than RMI as only one error parameter $\epsilon$ is required for PGMIndex. 
    On the other hand, RMI requires tuning of the error parameters, number of models, and model types for each level of the index. 
    \item In this experimental study, we focus more on evaluating the performance w.r.t.~different index design choices (e.g., data layout and projection function) under a unified environment, instead of a replaceable component that has a minor influence on the \emph{relative} performance. 
\end{itemize}

Besides learned indices, we also implement and compare 8 non-learned baselines that are commonly used in practice.

\noindent\underline{\ding{182} \texttt{FullScan}}: the na\"ive sequential scan algorithm. 
To speed up range query processing, we sort the points based on Z-order values to keep spatial locality and allow the early pruning of unnecessary points. 

\noindent\underline{\ding{183} \texttt{R*tree}}: the R*-tree index~\cite{DBLP:conf/sigmod/BeckmannKSS90} that optimizes the R-tree structure by minimizing the leaf node overlap. 

\noindent\underline{\ding{184} \texttt{STRTree}}: the R-tree index with the sort-tile-recursive (STR) bulk-loading strategy~\cite{DBLP:conf/icde/LeuteneggerEL97}, which is a simple yet effective R-tree variant. 
As the indexed geometries are multi-dimensional points, there is no overlap for the Minimum Bounding Regions (MBRs) of leaf nodes. 

Note that, for both \texttt{R*tree} and \texttt{STRTree}, we choose the implementations from the Boost geometry library~\cite{boost-geometry}, and the node capacity (i.e., fanout) is set to 128 based on a benchmark on popular spatial libraries~\cite{DBLP:journals/dase/PandeyRKK21}. 

\noindent\underline{\ding{185} \texttt{kdtree}}: the $k$d-tree index~\cite{DBLP:journals/cacm/Bentley75} that recursively partition the space by the median of each dimension in a round-robin fashion. 
We adopt the implementation in the nanoflann project~\cite{nanoflann}. 

\noindent\underline{\ding{186} \texttt{ANN}}: another $k$d-tree variant provided in \texttt{ANN}~\cite{ann}, a library for efficient exact and approximate $k$NN search. 
For both \texttt{ANN} and \texttt{kdtree}, the leaf node size (i.e., the threshold of a $k$d-tree turns to use brute-force search) is set to 16. 

\noindent\underline{\ding{187} \texttt{qdtree}}: the quad-tree index that can be viewed as an adaptive grid index. 
The implementation from GEOS library~\cite{geos} is adopted. 
Note that \texttt{qdtree} only supports range queries. 

\noindent\underline{\ding{188} \texttt{UG}}: a uniform grid index (a.k.a., equal-width grid index). The partition number of each dimension is set to $(N/2000)^{1/d}$. 

\noindent\underline{\ding{189} \texttt{EDG}}: an equal-depth grid index where each dimension is sorted and partitioned into $(N/2000)^{1/d}$ parts of equal size. 
\texttt{EDG} can be regarded as the non-learned version of the \texttt{Flood} index. 
Compared with the uniform grid, \texttt{EDG} can better deal with skewed data distribution with a sacrifice of higher construction cost. 

\begin{table}[t]
    \caption{Summary of datasets.}
    \label{tab:datasets}
    \centering
    \small
    \begin{tabular}{|c|c|c|c|c|}
        \hline
        Dataset & Size & \#Points & \#Dim. & Skewness \\\hline
        \textsf{Uniform}  & N.A. & 5--100M & 2--8 & low \\
        \textsf{Normal}  & N.A. & 5--100M & 2--8 & mid \\
        \textsf{Lognormal}  & N.A. & 5--100M & 2--8 & high \\\hline
        \textsf{FourSquare}  & 181 MB & 3.7M & 2 & mid \\
        \textsf{Toronto3d}  & 1.52 GB & 21M & 3 & high \\
        \textsf{OSM}  & 3.1 GB & 63M & 2 & mid \\\hline
    \end{tabular}
\end{table}

\begin{table*}
    \centering
    \caption{Index construction time (seconds) on real datasets and synthetic datasets of default configurations. \texttt{STR} is selected as the baseline index to compute the relative ratio. Note that, \texttt{RSMI} is not available on \texttt{Toronto3d} and fails to terminate construction in 5 hours for dataset \texttt{OSM}.}
    \label{tab:exp_construction}
    \begin{tabular}{|c|c|c|c|c|c|c|}
    \hline
        \backslashbox[5em]{Index}{Data}& \texttt{Uniform} & \texttt{Normal} & \texttt{Lognormal} & \texttt{Foursquare} & \texttt{Toronto3d} & \texttt{OSM} \\\hline
        \texttt{ZMI} & 1.89 ($\decrease{\blacktriangledown}1.76\times$) & 1.63 ($\decrease{\blacktriangledown}1.99\times$) & 1.13 ($\decrease{\blacktriangledown}2.85\times$) & 0.17 ($\decrease{\blacktriangledown}2.76\times$) & 1.09 ($\decrease{\blacktriangledown}3.0\times$) & 3.08 ($\decrease{\blacktriangledown}2.25\times$) \\\hline
        \texttt{MLI} & 93.54 ($\increase{\blacktriangle
}28.1\times$) & 90.49 ($\increase{\blacktriangle
}27.8\times$) & 83.99 ($\increase{\blacktriangle
}26.1\times$) & 14.48 ($\increase{\blacktriangle
}30.8\times$) & 83.13 ($\increase{\blacktriangle
}25.4\times$) & 209.22 ($\increase{\blacktriangle
}30.2\times$)\\\hline
        \texttt{IFI} & 4.88 ($\increase{\blacktriangle
}1.47\times$) & 5.04 ($\increase{\blacktriangle
}1.55\times$) & 5.08 ($\increase{\blacktriangle
}1.58\times$) & 0.71 ($\increase{\blacktriangle
}1.51\times$) & 4.61 ($\increase{\blacktriangle
}1.41\times$) & 10.15 ($\increase{\blacktriangle
}1.46\times$)\\\hline
        \texttt{RSMI} & 8978 ($\increase{\blacktriangle
}2696\times$) & 13017 ($\increase{\blacktriangle
}4005\times$) & 10311 ($\increase{\blacktriangle
}3202\times$) & 996.8 ($\increase{\blacktriangle
}2121\times$) & N.A. & N.A. \\\hline
        \texttt{LISA} & 7.46 ($\increase{\blacktriangle
}2.24\times$) & 7.45 ($\increase{\blacktriangle
}2.29\times$) & 7.48 ($\increase{\blacktriangle
}2.32\times$) & 1.25 ($\increase{\blacktriangle
}2.66\times$) & 7.55 ($\increase{\blacktriangle
}2.31\times$) & 15.62 ($\increase{\blacktriangle
}2.25\times$) \\\hline
        \texttt{Flood} & 5.34 ($\increase{\blacktriangle
}1.60\times$) & 5.44 ($\increase{\blacktriangle
}1.67\times$) & 5.57 ($\increase{\blacktriangle
}1.73\times$) & 0.99 ($\increase{\blacktriangle
}2.11\times$) & 9.80 ($\increase{\blacktriangle
}3.00\times$) & 18.05 ($\increase{\blacktriangle
}2.60\times$)\\\hline\hline
        \cellcolor[HTML]{DCDCDC}\texttt{STR} & \cellcolor[HTML]{DCDCDC}3.33 ($1.0\times$) & \cellcolor[HTML]{DCDCDC}3.25 ($1.0\times$) & \cellcolor[HTML]{DCDCDC}3.22 ($1.0\times$) & \cellcolor[HTML]{DCDCDC}0.47 ($1.0\times$) & \cellcolor[HTML]{DCDCDC}3.27 ($1.0\times$) & \cellcolor[HTML]{DCDCDC}6.93 ($1.0\times$) \\\hline\hline
        \texttt{R*tree} & 128.92 ($\increase{\blacktriangle
}38.7\times$) & 128.80 ($\increase{\blacktriangle
}39.6\times$) & 120.46 ($\increase{\blacktriangle
}37.4\times$) & 18.22 ($\increase{\blacktriangle
}38.8\times$) & 155.34 ($\increase{\blacktriangle
}47.5\times$) & 389.35 ($\increase{\blacktriangle
}56.2\times$)\\\hline
        \texttt{kdtree} & 28.61 ($\increase{\blacktriangle
}8.59\times$) & 27.96 ($\increase{\blacktriangle
}8.60\times$) & 29.03 ($\increase{\blacktriangle
}9.02\times$) & 3.87 ($\increase{\blacktriangle
}8.23\times$) & 14.66 ($\increase{\blacktriangle
}4.48\times$) & 30.53 ($\increase{\blacktriangle
}4.41\times$)\\\hline
        \texttt{qdtree} & 2.15 ($\decrease{\blacktriangledown
}1.55\times$) & 2.46 ($\decrease{\blacktriangledown
}1.32\times$) & 2.77 ($\decrease{\blacktriangledown
}1.16\times$) & 0.619 ($\increase{\blacktriangle
}1.32\times$) & 12.587 ($\increase{\blacktriangle
}3.85\times$) & 21.20 ($\increase{\blacktriangle
}3.06\times$) \\\hline
        \texttt{ANN} & 21.61 ($\increase{\blacktriangle
}6.49\times$) & 20.43 ($\increase{\blacktriangle
}6.29\times$) & 20.75 ($\increase{\blacktriangle
}6.44\times$) & 3.21 ($\increase{\blacktriangle
}6.83\times$) & 2.99 ($\decrease{\blacktriangledown
}1.09\times$) & 22.33 ($\increase{\blacktriangle
}3.22\times$) \\\hline
        \texttt{UG} & 0.73 ($\decrease{\blacktriangledown
}4.56\times$) & 0.51 ($\decrease{\blacktriangledown
}6.37\times$) & 0.34 ($\decrease{\blacktriangledown
}9.47\times$) & 0.07 ($\decrease{\blacktriangledown
}6.71\times$) & 0.48 ($\decrease{\blacktriangledown
}6.81\times$) & 1.24 ($\decrease{\blacktriangledown
}5.59\times$) \\\hline
        \texttt{EDG} & 4.17 ($\increase{\blacktriangle
}1.25\times$) & 4.16 ($\increase{\blacktriangle
}1.28\times$) & 4.15 ($\increase{\blacktriangle
}1.29\times$) & 0.63 ($\increase{\blacktriangle
}1.34\times$) & 4.53 ($\increase{\blacktriangle
}1.39\times$) & 8.44 ($\increase{\blacktriangle
}1.22\times$)\\\hline  
    \end{tabular}
\end{table*}

\subsection{Datasets and Query Generation}\label{subsec:data_query_prepare}
We evaluate all the compared methods in Section~\ref{subsec:impl_details} on both real and synthetic datasets, spanning a wide spectrum of data size, dimensionality, and distribution skewness. 
Table~\ref{tab:datasets} summarizes the dataset statistics. 

\noindent\underline{\textbf{Real Datasets.}} We adopt three commonly used real datasets in our evaluation. 
\begin{itemize}[leftmargin=*]
    \item \textbf{\texttt{FourSquare}} is a location dataset extracted from a geo-social network. 
    Each data point represents the coordinates of a co-location event of two users in the social network~\cite{foursquare}. 
    \item \textbf{\texttt{Toronto3d}} is a public dataset of LiDAR images of urban roadways~\cite{toronto3d}. 
    We extract the spatial coordinates (i.e., X, Y, Z) from the raw data to form a 3-D point cloud. 
    \item \textbf{\texttt{OSM}} is a recent dump of OpenStreatMap~\cite{osm-china}. We only use the point objects (i.e., pair of latitude and longitude), and other geometries like polygons and line-strings are excluded as querying these complicated geometries is not commonly supported by existing learned index studies. 
\end{itemize}

\begin{figure}[t]
    \centering
    \includegraphics[width=0.48\textwidth]{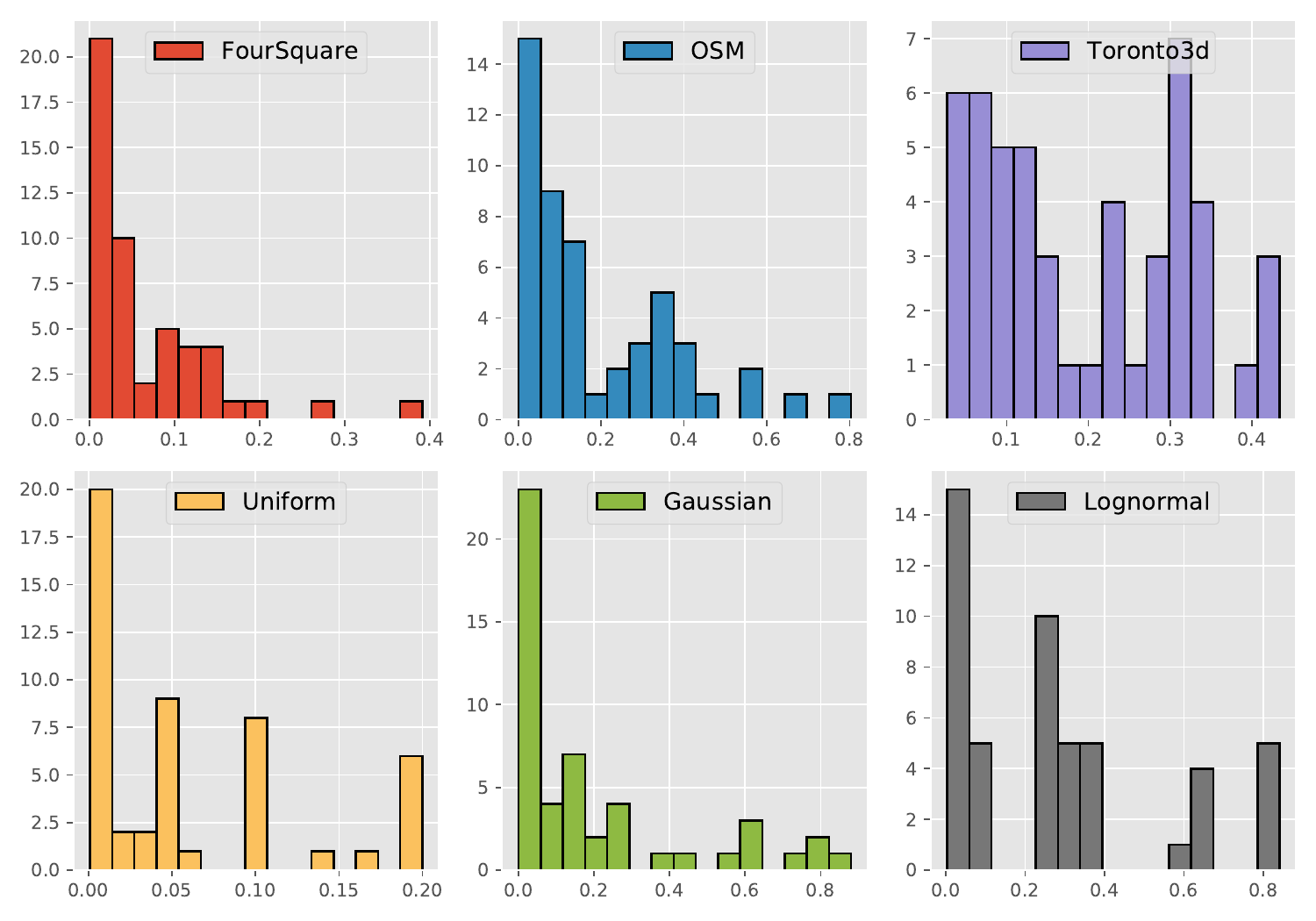}
    \caption{Selectivity distribution of generated range queries. The x-axis and y-axis refer to the query selectivity and number of queries, respectively.}
    \label{fig:range_query_dist}
\end{figure}

\noindent\underline{\textbf{Synthetic Datasets.}} 
To evaluate the scalability w.r.t. different data sizes and dimensions, we also sample random points from uniform, normal, and log-normal distributions where the dimension correlation is ignored. 
We vary the scale factors of each distribution (e.g., the standard deviation of normal distribution) to make sure that the density of generated data is similar. 
For all synthetic datasets, the default \#Points and \#Dim.~are set to 20 million and 2, respectively\footnote{\#Dim.~is set to 2 by default as some compared methods like \texttt{RSMI} only support 2-D datasets.}.  


\noindent\underline{\textbf{Range Query Generation.}} To generate range queries for evaluation, we first randomly sample $S=100$ points from a given dataset as the lower corners of query rectangles; then, for each dimension $i=1,\cdots,d$, a random width $\delta_i$ is added to the lower corner to generate the upper corner. 
For a generated query box $R$, the selectivity of $R$ is the percentage of points that $R$ is expected to cover, i.e., 
\begin{equation}\label{eq:query_sel}
    \text{Sel}(R) = \frac{\text{\#(Points covered by } R\text{)}}{\text{\#(Points)}}.
\end{equation}
Figure~\ref{fig:range_query_dist} presents the selectivity distributions of generated range queries on different datasets. 
The default selectivity in the subsequent evaluations is $1\%$. 

\noindent\underline{\textbf{$k$NN Query Generation.}} A $k$NN query is a pair of query point $q$ and a result size $k$. 
To generate $k$NN queries, for each $k\in$\{1, 10, 100, 1000, 10000\}, we randomly pick $S=20$ points from each dataset to generate a set of $k$NN queries. The default value of $k$ in the evaluation is set to $1000$.

In the subsequent evaluations, we report the \emph{average} query processing time of different selectivities for range queries and different $k$ for $k$NN queries. 
Note that, we do not iterate each query many times, which will warm the CPU cache and thus possibly reduce the query processing time. 
Such a setting can well simulate the index usage scenarios in real-world applications.

\begin{table*}
    \centering
    \caption{Index memory overhead (MB) on real datasets and synthetic datasets of default configurations. \texttt{STR} is chosen as the baseline to compute the relative comparison ratio. Note that, we exclude the memory cost for storing pointers to each data point from the overall index size where each pointer occupies 8 bytes on most modern architectures. }
    \label{tab:exp_index_memory}
    \begin{tabular}{|c|c|c|c|c|c|c|}
    \hline
       \backslashbox[5em]{\texttt{Index}}{\texttt{Data}} & \texttt{Uniform} & \texttt{Normal} & \texttt{Lognormal} & \texttt{Foursquare} & \texttt{Toronto3d} & \texttt{OSM} \\\hline
        \texttt{ZMI} & 0.02 ($\decrease{\blacktriangledown}7374\times$) & 0.05 ($\decrease{\blacktriangledown}3120\times$) & 0.03 ($\decrease{\blacktriangledown}5070\times$) & 0.03 ($\decrease{\blacktriangledown}907\times$) & 0.28 ($\decrease{\blacktriangledown}1244\times$) & 1.94 ($\decrease{\blacktriangledown}262\times$) \\\hline
        \texttt{MLI} & 0.04 ($\decrease{\blacktriangledown}3863\times$) & 0.04 ($\decrease{\blacktriangledown}3863\times$) & 0.04 ($\decrease{\blacktriangledown}3863\times$) & 0.05 ($\decrease{\blacktriangledown}564\times$) & 0.08 ($\decrease{\blacktriangledown}4518\times$) & 0.29 ($\decrease{\blacktriangledown}1749\times$) \\\hline
        \texttt{IFI} & 1.12 ($\decrease{\blacktriangledown}145\times$) & 1.12 ($\decrease{\blacktriangledown}145\times$) & 1.12 ($\decrease{\blacktriangledown}145\times$) & 0.21 ($\decrease{\blacktriangledown}140\times$) & 1.39 ($\decrease{\blacktriangledown}248\times$) & 3.50 ($\decrease{\blacktriangledown}145\times$) \\\hline
        \texttt{RSMI} & 21.5 ($\decrease{\blacktriangledown}7.55\times$) & 22.4 ($\decrease{\blacktriangledown}7.24\times$) & 29.8 ($\decrease{\blacktriangledown}5.44\times$) & 21.7 ($\decrease{\blacktriangledown}1.38\times$) & N.A. & N.A. \\\hline
        \texttt{LISA} & 0.53 ($\decrease{\blacktriangledown}305\times$) & 0.53 ($\decrease{\blacktriangledown}305\times$) & 0.53 ($\decrease{\blacktriangledown}305\times$) & 0.09 ($\decrease{\blacktriangledown}321\times$) & 0.30 ($\decrease{\blacktriangledown}1160\times$) & 1.97 ($\decrease{\blacktriangledown}258\times$) \\\hline
        \texttt{Flood} & 0.33 ($\decrease{\blacktriangledown}489\times$) & 0.37 ($\decrease{\blacktriangledown}436\times$) & 0.37 ($\decrease{\blacktriangledown}436\times$) & 0.28 ($\decrease{\blacktriangledown}106\times$) & 2.81 ($\decrease{\blacktriangledown}122\times$) & 4.62 ($\decrease{\blacktriangledown}110\times$) \\\hline\hline
        \cellcolor[HTML]{DCDCDC}\texttt{STR} & \cellcolor[HTML]{DCDCDC}162.2 ($1.0\times$) & \cellcolor[HTML]{DCDCDC}162.2 ($1.0\times$) & \cellcolor[HTML]{DCDCDC}162.2 ($1.0\times$) & \cellcolor[HTML]{DCDCDC}29.9 ($1.0\times$) & \cellcolor[HTML]{DCDCDC}343.4 ($1.0\times$) & \cellcolor[HTML]{DCDCDC}508.9 ($1.0\times$) \\\hline\hline
        \texttt{R*tree} & 295.5 ($\increase{\blacktriangle}1.82\times$) & 292.5 ($\increase{\blacktriangle}1.80\times$) & 287.5 ($\increase{\blacktriangle}1.77\times$) & 56.0 ($\increase{\blacktriangle}1.88\times$) & 581.6 ($\increase{\blacktriangle}1.69\times$) & 1021.9 ($\increase{\blacktriangle}2.01\times$) \\\hline
        \texttt{kdtree} & 120.5 ($\decrease{\blacktriangledown}1.35\times$) & 118.6 ($\decrease{\blacktriangledown}1.37\times$) & 116.1 ($\decrease{\blacktriangledown}1.40\times$) & 16.7 ($\decrease{\blacktriangledown}1.79\times$) & 117.0 ($\decrease{\blacktriangledown}2.94\times$) & 200.0 ($\decrease{\blacktriangledown}2.54\times$) \\\hline
        \texttt{qdtree} & 1516.1 ($\increase{\blacktriangle}9.35\times$) & 1542.8 ($\increase{\blacktriangle}9.51\times$) & 1570.4 ($\increase{\blacktriangle}9.68\times$) & 266.5 ($\increase{\blacktriangle}8.93\times$) & 121.0 ($\decrease{\blacktriangledown}2.84\times$) & 4583.0 ($\increase{\blacktriangle}9.01\times$) \\\hline
        \texttt{ANN} & 123.8 ($\decrease{\blacktriangledown}1.31\times$) & 122.2 ($\decrease{\blacktriangledown}1.33\times$) & 118.9 ($\decrease{\blacktriangledown}1.36\times$) & 16.1 ($\decrease{\blacktriangledown}1.85\times$) & 126.7 ($\decrease{\blacktriangledown}2.71\times$) & 204.7 ($\decrease{\blacktriangledown}2.49\times$) \\\hline
        \texttt{UG} & 0.22 ($\decrease{\blacktriangledown}731\times$) & 0.22 ($\decrease{\blacktriangledown}731\times$) & 0.22 ($\decrease{\blacktriangledown}731\times$) & 0.08 ($\decrease{\blacktriangledown}360\times$) & 0.18 ($\decrease{\blacktriangledown}1951\times$) & 1.43 ($\decrease{\blacktriangledown}356\times$) \\\hline
        \texttt{EDG} & 0.22 ($\decrease{\blacktriangledown}731\times$)  & 0.22 ($\decrease{\blacktriangledown}731\times$) & 0.22 ($\decrease{\blacktriangledown}731\times$) & 0.08 ($\decrease{\blacktriangledown}360\times$) & 0.18 ($\decrease{\blacktriangledown}1951\times$) & 1.43 ($\decrease{\blacktriangledown}356\times$) \\\hline
    \end{tabular}
\end{table*}

\section{Experiment Results}\label{sec:exp_results}
In this section, we report the evaluation results based on the above configurations to answer the core question: \emph{How good are multi-dimensional learned indices?} 
In particular, we are interested in the following 8 sub-questions.
\begin{enumerate}[leftmargin=*,label={\bfseries Q\arabic*:}]
    \item \textbf{Construction Time}: whether the construction time of learned indices is as fast as the non-learned baselines ($\triangleright
$ Section~\ref{subsec:exp_construction_time});
    \item \textbf{Space Cost}: whether the space cost of learned indices is significantly lower than the non-learned baselines ($\triangleright
$ Section~\ref{subsec:exp_memory_cost});
    \item \textbf{Range Query Efficiency}: whether the learned indices can outperform baselines in terms of range query processing ($\triangleright
$ Section~\ref{subsec:exp_range_time});
    \item \textbf{$k$NN Query Efficiency}: whether the learned indices can outperform baselines in terms of $k$NN query processing ($\triangleright
$ Section~\ref{subsec:exp_knn_analysis});
    \item \textbf{Scalability}: whether the learned indices can well scale to datasets of larger sizes and higher dimensions ($\triangleright
$ Section~\ref{subsec:exp_scalability});
    \item \textbf{Parameter Setting:} how to properly configure the learned index hyper-parameters (e.g., the error parameter $\epsilon$) ($\triangleright
$ Section~\ref{subsec:exp_error});
    \item \textbf{Dynamic Update Efficiency:} whether existing learned indices can support efficient dynamic updates ($\triangleright
$ Section~\ref{subsec:exp_dynamic});
    \item \textbf{Explanation of Performance:} why learned indices are effective (or ineffective) ($\triangleright
$ Section~\ref{subsec:exp_cache}).
\end{enumerate}

\begin{figure*}
    \centering
    \begin{subfigure}[b]{0.48\textwidth}
        \includegraphics[width=\textwidth]{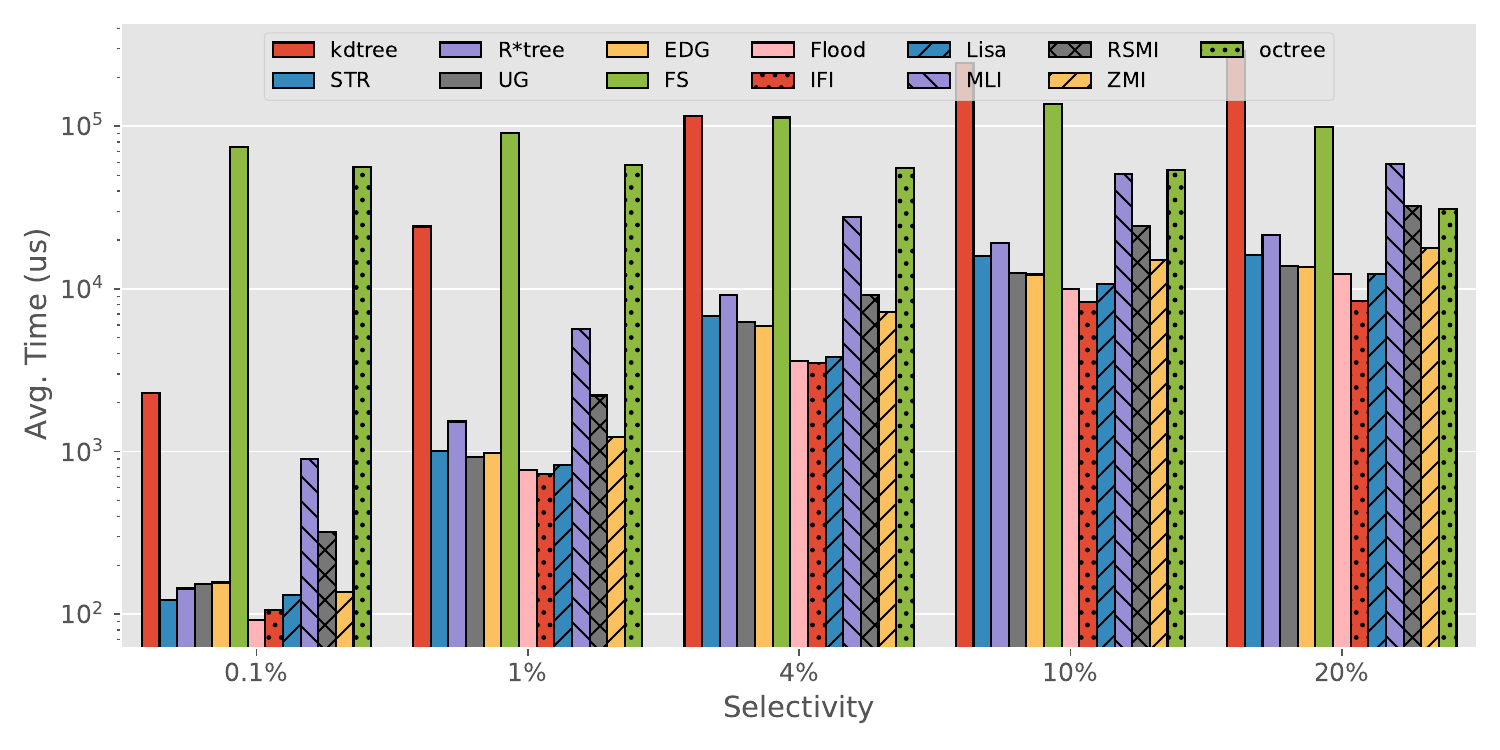}
        \caption{Uniform: Time v.s. Selectivity}
    \end{subfigure}
    \begin{subfigure}[b]{0.48\textwidth}
        \includegraphics[width=\textwidth]{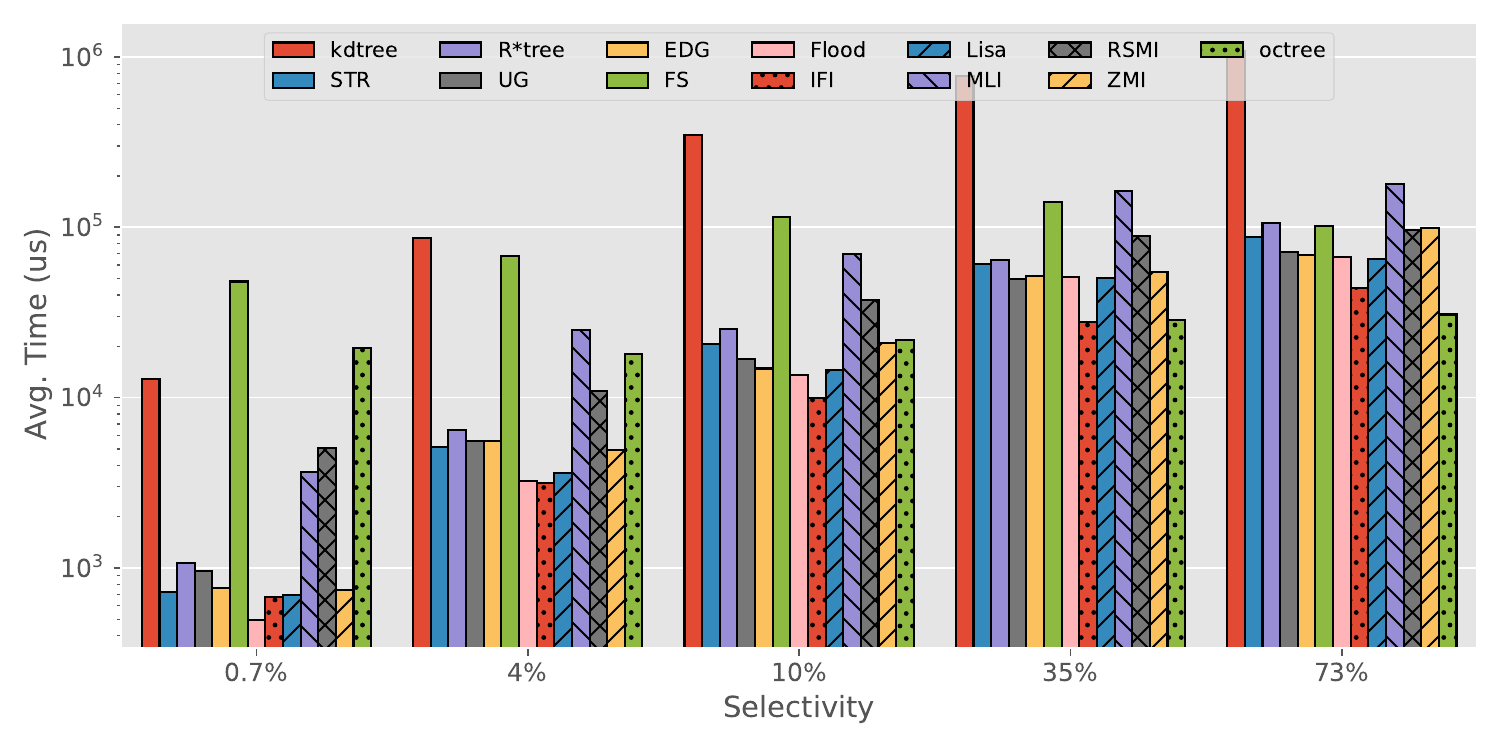}
        \caption{Gaussian: Time v.s. Selectivity}
    \end{subfigure}
    
    \begin{subfigure}[b]{0.48\textwidth}
        \includegraphics[width=\textwidth]{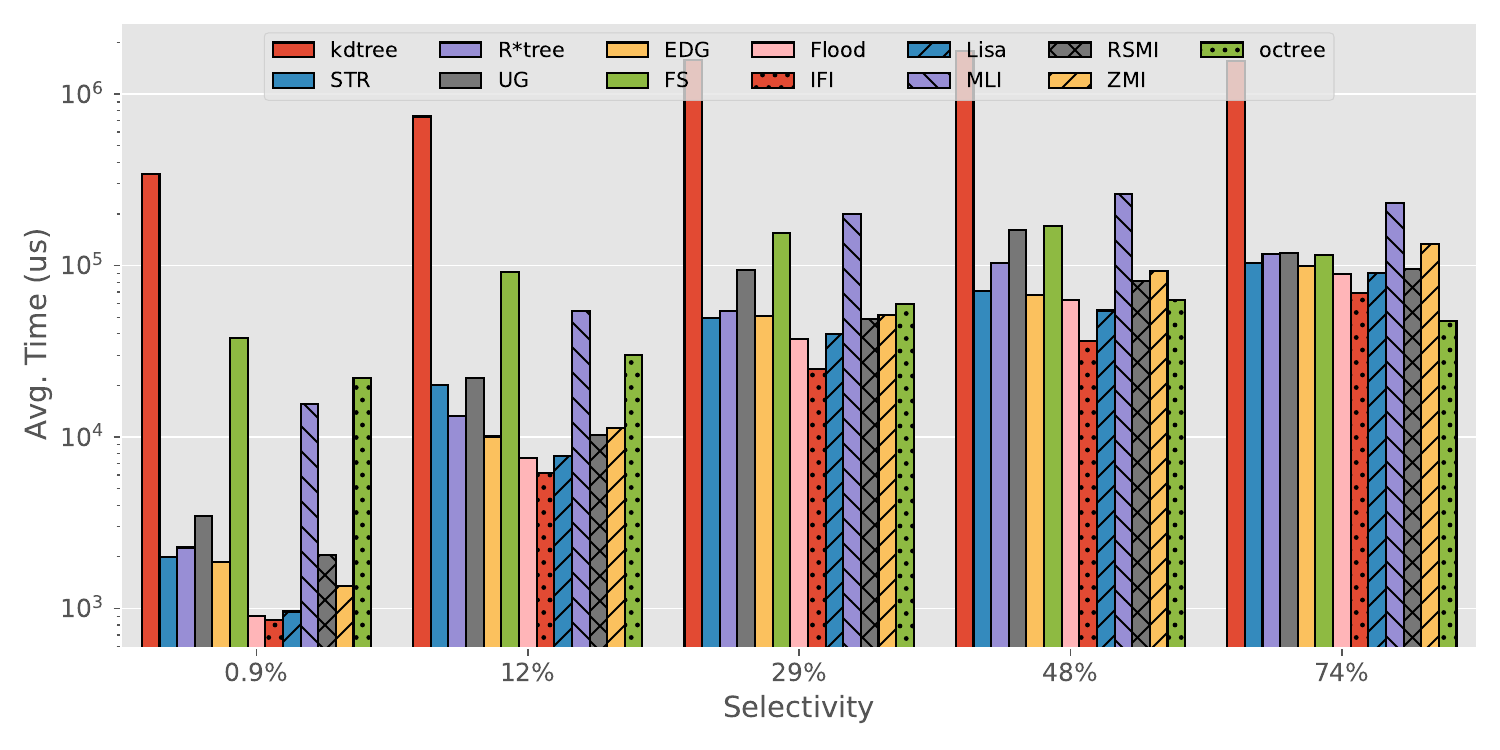}
        \caption{Lognormal: Time v.s. Selectivity}
    \end{subfigure}
    \begin{subfigure}[b]{0.48\textwidth}
        \includegraphics[width=\textwidth]{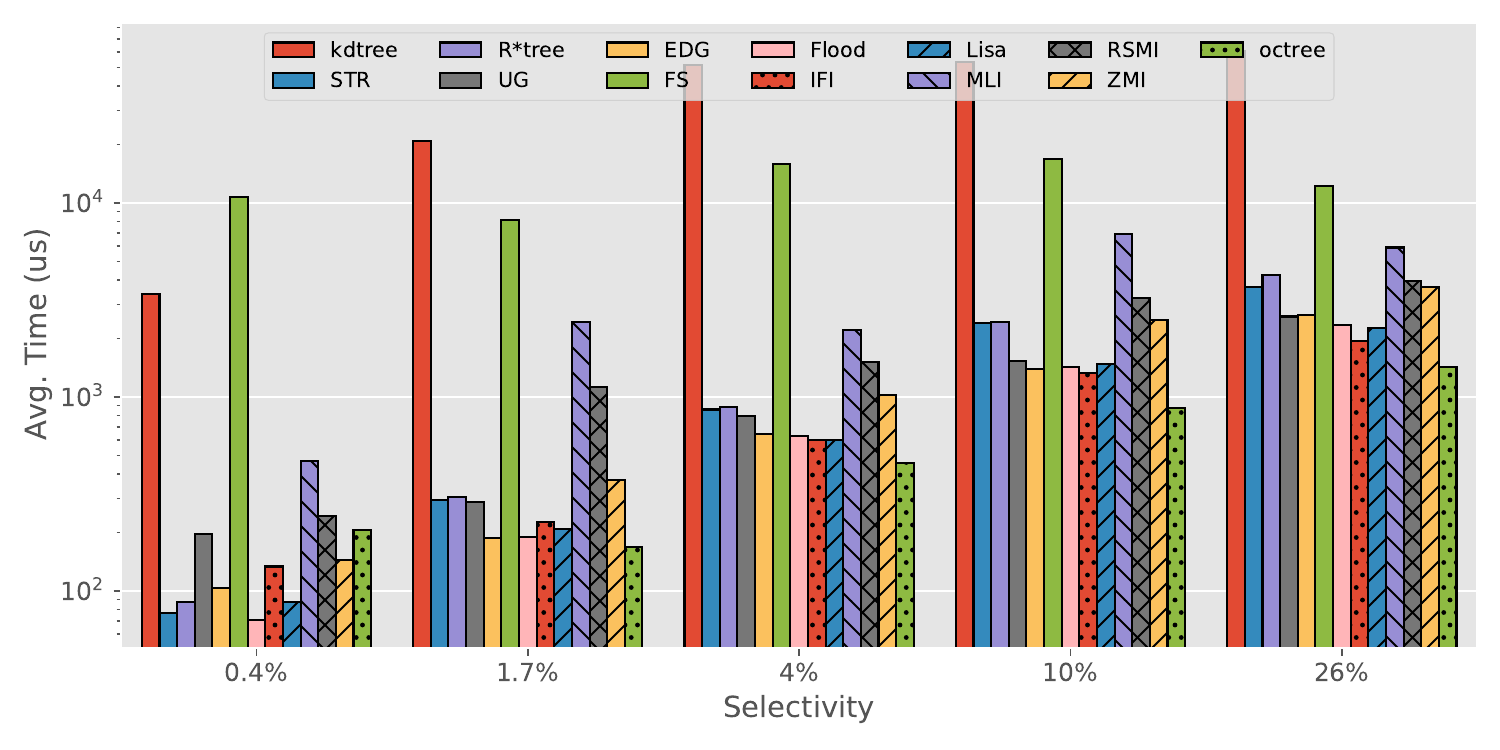}
        \caption{FourSquare: Time v.s. Selectivity}
    \end{subfigure}
    
    \begin{subfigure}[b]{0.48\textwidth}
        \includegraphics[width=\textwidth]{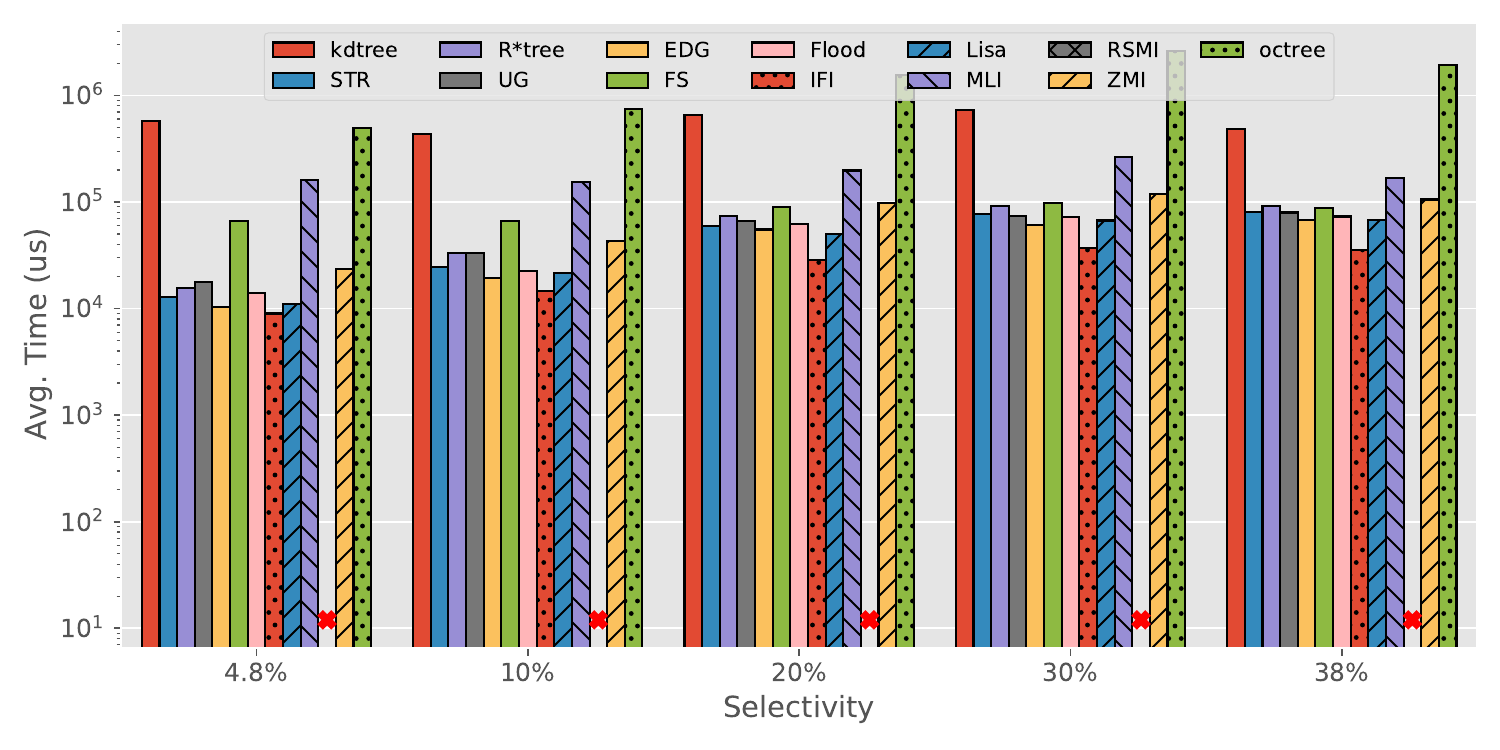}
        \caption{Toronto3d: Time v.s. Selectivity}
    \end{subfigure}
    \begin{subfigure}[b]{0.48\textwidth}
        \includegraphics[width=\textwidth]{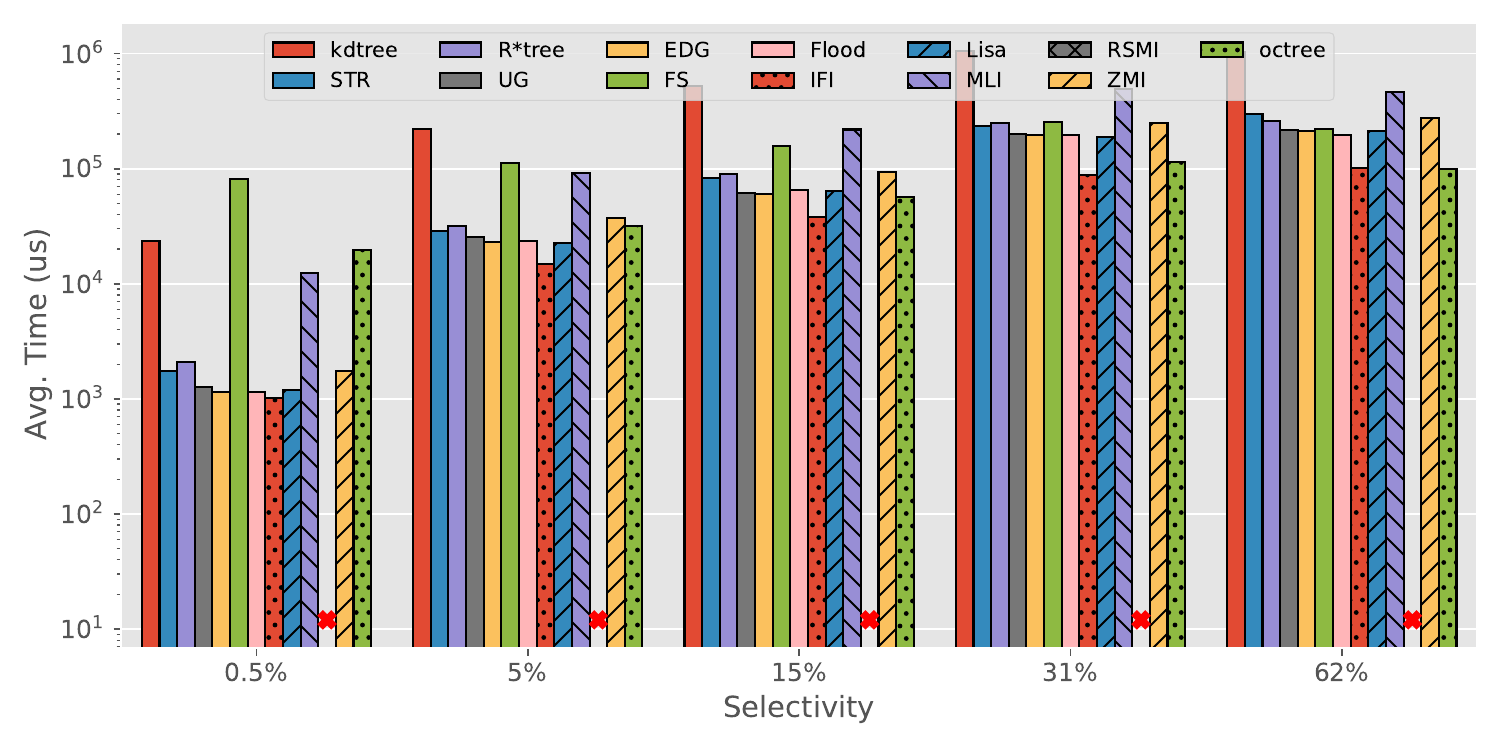}
        \caption{OSM: Time v.s. Selectivity}
    \end{subfigure}
    \caption{Range query evaluation results on real datasets and synthetic datasets of the default configuration (\#Points=20M and \#Dim=2) where the y-axis and x-axis refer to the \emph{average} query processing time and \emph{average} selectivity of test queries, respectively. Note that, \texttt{RSMI} is not available on \texttt{Toronto3d} and fails to terminate training on \texttt{OSM}. } 
    \label{fig:range_default}
\end{figure*}

\subsection{Index Construction Time (Q1)}\label{subsec:exp_construction_time}
We first report the index construction time of the default configuration (\#Points=20M and \#Dim=2) in Table~\ref{tab:exp_construction}. We set a threshold of 5 hours, and if the program cannot finish in 5 hours, it will be killed and N.A.~will be reported. 

As shown in Table~\ref{tab:exp_construction}, the projection-based indices except \texttt{MLI} (i.e., \texttt{ZMI} and \texttt{LISA}) are comparable with or even faster than the non-learned baselines in terms of index construction time. 
This is because we unify the underlying 1-D learned index of these indices as the PGMIndex, which requires only one pass of data (i.e., $O(N)$ time) to obtain the minimized number of error-bounded line segments~\cite{DBLP:journals/pvldb/FerraginaV20,DBLP:journals/pvldb/MarcusKRSMK0K20}. 
\texttt{MLI} is slower because finding reference points via $k$-means is costly, which occupies $>95\%$ of the total index building time. 
Surprisingly, \texttt{ZMI} is $\sim2\times$ faster than the bulk-loading R-tree index \texttt{STR} and is only slower than the simple uniform grid index \texttt{UG}. 
This is because, in addition to adopting the PGMIndex, \texttt{ZMI} also benefits from the bit manipulation instruction set provided by modern CPUs, which can significantly accelerate the computation of Z-order curve values~\cite{zvalue_comp}. 

Indices that adopt the grid layouts, i.e., \texttt{LISA} and \texttt{Flood}, take slightly longer construction time due to the generation of internal equal-depth grid cells. 
The construction time of \texttt{IFI} is $1.5\times$ of \texttt{STR} but much faster than \texttt{R*tree} (about $30\times$ faster). 
This is because, in our implementation of \texttt{IFI}, we adopt a similar bulking loading strategy to find the leaf nodes. 
Besides, the leaf node capacity of \texttt{IFI} is generally larger (e.g., 1000) than ordinary R-tree indices (e.g., 128), which also alleviates the extra overhead caused by fitting linear models. 
Compared with all the other learned indices, \texttt{RSMI} takes a much longer time to build ($10^4\times$ slower than \texttt{ZMI}) and even fails to terminate on the largest dataset \texttt{OSM}. 
This is because the deep learning models adopted by \texttt{RSMI} are much more complex than the other indices and the CPU-only training is generally inefficient compared with using GPUs. 
Though accelerating query processing using novel hardware like GPU and TPU is getting popular recently~\cite{rosenfeld2022query}, in this paper, we focus on the general application scenarios of multi-dimensional indices where a powerful GPU is usually not available, thus the CPU-based \texttt{RSMI} is evaluated. 

\noindent\underline{\textbf{Takeaways.}} 
The construction time highly depends on the internal model choices of multi-dimensional learned indices. 
Indices based on PGMIndex are generally as fast as efficient non-learned baselines; on the other hand, indices employing deep learning techniques are much more slower.

\subsection{Index Memory Cost (Q2)}\label{subsec:exp_memory_cost}
In this section, we report the memory cost of each compared method. 
Note that, for indices \texttt{RSMI}, \texttt{kdtree}, \texttt{qdtree}, and \texttt{ANN}, we cannot programmatically retrieve their memory cost at runtime. 
Thus, we perform heap profiling for these indices via \texttt{gperftools}~\cite{gperf} starting before building the index and ending after the index is constructed\footnote{Such a method for measuring memory cost at runtime is also adopted in a recent benchmark on modern spatial libraries~\cite{DBLP:journals/dase/PandeyRKK21}.}.

Table~\ref{tab:exp_index_memory} shows the memory cost evaluation results. 
As we fix the error threshold $\epsilon$ of the underlying PGMIndex to 64, the resulting learned multi-dimensional indices except \texttt{RSMI} share a similar level of memory cost. 
And all the learned indices achieve significant improvement of memory overhead compared with popular non-learned indices (e.g., on \texttt{Toronto3d} dataset, \texttt{LISA} is $\sim 1000\times$ smaller than \texttt{STR}, $\sim 2000\times$ smaller than \texttt{R*tree}, and $\sim 400\times$ smaller than \texttt{kdtree}). 

Note that, grid indices like \texttt{UG} and \texttt{EDG} also have low space cost (sometimes even better than learned indices). 
This is because these grid indices only require to store the partition boundaries on each dimension. 
However, the grid indices generally perform badly on query processing, especially for skewed and high-dimensional datasets, which will be discussed in Section~\ref{subsec:exp_range_time}. 

As a learned index augmented on top of R-tree, \texttt{IFI}
can also achieve remarkable memory cost as its leaf node capacity is much larger (i.e., 1000) than an ordinary R-tree (i.e., 64), thus requiring much fewer tree nodes to store. 
As for \texttt{RSMI}, though its index construction time is significantly higher than that of other indices ($9124\times$ larger than \texttt{ZMI} on dataset \texttt{Lognormal}, see Table~\ref{tab:exp_construction}), the trained model is generally more compact for storage and not sensitive to the size of various datasets (about 20MB for all datasets). 
Thus, an ideal use-case of \texttt{RSMI} is to perform offline training using machines equipped with powerful GPUs and then deploy the trained models for subsequent query processing. 

\noindent\underline{\textbf{Takeaways.}} All evaluated learned indices can obtain a much more \textbf{compact} structure (up to $7374\times$ reduction) than the commonly used non-learned indices with a sacrifice of affordable index construction cost (except \texttt{RSMI}, which has a much higher construction overhead caused by training deep models).

\begin{figure*}
    \centering
    \begin{subfigure}[b]{0.47\textwidth}
        \includegraphics[width=\textwidth]{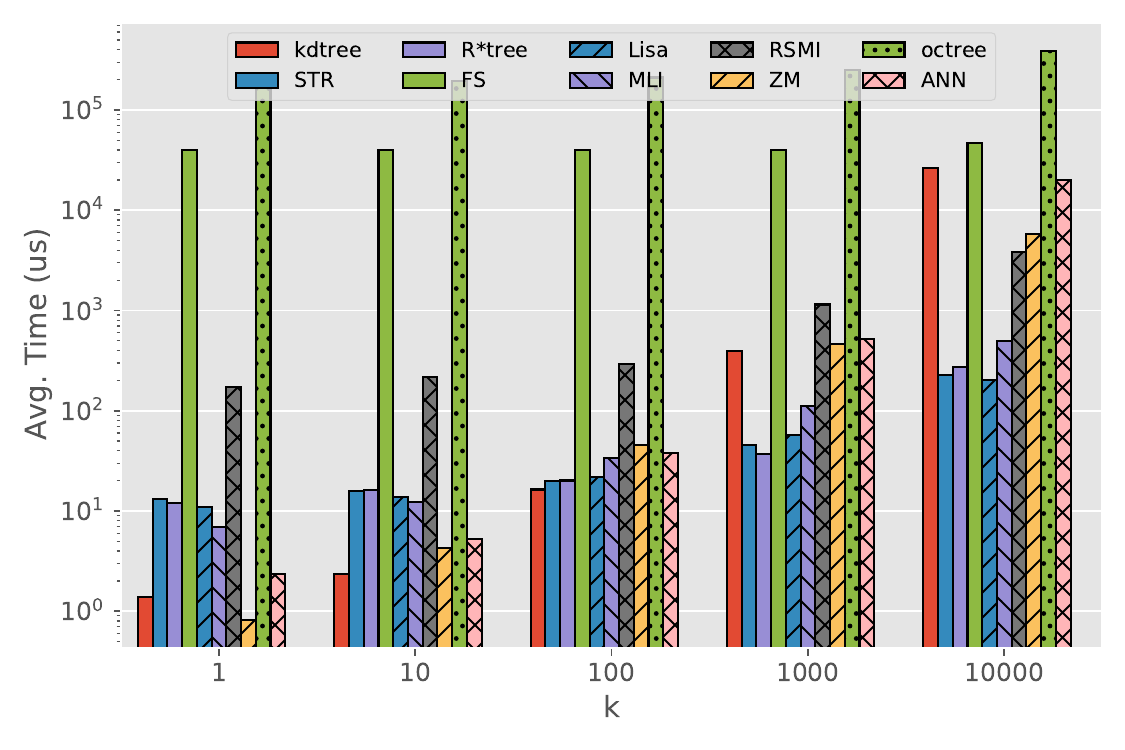}
        \caption{Uniform: Time v.s. $k$}
    \end{subfigure}
    \begin{subfigure}[b]{0.47\textwidth}
        \includegraphics[width=\textwidth]{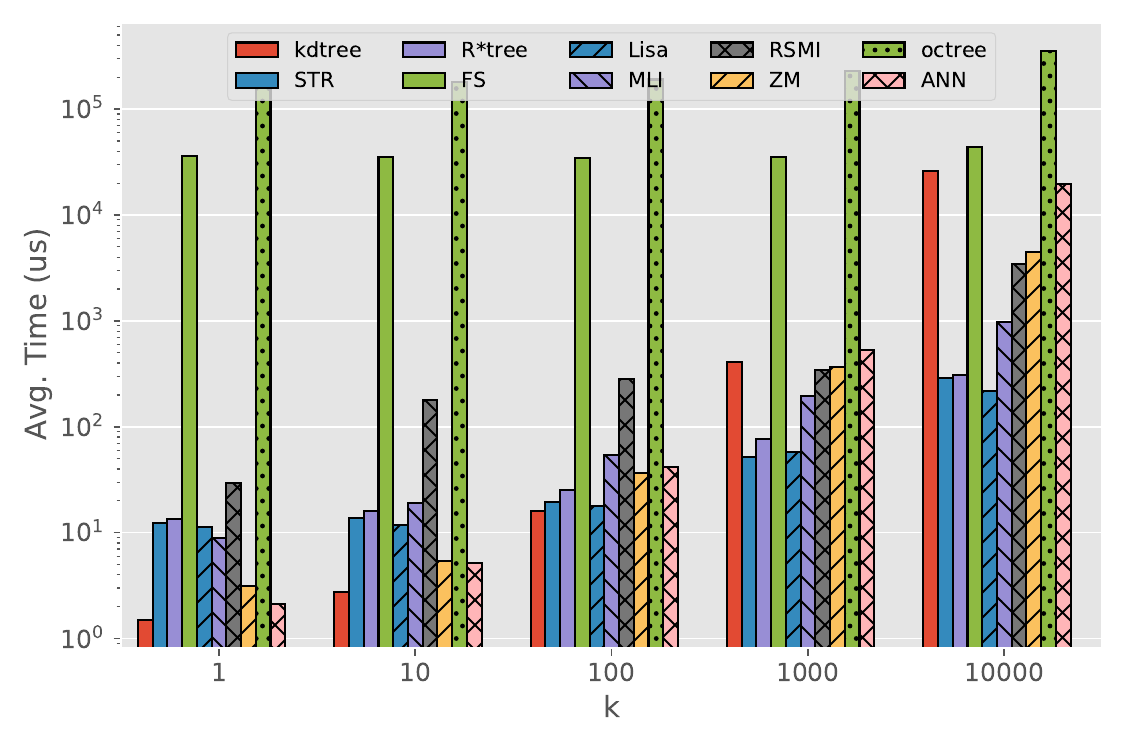}
        \caption{Gaussian: Time v.s. $k$}
    \end{subfigure}
    
    \begin{subfigure}[b]{0.47\textwidth}
        \includegraphics[width=\textwidth]{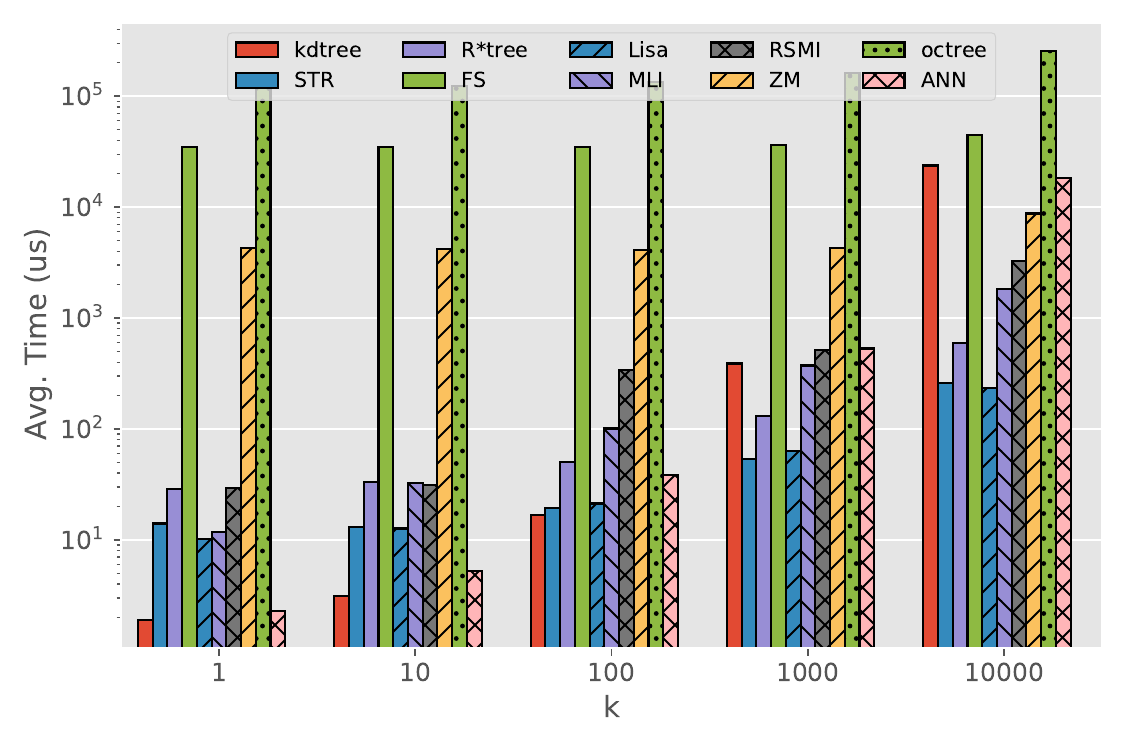}
        \caption{Lognormal: Time v.s. $k$}
    \end{subfigure}
    \begin{subfigure}[b]{0.47\textwidth}
        \includegraphics[width=\textwidth]{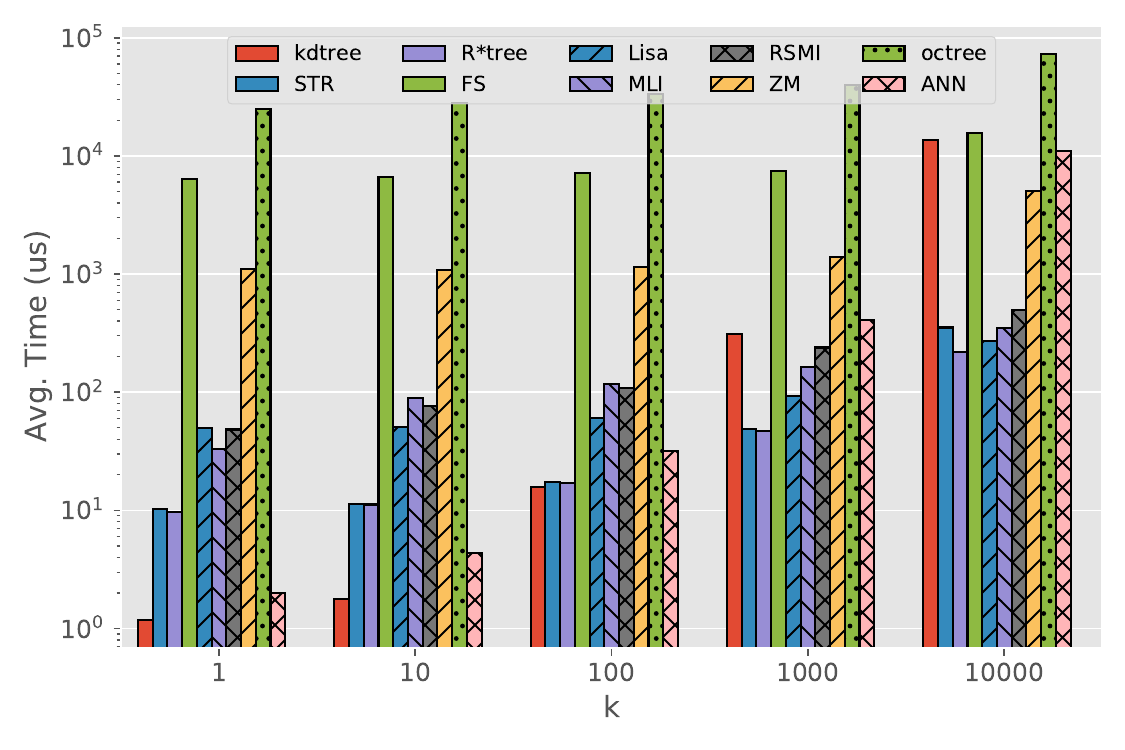}
        \caption{FourSquare: Time v.s. $k$}
    \end{subfigure}
    
    \begin{subfigure}[b]{0.47\textwidth}
        \includegraphics[width=\textwidth]{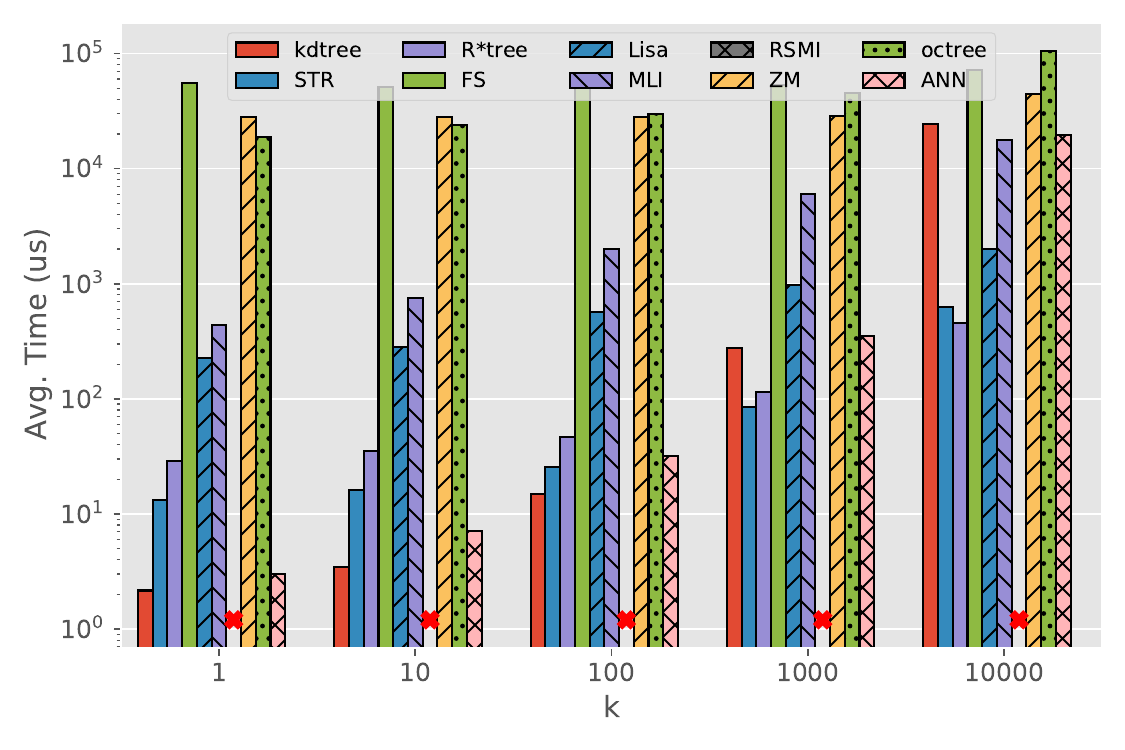}
        \caption{Toronto3d: Time v.s. $k$}
    \end{subfigure}
    \begin{subfigure}[b]{0.47\textwidth}
        \includegraphics[width=\textwidth]{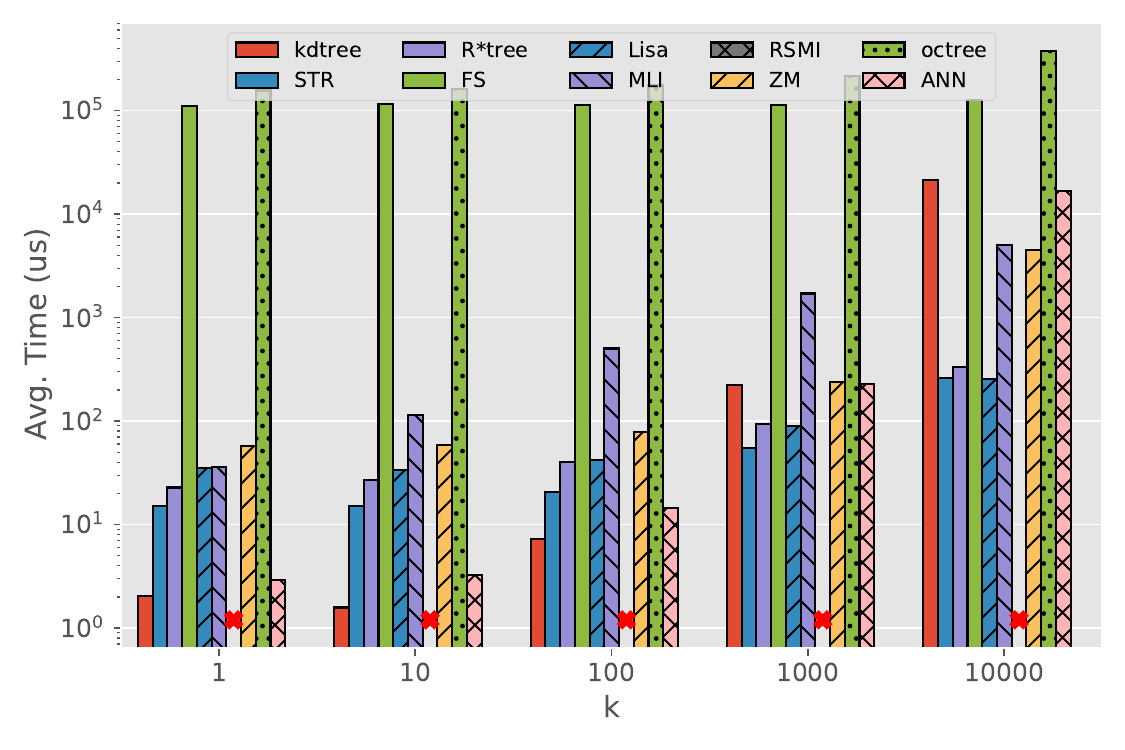}
        \caption{OSM: Time v.s. $k$}
    \end{subfigure}
    \caption{$k$NN query evaluation results on default settings. The y-axis refers to the \emph{average} query processing time for each $k$. 
    Note that \texttt{IFI} and \texttt{Flood} do not support $k$NN queries, and \texttt{RSMI} is not available on \texttt{Toronto3d} and fails to terminate training on \texttt{OSM}.}
    \label{fig:knn_default}
\end{figure*}

\begin{figure*}
    \centering
    \begin{subfigure}[b]{0.47\textwidth}
        \includegraphics[width=\textwidth]{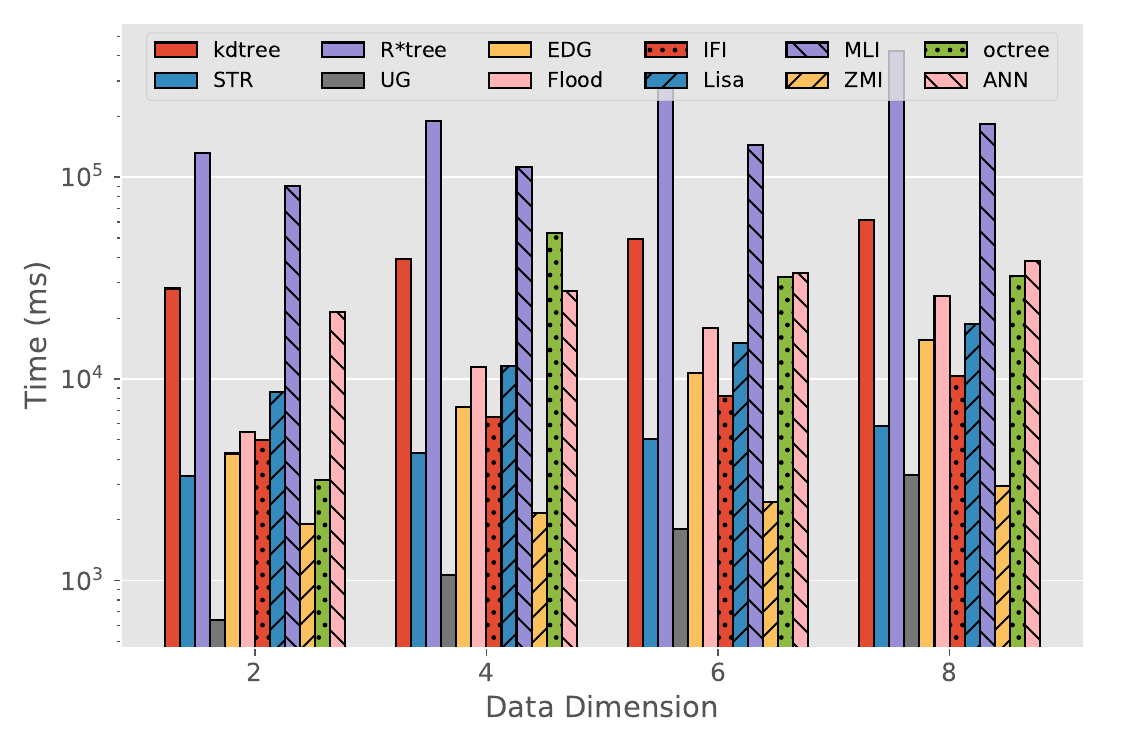}
        \caption{Uniform: Construction Time v.s. $d$}
        \label{fig:uniform_build_d}
    \end{subfigure}
    \begin{subfigure}[b]{0.47\textwidth}
        \includegraphics[width=\textwidth]{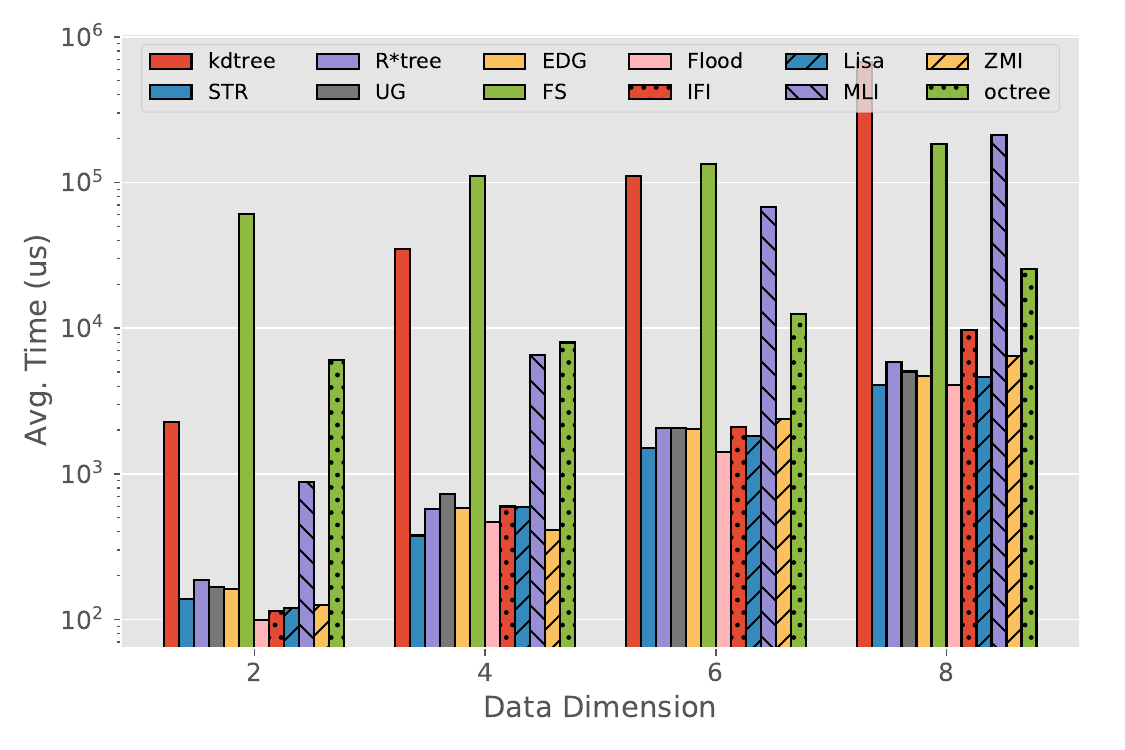}
        \caption{Uniform: Range Query Time v.s. $d$}
        \label{fig:uniform_range_d}
    \end{subfigure}
    
    \begin{subfigure}[b]{0.47\textwidth}
        \includegraphics[width=\textwidth]{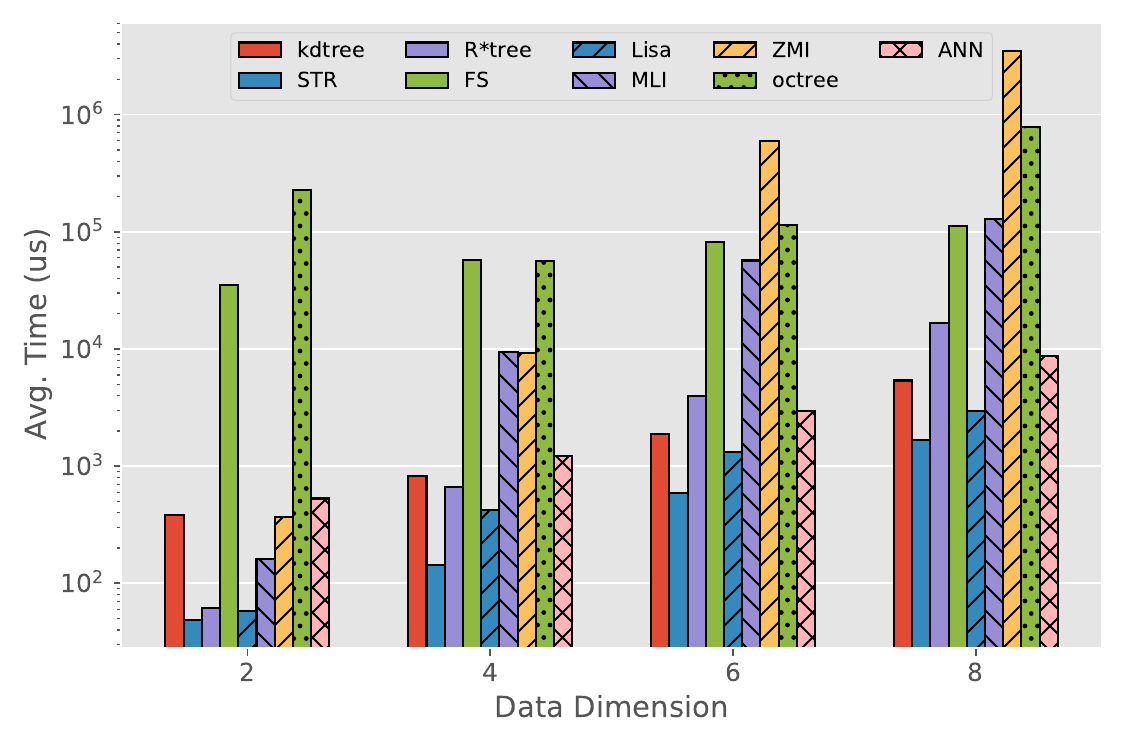}
        \caption{Uniform: $k$NN Query Time v.s. $d$}
        \label{fig:uniform_knn_d}
    \end{subfigure}
    \begin{subfigure}[b]{0.47\textwidth}
        \includegraphics[width=\textwidth]{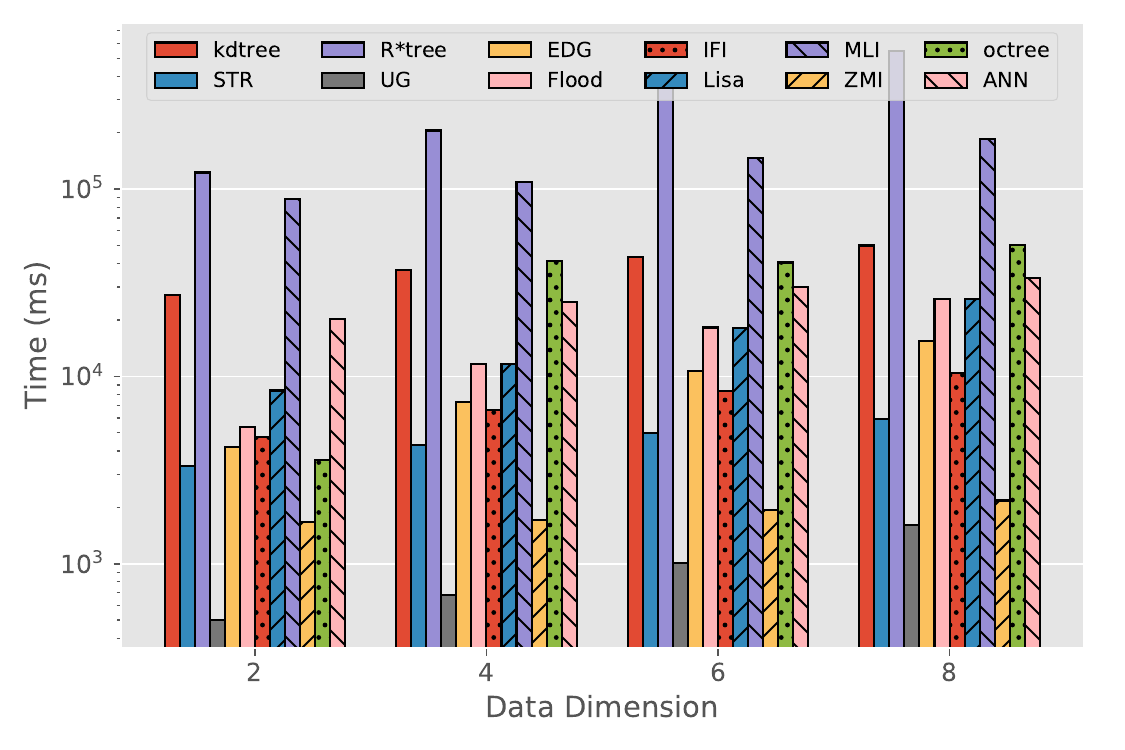}
        \caption{Gaussian: Construction Time v.s. $d$}
        \label{fig:gaussian_build_d}
    \end{subfigure}
    
    \begin{subfigure}[b]{0.47\textwidth}
        \includegraphics[width=\textwidth]{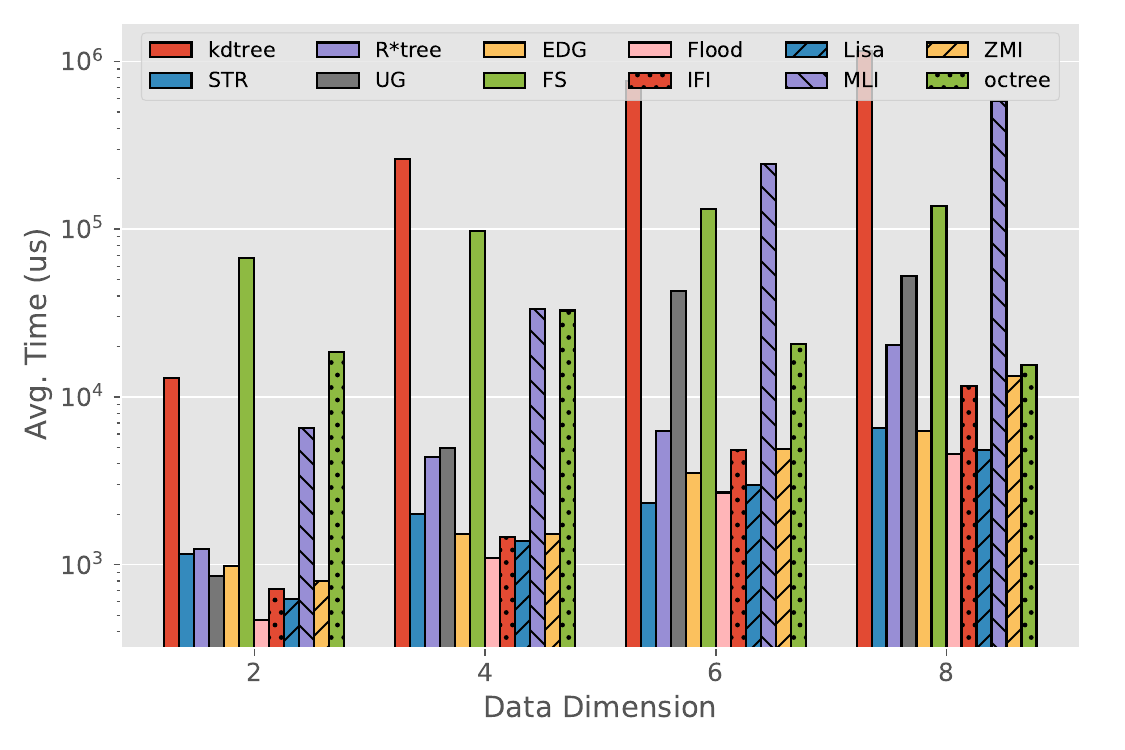}
        \caption{Gaussian: Range Query Time v.s. $d$}
        \label{fig:gaussian_range_d}
    \end{subfigure}
    \begin{subfigure}[b]{0.47\textwidth}
        \includegraphics[width=\textwidth]{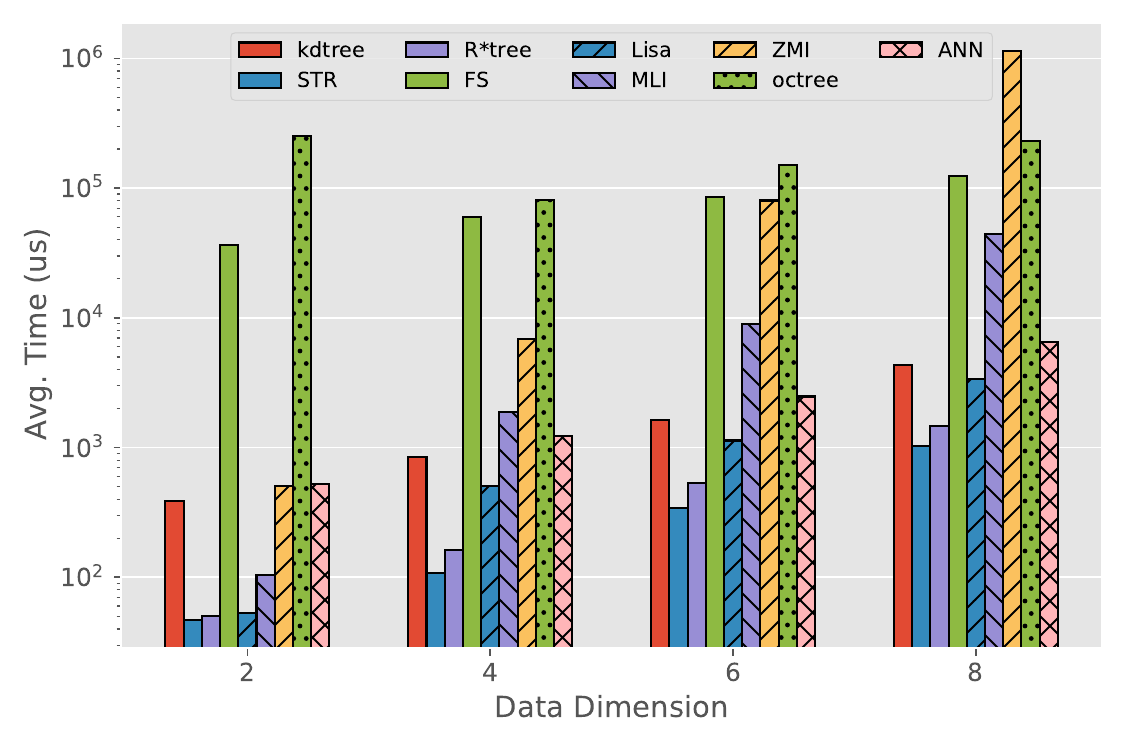}
        \caption{Gaussian: $k$NN Query Time v.s. $d$}
        \label{fig:gaussian_knn_d}
    \end{subfigure}
    \caption{Scalability evaluation w.r.t.~dimension $d$. The range query selectivity is in $[10^{-5}, 10^{-2}]$, and $k=1000$ for $k$NN queries. }
    \label{fig:exp_scalability_dim}
\end{figure*}

\begin{figure*}
    \centering
    \begin{subfigure}[b]{0.47\textwidth}
        \includegraphics[width=\textwidth]{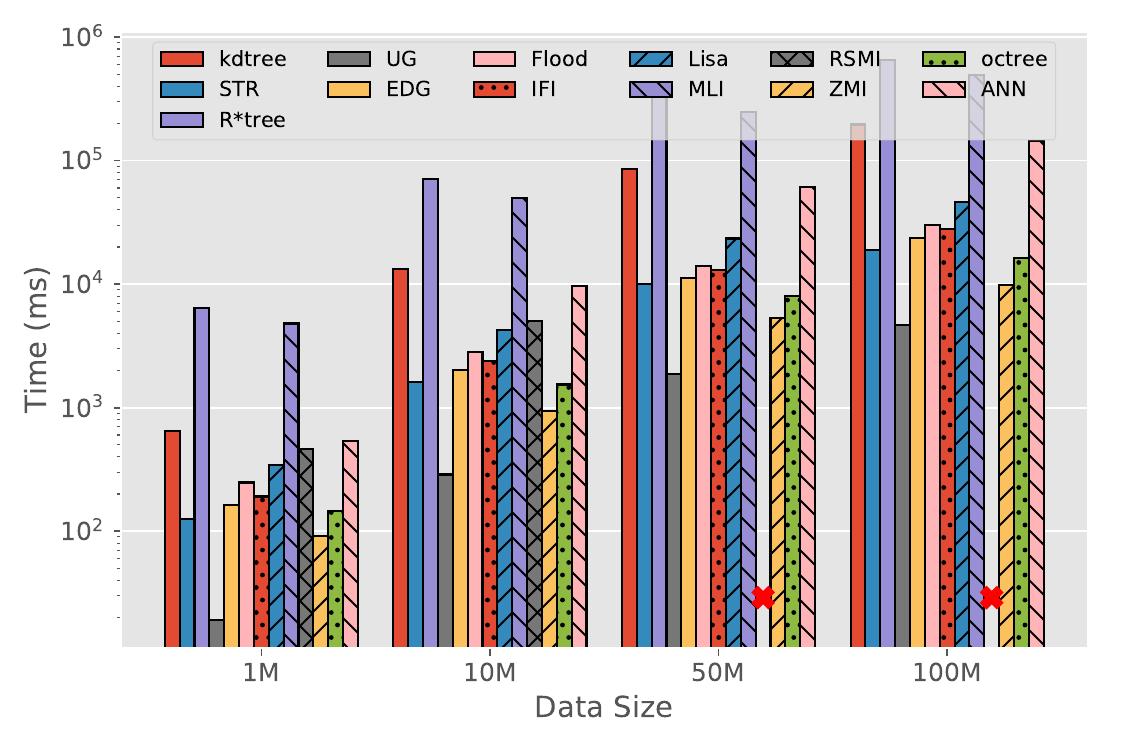}
        \caption{Uniform: Construction time v.s.~$N$}
    \end{subfigure}
    \begin{subfigure}[b]{0.47\textwidth}
        \includegraphics[width=\textwidth]{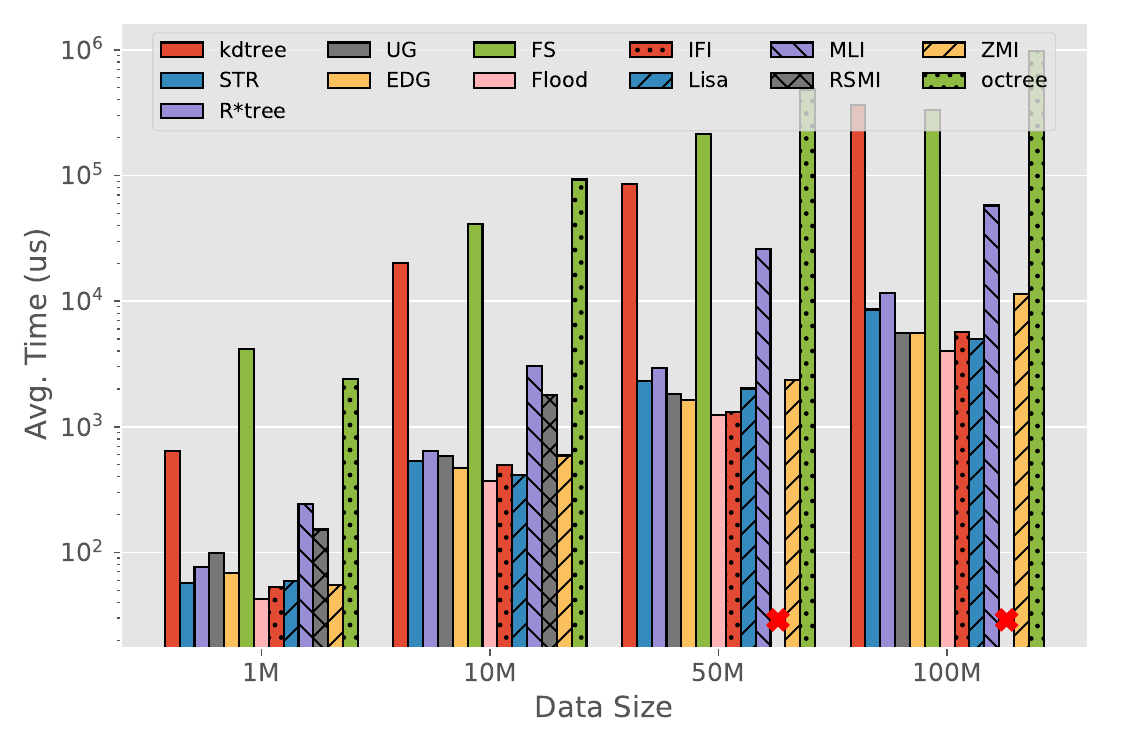}
        \caption{Uniform: Range Query time v.s.~$N$}
    \end{subfigure}
    
    \begin{subfigure}[b]{0.47\textwidth}
        \includegraphics[width=\textwidth]{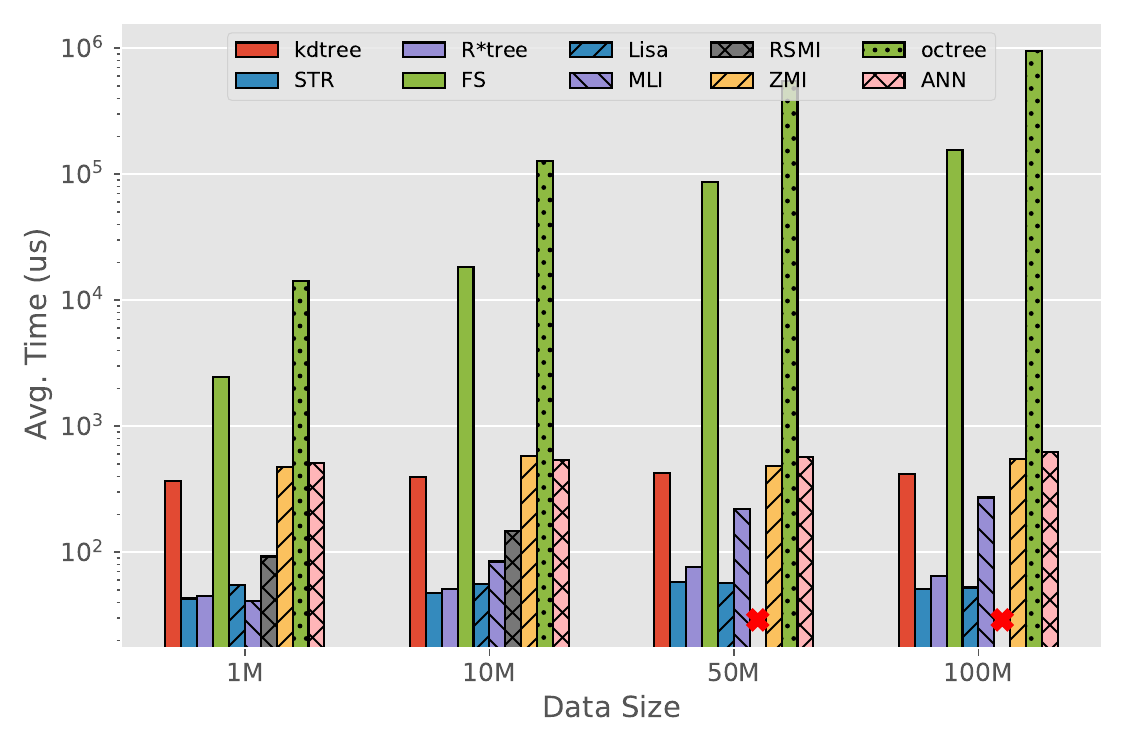}
        \caption{Uniform: $k$NN Query time v.s.~$N$}
    \end{subfigure}
    \begin{subfigure}[b]{0.47\textwidth}
        \includegraphics[width=\textwidth]{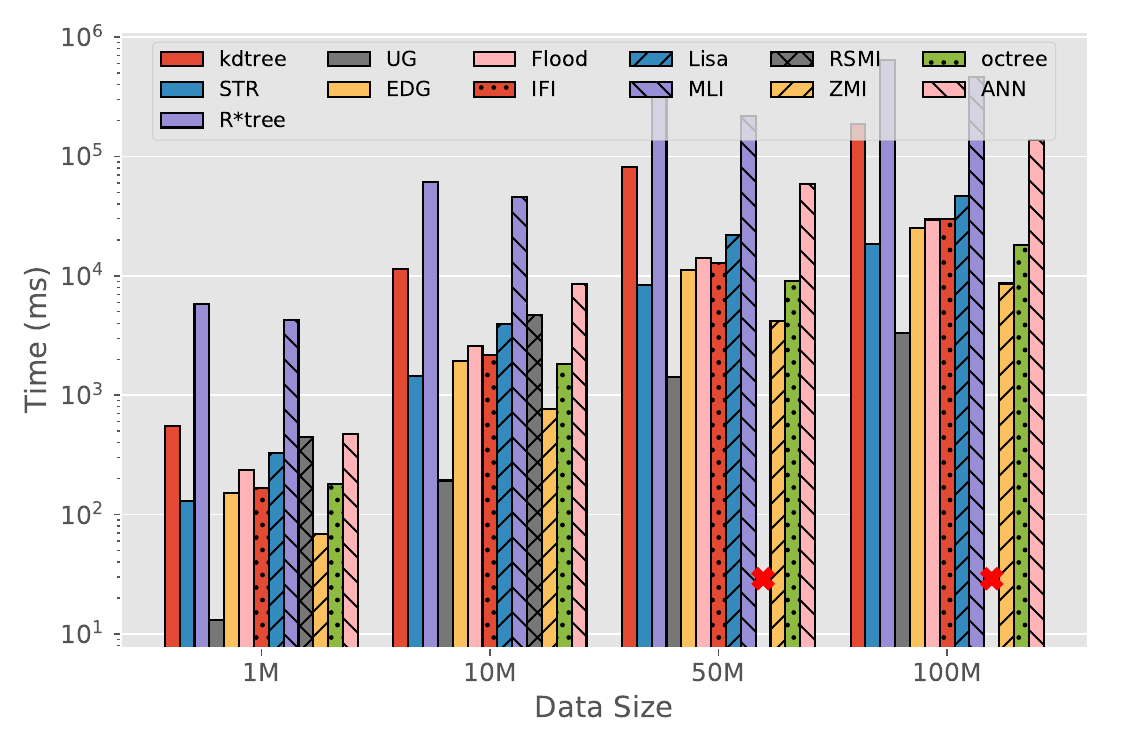}
        \caption{Normal: Construction time v.s.~$N$}
    \end{subfigure}
    
    \begin{subfigure}[b]{0.47\textwidth}
        \includegraphics[width=\textwidth]{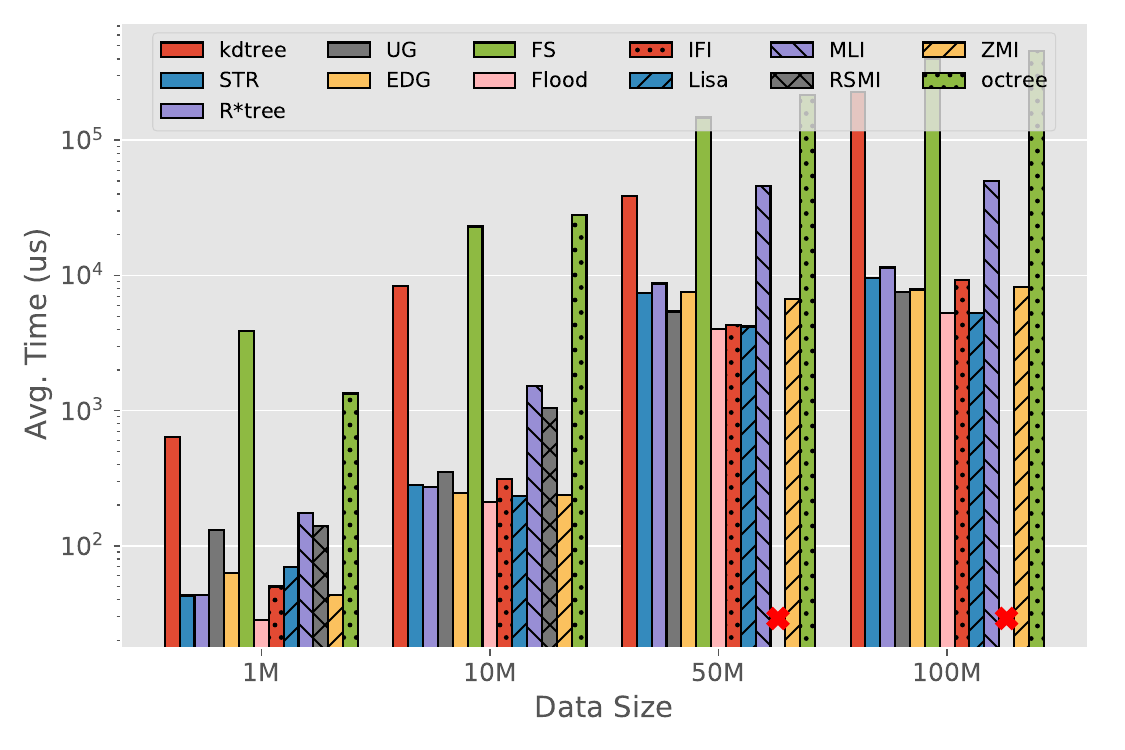}
        \caption{Normal: Range Query time v.s.~$N$}
    \end{subfigure}
    \begin{subfigure}[b]{0.47\textwidth}
        \includegraphics[width=\textwidth]{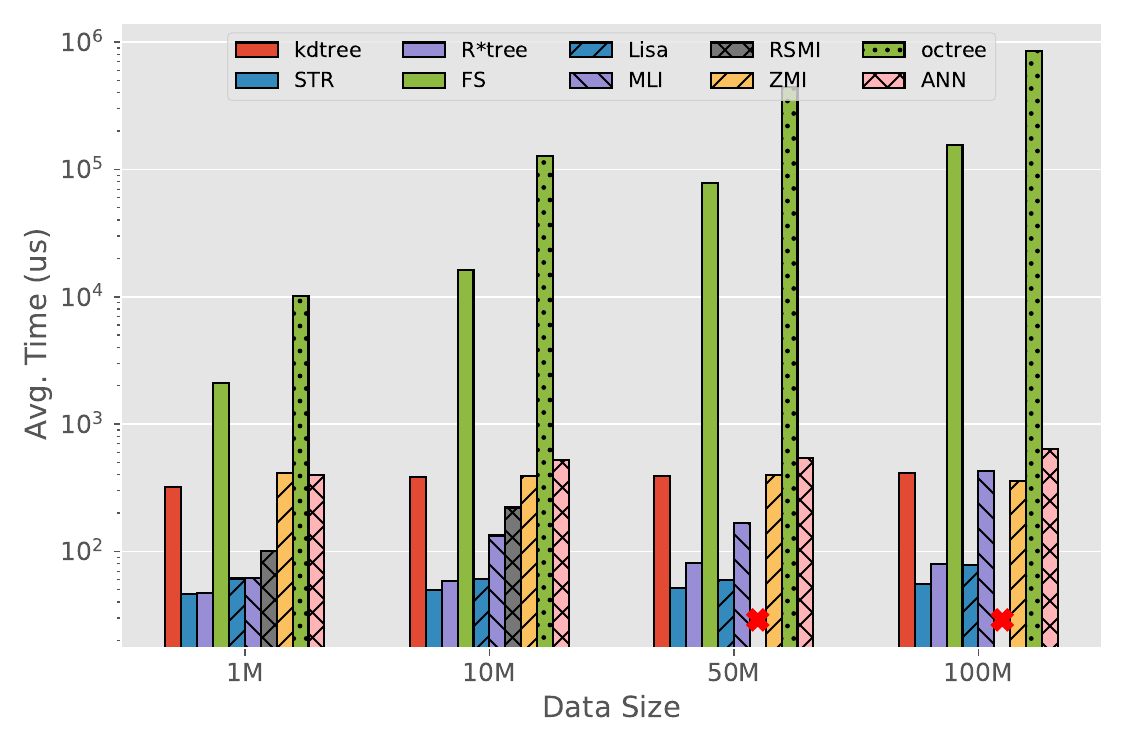}
        \caption{Normal: $k$NN Query time v.s.~$N$}
    \end{subfigure}
    \caption{Scalability evaluation w.r.t.~data size $N$. The range query selectivity is in $[10^{-4}, 10^{-2}]$, and $k=1000$ for $k$NN queries.}
    \label{fig:exp_scalability_n}
\end{figure*}

\subsection{Range Query Processing (Q3)}\label{subsec:exp_range_time}
We then report the range query evaluation results using the queries generated in Section~\ref{subsec:data_query_prepare}. 
Figure~\ref{fig:range_default} shows the average query processing time w.r.t.~different levels of query selectivities. 

Among non-learned baselines, \texttt{STR} and \texttt{EDG} are the fastest indices on range query processing. 
Compared with these two methods, learned indices \texttt{IFI}, \texttt{Flood}, and \texttt{LISA} can always achieve better performance regardless of data skewness and query selectivity. 
Not surprisingly, learned indices perform much better for queries of lower selectivity. 
For example, on dataset \texttt{Lognormal}, \texttt{Flood} is $2.19\times$ faster than \texttt{STR} when query selectivity is less than $1\%$; on the other hand, the speed-up ratio decreases to $1.31$ when the selectivity ratio increases to $48\%$. 
This is because, the pruning power of learned models becomes more significant for small query ranges, and in the worst case, all the indices (learned and non-learned) are no better than a linear scan. 

\texttt{RSMI} does not show satisfactory performance, especially on non-uniform data. This is because \texttt{RSMI} aims to design a disk-based index structure where page access is the major optimized objective. 
Besides, \texttt{RSMI} also supports approximate range query processing, which is about an order of magnitude faster than the exact query processing when query selectivity is lower than $5\%$. 
The performance of \texttt{MLI} is also not satisfactory as its metric-based projection function (Eq.~\eqref{eq:mlindex_projection}) cannot well encode locality information and thus is not suitable for orthogonal range query processing. 

\begin{figure*}[t]
    \centering
    \begin{subfigure}[b]{0.32\textwidth}
        \includegraphics[width=\textwidth]{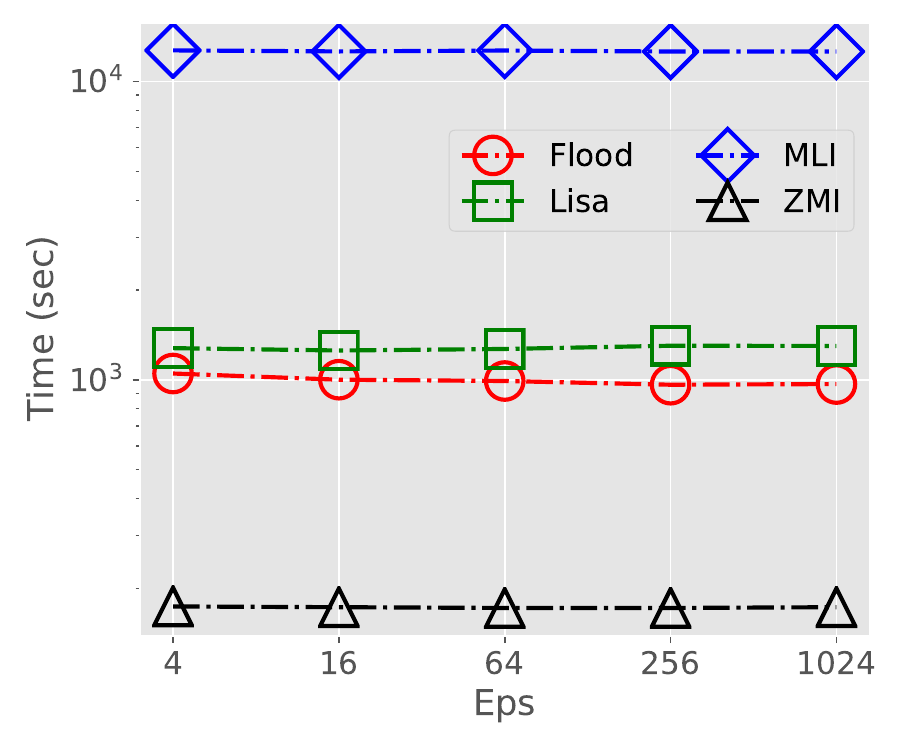}
        \caption{Foursquare: construction time v.s.~$\epsilon$}
    \end{subfigure}
    \begin{subfigure}[b]{0.32\textwidth}
        \includegraphics[width=\textwidth]{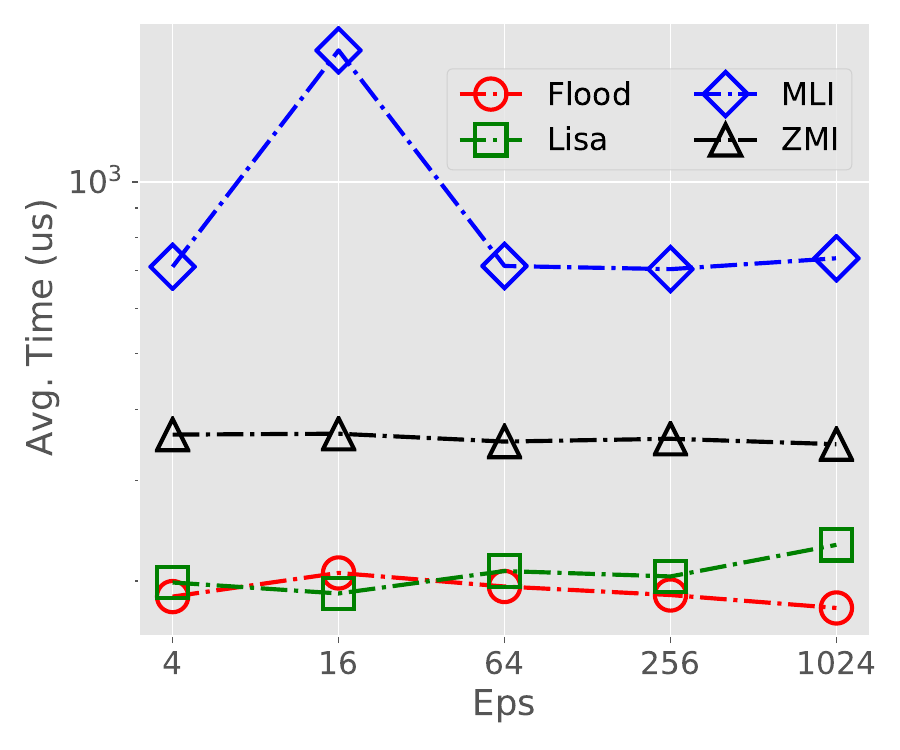}
        \caption{Foursquare: range query v.s.~$\epsilon$}
    \end{subfigure}
    \begin{subfigure}[b]{0.32\textwidth}
        \includegraphics[width=\textwidth]{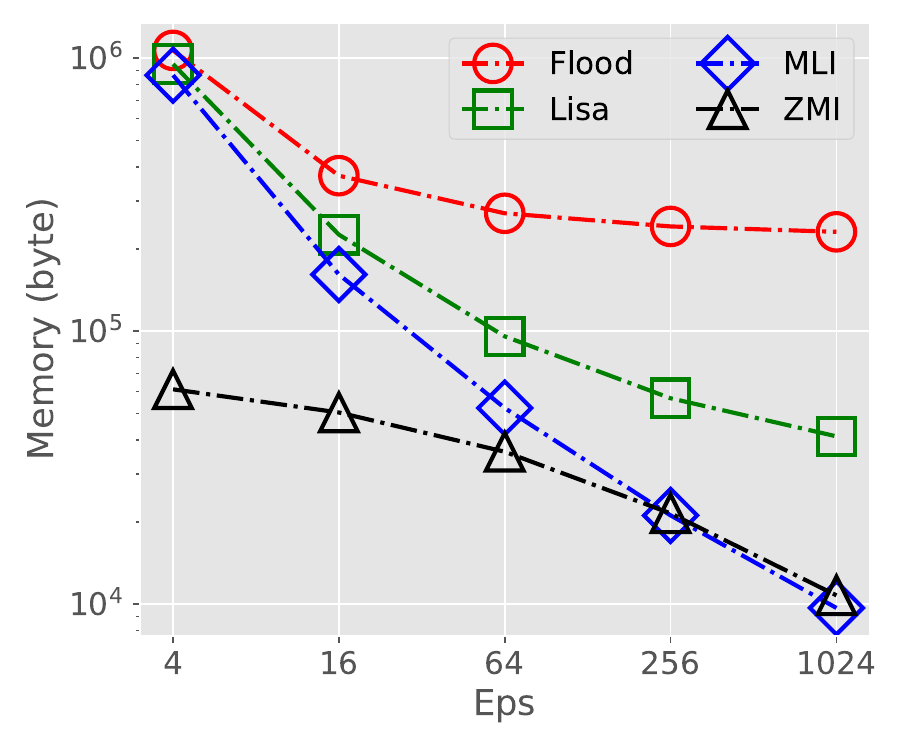}
        \caption{Foursquare: memory cost v.s.~$\epsilon$}
    \end{subfigure}
    
    \begin{subfigure}[b]{0.32\textwidth}
        \includegraphics[width=\textwidth]{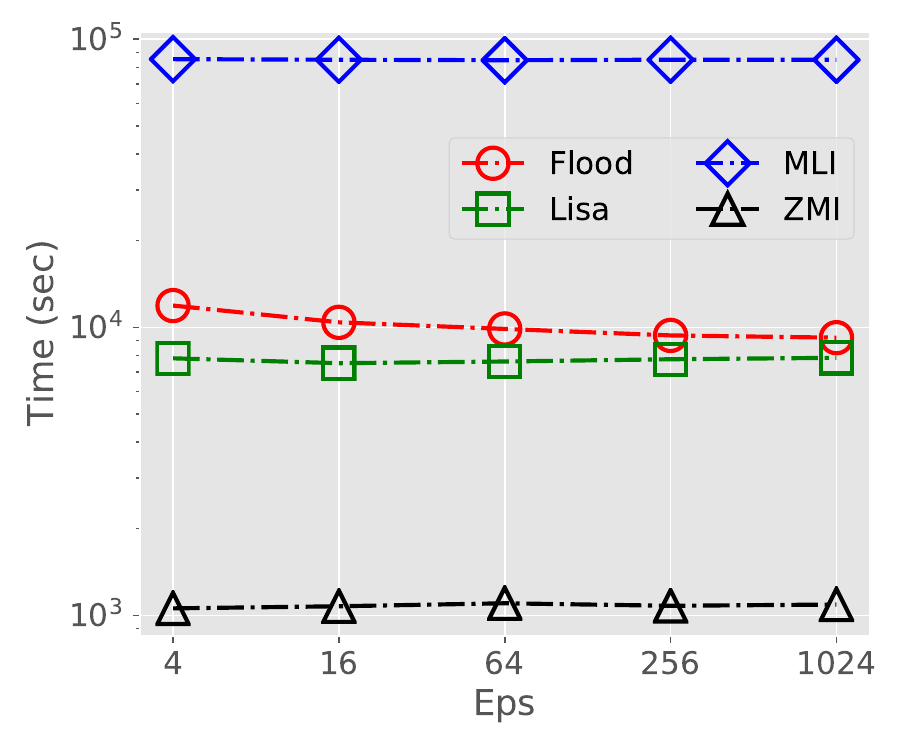}
        \caption{Toronto3d: construction time v.s.~$\epsilon$}
    \end{subfigure}
    \begin{subfigure}[b]{0.32\textwidth}
        \includegraphics[width=\textwidth]{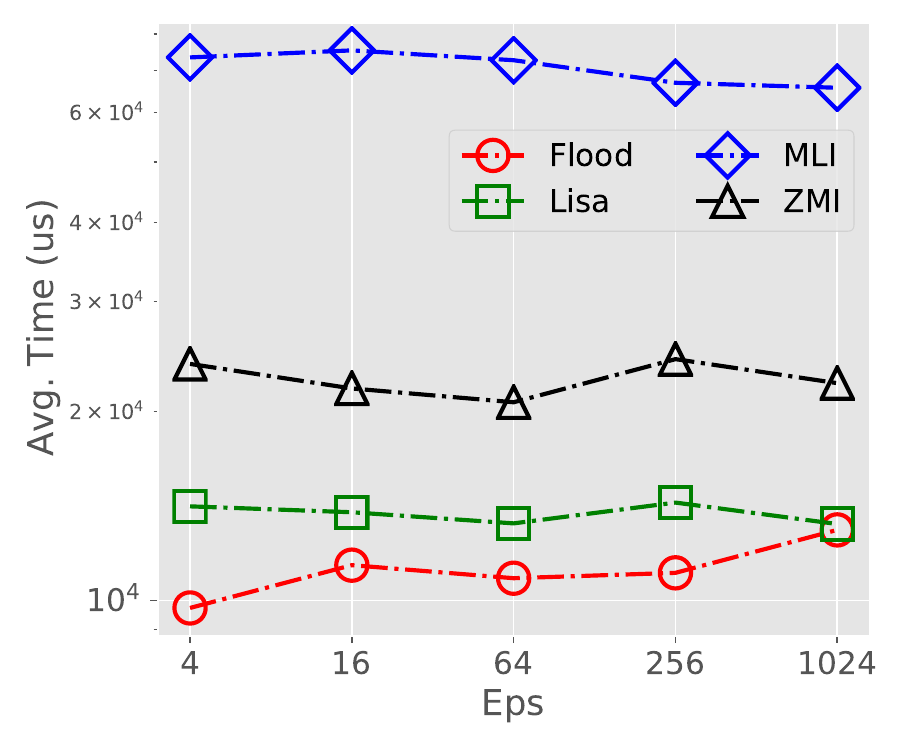}
        \caption{Toronto3d: range query v.s.~$\epsilon$}
    \end{subfigure}
    \begin{subfigure}[b]{0.32\textwidth}
        \includegraphics[width=\textwidth]{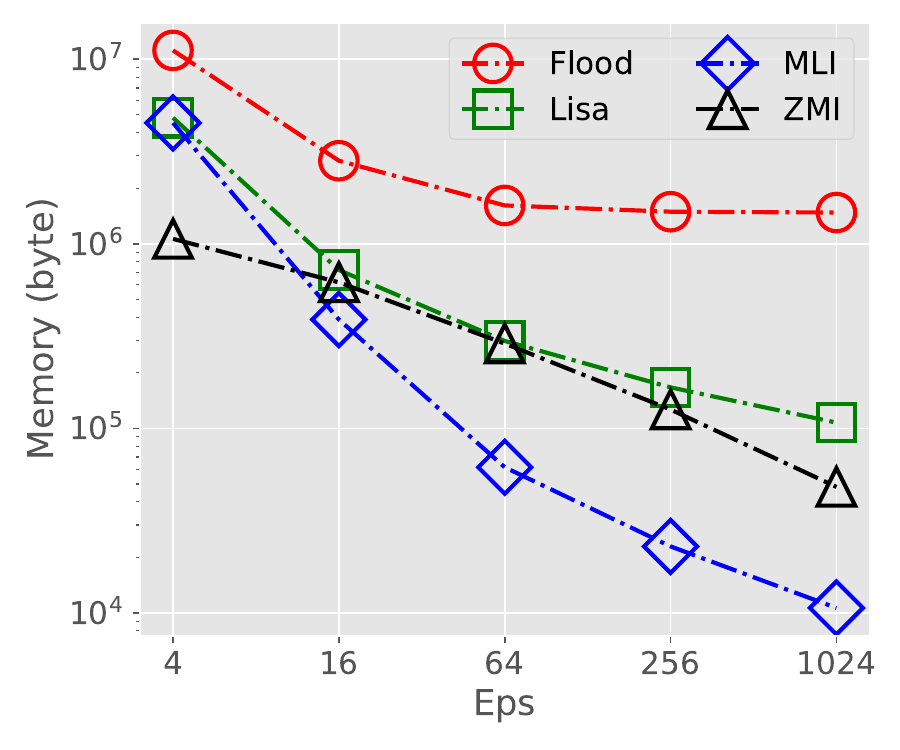}
        \caption{Toronto3d: memory cost v.s.~$\epsilon$}
    \end{subfigure}
    \caption{Evaluation results w.r.t.~error parameter $\epsilon$ (for learned indices that are based on PGM-Index) on two real datasets \texttt{Foursquare} and \texttt{Toronto3d}.}
    \label{fig:exp_error}
\end{figure*}

An interesting point is that, as the first learned multi-dimensional index, \texttt{ZMI} is usually regarded as a weak baseline in previous studies~\cite{DBLP:journals/pvldb/QiLJK20,DBLP:conf/sigmod/Li0ZY020,DBLP:conf/sigmod/NathanDAK20}. 
However, from our evaluation, the performance of \texttt{ZMI} is not considerably worse and is comparable with \texttt{STR} (recalling that \texttt{ZMI} has very low construction time and memory cost). 
The major reason is that the original implementation of \texttt{ZMI}~\cite{DBLP:conf/mdm/WangFX019} adopts a coarse pruning strategy where the whole range of [Z(min\_corner), Z(max\_corner)] should be searched. 
In this work, we use the BigMin algorithm~\cite{DBLP:conf/vldb/RamsakMFZEB00} to divide a query box into fine-grained candidate Z-value ranges and perform a learned index-based search over the points sorted by Z-values, leading to a significant performance improvement.

\noindent\underline{\textbf{Takeaways.}} 
Learned indices \texttt{IFI}, \texttt{Flood}, and \texttt{LISA} can robustly outperform the non-learned baselines across all levels of query selectivity. 
The speedup ratio is especially significant for \textbf{uniformly} distributed datasets and \textbf{low-selectivity} queries. 
On the other hand, other indices like \texttt{ZMI}, \texttt{MLI}, and \texttt{RSMI} cannot systematically outperform non-learned indices. 

\begin{figure*}[t]
    \centering
    \begin{subfigure}[b]{0.24\textwidth}
        \includegraphics[width=\textwidth]{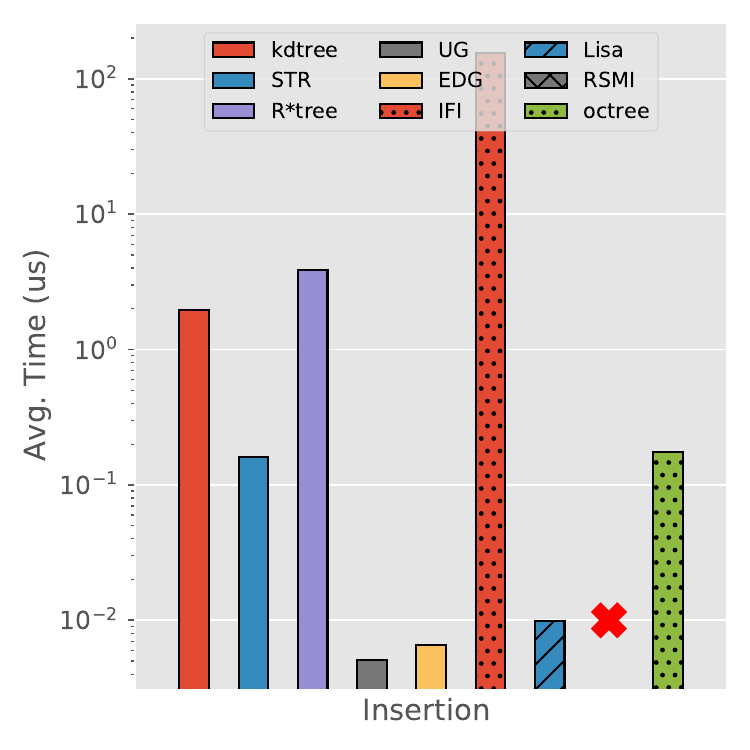}\label{fig:fs_insert}
        \caption{Foursquare: insertion Time}
    \end{subfigure}
	\begin{subfigure}[b]{0.24\textwidth}
        \includegraphics[width=\textwidth]{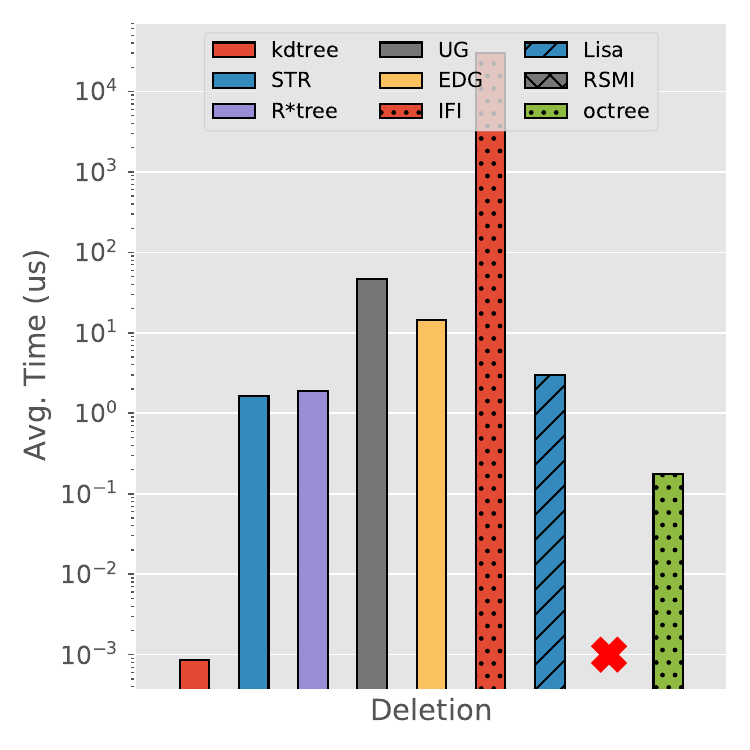}\label{fig:fs_delete}
        \caption{Foursquare: deletion Time}
    \end{subfigure}
    \begin{subfigure}[b]{0.24\textwidth}
        \includegraphics[width=\textwidth]{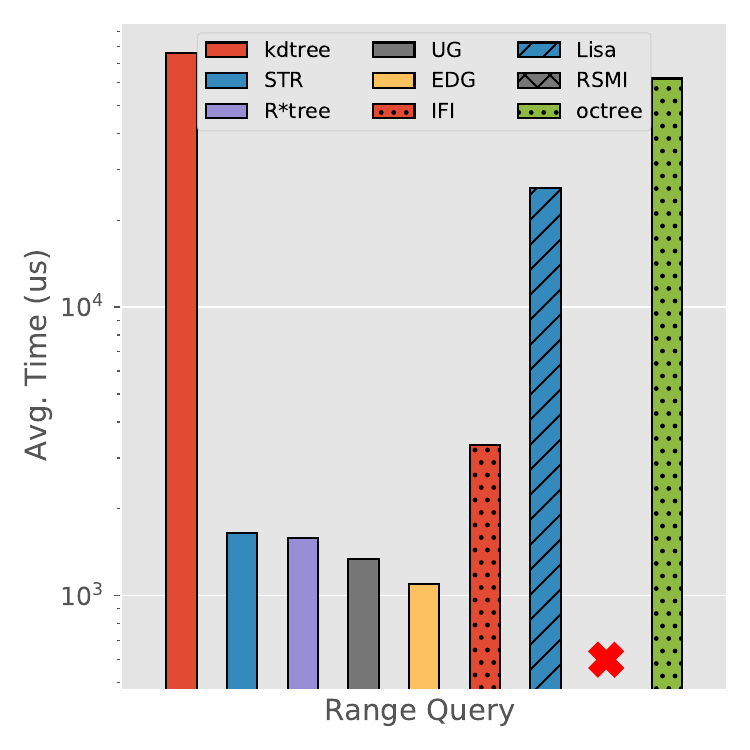}\label{fig:fs_range_update}
        \caption{Foursquare: range Time}
    \end{subfigure}
    \begin{subfigure}[b]{0.24\textwidth}
        \includegraphics[width=\textwidth]{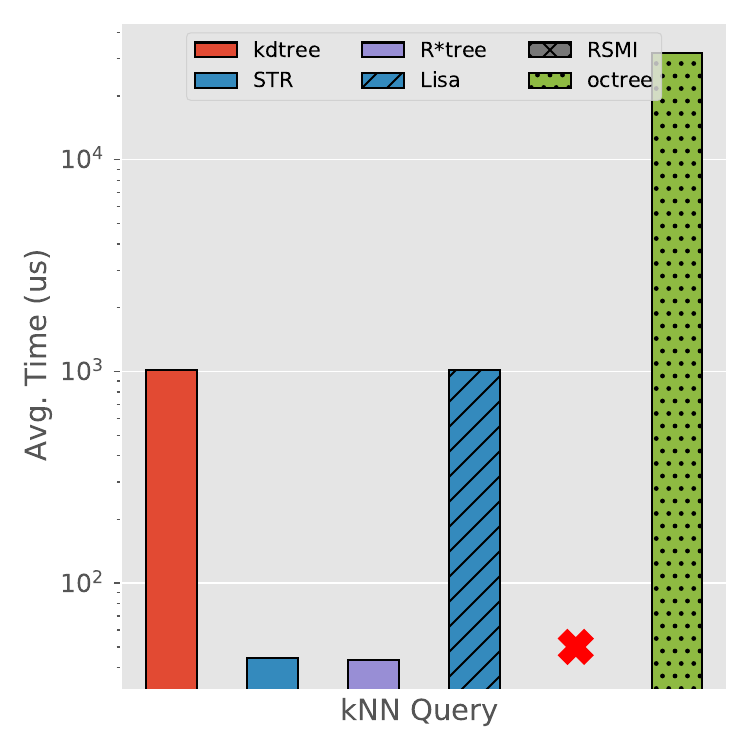}\label{fig:fs_knn_update}
        \caption{Foursquare: $k$nn Time}
    \end{subfigure}

	\begin{subfigure}[b]{0.24\textwidth}
        \includegraphics[width=\textwidth]{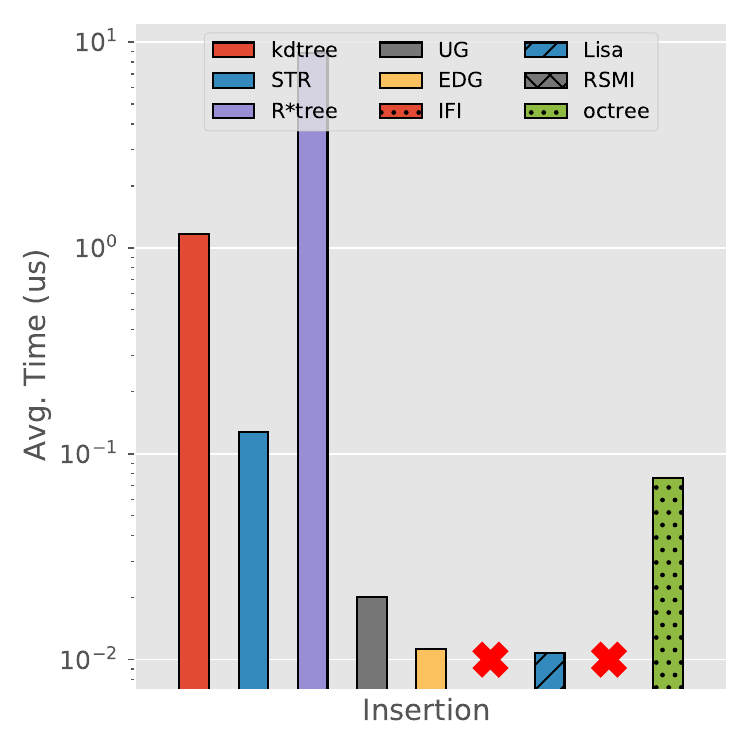}\label{fig:toronto_insert}
        \caption{Toronto3d: insertion Time}
    \end{subfigure}
	\begin{subfigure}[b]{0.24\textwidth}
        \includegraphics[width=\textwidth]{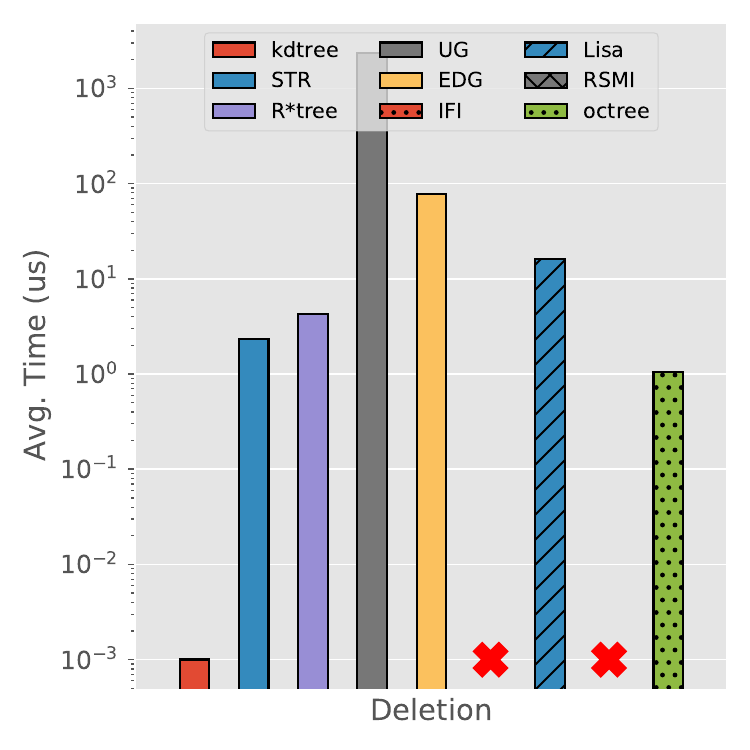}\label{fig:toronto_delete}
        \caption{Toronto3d: deletion Time}
    \end{subfigure}
    \begin{subfigure}[b]{0.24\textwidth}
        \includegraphics[width=\textwidth]{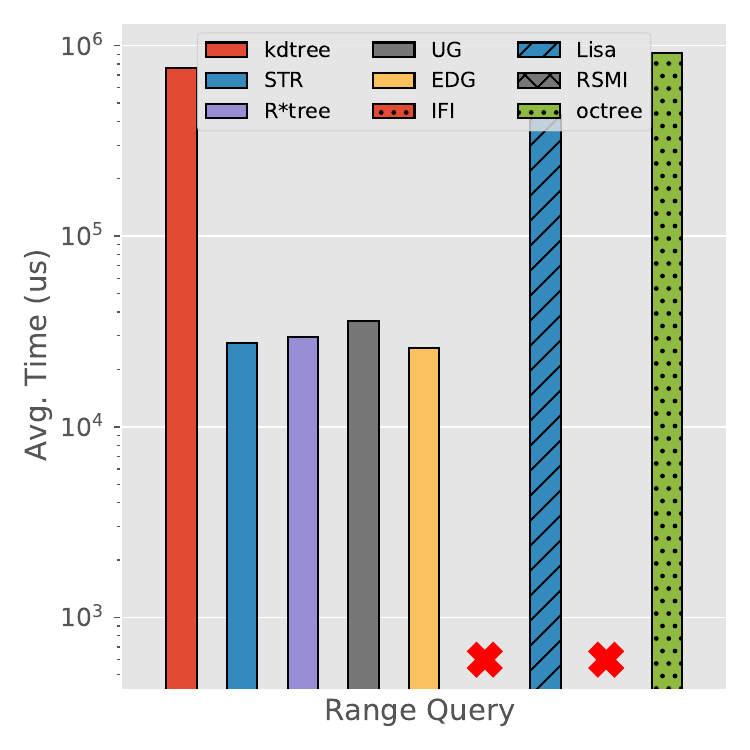}\label{fig:toronto_range_update}
        \caption{Toronto3d: range Time}
    \end{subfigure}
    \begin{subfigure}[b]{0.24\textwidth}
        \includegraphics[width=\textwidth]{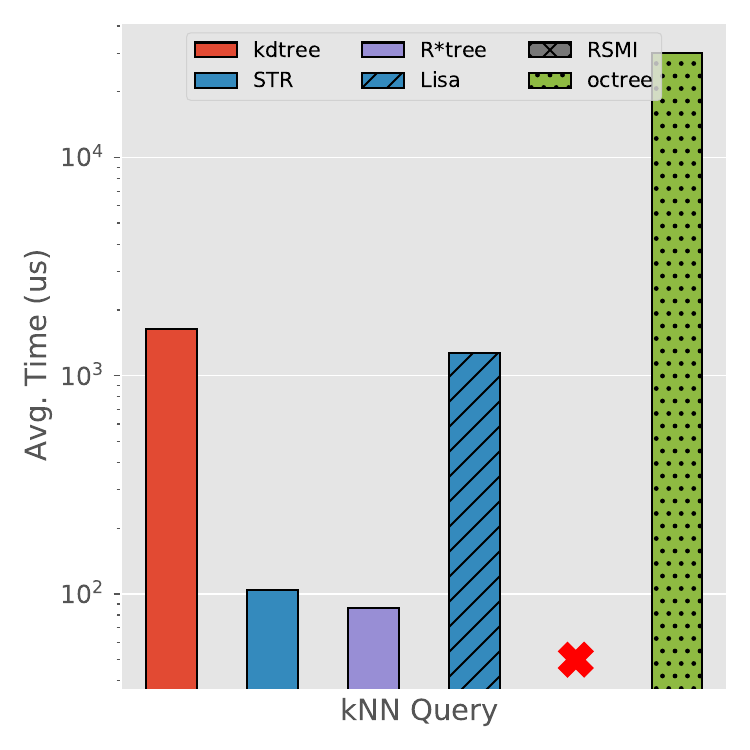}\label{fig:toronto_knn_update}
        \caption{Toronto3d: $k$nn Time}
    \end{subfigure}
    \caption{Data update evaluation results. Note that, \texttt{RSMI} fails to terminate on both datasets and \texttt{IFI} fails to terminate on the \texttt{Toronto3d} dataset.}
    \label{fig:update}
\end{figure*} 

\subsection{kNN Query Processing (Q4)}\label{subsec:exp_knn_analysis}
We further evaluate the $k$NN query processing efficiency. 
Note that, learned indices \texttt{IFI} and \texttt{Flood} focus on range query processing and do not support $k$NN queries, thus are excluded from this set of experiments. 
Figure~\ref{fig:knn_default} reports the average $k$NN query processing time w.r.t.~different $k\in\{1, 10, 100, 1000, 10000\}$. 

For non-learned indices, two $k$d-tree variants \texttt{kdtree} and \texttt{ANN} perform the best for $k\leq 100$; and R-tree variants  \texttt{STR} and \texttt{R*tree} are more efficient for $k>100$. 
Different from range queries where learned indices can generally outperform non-learned baselines, only \texttt{MLI}, \texttt{LISA}, and \texttt{ZM} can outperform the highly optimized R-tree and $k$d-tree libraries on \emph{some} datasets and \emph{some} $k$. 
For example, on synthetic datasets, \texttt{MLI} is $1.1\sim 1.3\times$ faster than \texttt{STR} when $k\leq 10$; however, \texttt{MLI} can be up to $3.6\times$ slower compared with \texttt{STR} when $k$ increases to 10000. 
And unfortunately, no multi-dimensional learned index can beat the best non-learned baselines on all three real-world datasets. 

The intrinsic reason is that, most of the existing learned indices process $k$NN queries by repeatedly invoking \emph{range queries} of progressively increased search radius until the nearest-$k$ results are found. 
Thus, the $k$NN processing time highly depends on the proper setting of the initial search range. 
In the worst case, such a progressive search method requires invoking range queries from a very small range to the whole data space, where a large portion of range queries are unnecessary to retrieve $k$NN results. 
Besides, different from R-tree or $k$d-tree variants that are based on recursive space partitioning, it is generally intractable to inject local aggregation information like range count values into the learned index structures, also making it hard to perform efficient candidate pruning. 

\noindent\underline{\textbf{Takeaways.}} Different from range queries, the nature of progressive range search prevents efficient $k$NN query processing for existing multi-dimensional learned indices. 
The $k$NN processing efficiency for learned indices \textbf{cannot} systematically outperform non-learned baselines, especially on datasets of high dimensions.

\subsection{Scalability Evaluation (Q5)}\label{subsec:exp_scalability}
This section studies the index performance (including index construction, range query processing, and $k$NN query processing) when scaling the size and dimension of datasets. 

Figure~\ref{fig:exp_scalability_dim} shows the scalability evaluation results by varying data dimension $d\in\{2, 4, 6, 8\}$ for two synthetic datasets. 
The construction time for all compared indices grows as $d$ increases. 
For all grid-based learned indices (i.e., \texttt{Flood} and \texttt{LISA}), we set the partition number of each dimension to $(N/B)^{1/d}$, which avoids the exponentially growing (w.r.t.~$d$) of the grid size and makes each grid cell contains roughly $B$ points. 
Note that, \texttt{ZMI} is insensitive to the increase of $d$ as increasing $d$ slightly slows down the computation of Z-curve values which occupies a relatively small portion of the whole index construction. 
For query processing efficiency, the results are similar to those discussed in Section~\ref{subsec:exp_range_time} and Section~\ref{subsec:exp_knn_analysis}. 
Specifically, \texttt{Flood} and \texttt{LISA} can robustly achieve better range query performance when $d$ scales.  
As for $k$NN query processing, \texttt{MLI} is the best learned index on low dimensional data ($d\leq 3$) as \texttt{MLI} can be viewed as a learned version of ``iDistance + B-tree'', which encodes the distance information and thus is more suitable for $k$NN query processing. 
However, the performance of \texttt{MLI} is worse than that of \texttt{Flood} and \texttt{LISA} when $d$ increases. 
This is because the projection function based on Euclidean distance (i.e., Eq.~\eqref{eq:mlindex_projection}) cannot well distinguish the differences of two points to a query point when $d$ is large~\cite{aggarwal2001surprising}. 
Besides, the cost of computing Eq.~\eqref{eq:mlindex_projection} also increases as $d$ increases, becoming a non-neglectable part of the total $k$NN query processing overhead.

Figure~\ref{fig:exp_scalability_n} also shows the scalability evaluation results by varying the size of synthetic datasets from 1M to 100M. 
The index construction time for learned indices is basically proportional to the increase of data size $N$ considering that all the learned indices require sorting the data based on some attributes and then training models on the sorted data. 
\texttt{RSMI} fails to terminate training for data sizes larger than 50M (i.e., exceeding 5 hours). 
For range queries of a fixed selectivity, the processing time of all indices increases as there are more points that need to be reported; 
however, the relative ratio between learned and non-learned indices is similar to that of the default setting (Figure~\ref{fig:range_default}). 


\noindent\underline{\textbf{Takeaways.}} Existing multi-dimensional learned indices can generally scale to large-scale datasets (up to 100M) and a medium level of dimensions (up to 8). 
Specifically, indices \texttt{Flood} and \texttt{LISA} can robustly outperform conventional indices in range query processing time. 
Index \texttt{RSMI} fails to scale to datasets of size up to 50M as the model training cannot be terminated within 5 hours.

\begin{table*}
\centering
\caption{CPU performance counter statistics (obtained by \texttt{perf}) when processing range queries on real and synthetic datasets of default configurations. 
\texttt{\#CR}, \texttt{CMR\%}, \texttt{\#BR}, \texttt{BMR\%} refer to the count of total cache references, cache miss rate, count of CPU branches, and CPU branch predication miss rate, respectively. 
The cell colors from left to right {\color[HTML]{CB7373}{$\blacksquare$}}{\color[HTML]{FFCCC9}{$\blacksquare$}}{\color[HTML]{FFECEC}{$\blacksquare$}}{\color[HTML]{B4FFB4}{$\blacksquare$}}{\color[HTML]{329832}{$\blacksquare$}} represent range query processing costs ranging from high to low.}
\label{tab:exp_cache}
\begin{tabular}{|c|c|c|c|c|c|c|}
\hline
\multirow{3}{*}{Index} & \texttt{Uniform} & \texttt{Normal} & \texttt{Lognormal} & \texttt{Foursquare} & \texttt{Toronto3d} & \texttt{OSM} \\\cline{2-7}
& \texttt{\#CR (CMR\%)} & \texttt{\#CR (CMR\%)} & \texttt{\#CR (CMR\%)} & \texttt{\#CR (CMR\%)} & \texttt{\#CR (CMR\%)} & \texttt{\#CR (CMR\%)} \\
 & \texttt{\#BR (BMR\%)} & \texttt{\#BR (BMR\%)} & \texttt{\#BR (BMR\%)} & \texttt{\#BR (BMR\%)} & \texttt{\#BR (BMR\%)} & \texttt{\#BR (BMR\%)} \\\hline
\multirow{2}{*}{\texttt{ZMI}}  & \cellcolor[HTML]{FFCCC9}   229M (33.86\%)   & \cellcolor[HTML]{FFCCC9} 136M (31.51\%)  & \cellcolor[HTML]{FFECEC} 81M (38.48\%)      & \cellcolor[HTML]{FFCCC9} 51M (11.37\%)      & \cellcolor[HTML]{FFCCC9} 288M (48.45\%)     & \cellcolor[HTML]{FFCCC9} 581M (53.11\%) \\
              & \cellcolor[HTML]{FFCCC9}  979M (0.10\%)    & \cellcolor[HTML]{FFCCC9} 544M (0.10\%)   & \cellcolor[HTML]{FFECEC} 331M (0.10\%)      & \cellcolor[HTML]{FFCCC9} 207M (0.09\%)      & \cellcolor[HTML]{FFCCC9} 1.15B (0.07\%)     & \cellcolor[HTML]{FFCCC9} 2.32B (0.04\%) \\
\hline
\multirow{2}{*}{\texttt{MLI}}  & \cellcolor[HTML]{CB7373} 557M (45.33\%)   & \cellcolor[HTML]{CB7373} 362M (46.62\%)  & \cellcolor[HTML]{CB7373} 351M (50.12\%)     & \cellcolor[HTML]{CB7373} 127M (30.12\%)     & \cellcolor[HTML]{CB7373} 2.34B (54.55\%)    & \cellcolor[HTML]{CB7373} 1.50B (58.83\%) \\
              & \cellcolor[HTML]{CB7373} 1.36B (6.60\%)   & \cellcolor[HTML]{CB7373} 899M (7.14\%)   & \cellcolor[HTML]{CB7373} 842M (3.47\%)      & \cellcolor[HTML]{CB7373} 299M (1.45\%)      & \cellcolor[HTML]{CB7373} 4.14B (4.12\%)     & \cellcolor[HTML]{CB7373} 3.39B (6.30\%) \\
\hline
\multirow{2}{*}{\texttt{IFI}}  & \cellcolor[HTML]{329832}  149M (41.54\%)   & \cellcolor[HTML]{B4FFB4} 114M (49.37\%)  & \cellcolor[HTML]{329832} 54M (49.08\%)     & \cellcolor[HTML]{B4FFB4} 42M (24.45\%)     & \cellcolor[HTML]{FFECEC} 225M (56.71\%)    & \cellcolor[HTML]{329832} 356M (54.24\%) \\
              & \cellcolor[HTML]{329832}  197M (1.23\%)   & \cellcolor[HTML]{B4FFB4} 314M (0.90\%)   & \cellcolor[HTML]{329832} 76M (1.54\%)      & \cellcolor[HTML]{B4FFB4} 130M (0.33\%)      & \cellcolor[HTML]{FFECEC} 528M (1.03\%)     & \cellcolor[HTML]{329832} 453M (0.53\%) \\
\hline
\multirow{2}{*}{\texttt{LISA}} & \cellcolor[HTML]{B4FFB4} 153M (41.82\%)   & \cellcolor[HTML]{329832} 98M (35.39\%)  & \cellcolor[HTML]{B4FFB4} 55M (43.71\%)     & \cellcolor[HTML]{B4FFB4} 37M (18.40\%)     & \cellcolor[HTML]{B4FFB4} 215M (52.58\%)    & \cellcolor[HTML]{329832} 421M (54.67\%) \\
              & \cellcolor[HTML]{B4FFB4} 451M (0.88\%)   & \cellcolor[HTML]{329832} 275M (1.21\%)  & \cellcolor[HTML]{B4FFB4} 147M (0.92\%)      & \cellcolor[HTML]{B4FFB4} 107M (1.00\%)      & \cellcolor[HTML]{B4FFB4} 592M (2.37\%)     & \cellcolor[HTML]{329832} 1.14B (0.38\%) \\
\hline
\multirow{2}{*}{\texttt{Flood}} & \cellcolor[HTML]{329832} 152M (40.77\%)   & \cellcolor[HTML]{329832} 97M (33.81\%)  & \cellcolor[HTML]{329832} 55M (43.99\%)     & \cellcolor[HTML]{329832} 34M (15.16\%)      & \cellcolor[HTML]{329832} 198M (52.02\%)    & \cellcolor[HTML]{B4FFB4} 419M (56.44\%) \\
              & \cellcolor[HTML]{329832} 426M (0.65\%)   & \cellcolor[HTML]{329832} 253M (0.83\%)   & \cellcolor[HTML]{329832} 139M (0.69\%)     &  \cellcolor[HTML]{329832} 92M (0.70\%)      & \cellcolor[HTML]{329832} 486M (2.16\%)     & \cellcolor[HTML]{B4FFB4} 1.10B (0.29\%) \\
\hline
\multirow{2}{*}{\texttt{STR}}  & \cellcolor[HTML]{FFECEC} 175M (47.95\%)   &  \cellcolor[HTML]{FFECEC} 106M (45.06\%)  & \cellcolor[HTML]{B4FFB4} 64M (54.49\%)     &  \cellcolor[HTML]{FFECEC} 35M (30.53\%)      & \cellcolor[HTML]{B4FFB4} 205M (48.99\%)    & \cellcolor[HTML]{FFECEC} 470M (57.19\%) \\
              &  \cellcolor[HTML]{FFECEC} 499M (0.42\%)   &  \cellcolor[HTML]{FFECEC} 281M (0.57\%)  & \cellcolor[HTML]{B4FFB4} 171M (0.50\%)     &  \cellcolor[HTML]{FFECEC} 88M (0.48\%)      & \cellcolor[HTML]{B4FFB4} 487M (0.71\%)     & \cellcolor[HTML]{FFECEC} 1.25B (0.27\%) \\
\hline
\multirow{2}{*}{\texttt{R*tree}}  & \cellcolor[HTML]{FFCCC9} 195M (50.74\%)   & \cellcolor[HTML]{FFCCC9} 117M (49.81\%)  & \cellcolor[HTML]{FFCCC9} 70M (56.06\%)     & \cellcolor[HTML]{FFECEC} 38M (32.88\%)      & \cellcolor[HTML]{FFECEC} 214M (52.47\%)    & \cellcolor[HTML]{FFCCC9} 505M (59.04\%) \\
              & \cellcolor[HTML]{FFCCC9} 505M (0.54\%)   & \cellcolor[HTML]{FFCCC9} 278M (0.67\%)  & \cellcolor[HTML]{FFCCC9} 174M (0.73\%)     & \cellcolor[HTML]{FFECEC} 89M (0.60\%)      & \cellcolor[HTML]{FFECEC} 477M (0.58\%)     & \cellcolor[HTML]{FFCCC9} 1.25B (0.30\%) \\
\hline

\multirow{2}{*}{\texttt{qdtree}}  & \cellcolor[HTML]{CB7373} 50.4B (90.50\%)   & \cellcolor[HTML]{CB7373} 19.2B (95.25\%)  & \cellcolor[HTML]{CB7373} 7.20B (93.71\%)     & \cellcolor[HTML]{CB7373} 4.79B (94.60\%)      & \cellcolor[HTML]{CB7373} 13.6B (88.43\%)    & \cellcolor[HTML]{CB7373} 10.4B (80.44\%) \\
              & \cellcolor[HTML]{CB7373} 137B (1.50\%)   & \cellcolor[HTML]{CB7373} 43.8B (1.00\%)  & \cellcolor[HTML]{CB7373} 18.7B (0.67\%)     & \cellcolor[HTML]{CB7373} 12.0B (0.88\%)      &  \cellcolor[HTML]{CB7373} 45.8B (0.38\%)     & \cellcolor[HTML]{CB7373} 81.8B (0.03\%) \\
\hline

\multirow{2}{*}{\texttt{UG}}  &  \cellcolor[HTML]{FFECEC} 169M (40.67\%) & \cellcolor[HTML]{B4FFB4} 104M (42.01\%) & \cellcolor[HTML]{FFCCC9} 94M (57.05\%) & \cellcolor[HTML]{FFCCC9} 43M (24.68\%) & \cellcolor[HTML]{FFCCC9} 350M (57.75\%) & \cellcolor[HTML]{FFECEC} 434M (56.57\%) \\
              & \cellcolor[HTML]{FFECEC} 518M (1.10\%) & \cellcolor[HTML]{B4FFB4} 304M (3.32\%) & \cellcolor[HTML]{FFCCC9} 338M (7.20\%) & \cellcolor[HTML]{FFCCC9} 135M (3.95\%) & \cellcolor[HTML]{FFCCC9} 1.09B (0.85\%) & \cellcolor[HTML]{FFECEC} 1.18B (0.16\%)  \\
\hline

\multirow{2}{*}{\texttt{EDG}}  & \cellcolor[HTML]{B4FFB4} 171M (37.04\%) & \cellcolor[HTML]{FFECEC} 104M (38.17\%) & \cellcolor[HTML]{FFECEC} 62M (39.31\%) & \cellcolor[HTML]{329832} 37M (17.57\%) & \cellcolor[HTML]{329832} 212M (51.65\%) & \cellcolor[HTML]{B4FFB4} 423M (57.92\%) \\
              & \cellcolor[HTML]{B4FFB4} 522M (1.10\%) & \cellcolor[HTML]{FFECEC} 297M (1.61\%) & \cellcolor[HTML]{FFECEC} 176M (1.16\%) & \cellcolor[HTML]{329832} 106M (2.20\%) & \cellcolor[HTML]{329832} 589M (1.00\%) & \cellcolor[HTML]{B4FFB4} 1.13B (0.13\%)  \\
\hline

\multirow{2}{*}{\texttt{Fullscan}}  & \cellcolor[HTML]{CB7373} 3.2B (66.21\%) & \cellcolor[HTML]{CB7373} 1.7B (68.53\%) & \cellcolor[HTML]{CB7373} 620M (65.63\%) & \cellcolor[HTML]{CB7373} 501M (63.78\%) & \cellcolor[HTML]{CB7373} 2.50B (60.56\%) & \cellcolor[HTML]{CB7373} 5.67B (61.88\%) \\
              & \cellcolor[HTML]{CB7373} 14.4B (16.56\%) & \cellcolor[HTML]{CB7373} 8.3B (15.23\%) & \cellcolor[HTML]{CB7373} 2.7B (9.82\%) & \cellcolor[HTML]{CB7373} 2.3B (11.93\%) & \cellcolor[HTML]{CB7373} 9.6B (0.51\%) & \cellcolor[HTML]{CB7373} 28.3B (0.19\%)  \\
\hline
\end{tabular}
\end{table*}

\vspace{-2ex}
\subsection{Effects of Error Parameter (Q6)}\label{subsec:exp_error}
In this section, we study the effect of the error threshold parameter $\epsilon$ for the indices internally using PGM-Index as the 1-D learned index implementation (i.e., \texttt{Flood}, \texttt{LISA}, \texttt{MLI}, and \texttt{ZMI}). 
Figure~\ref{fig:exp_error} shows the index construction time, range query processing time, and index memory cost when varying $\epsilon$ in the range [4, 1024] on two real datasets. 

The index construction time is generally insensitive to $\epsilon$ as the optimal piece-wise linear fitting algorithm adopted by PGM-Index requires only one pass of data (i.e., $O(N)$ time), which contributes a minor part compared with the overhead caused by data partitioning and projection function computation. 

As for range query processing, different from 1-D learned indices where a small $\epsilon$ definitely leads to less search time, the relationship between query time and $\epsilon$ is not monotonic. 
The reason is that a smaller $\epsilon$ can only reduce the overhead of a single point search; however, the most costly part of range query processing is whether the learned models and the underlying data layout can well filter unnecessary result points. 

As for the space cost, our results are similar to the 1-D case reported in~\cite{DBLP:journals/pvldb/FerraginaV20} where decreasing $\epsilon$ will enlarge the space overhead as more line segments are required to satisfy the maximum error constraint. 
According to a recent theoretical analysis of PGM-Index~\cite{DBLP:conf/icml/FerraginaLV20}, the space cost for a PGM-Index of error parameter $\epsilon$ is bounded by $O(N/\epsilon^2)$, which coincides with the memory cost evaluation results as shown in Figure~\ref{fig:exp_error}.

\noindent{\underline{\textbf{Takeaways.}}} For multi-dimensional learned indices internally employing PGM-Index, much different from the 1-D case, both index construction time and query processing time are \textbf{insensitive} to the setting of $\epsilon$. 
In summary, we suggest $\epsilon=64$ for all the learned indices that employ the PGM-Index in their structure, as such an $\epsilon$ can well trade-off the index building time, memory overhead, and query processing efficiency.

\subsection{Index Update Efficiency Evaluation (Q7)}\label{subsec:exp_dynamic}
In this section, we report the update efficiency evaluation results for the learned indices that support dynamic operations (i.e., \texttt{LISA}, \texttt{RSMI}, and \texttt{IFI}). 
Specifically, for each dataset, we first use the first 80\% points to construct an index. 
Then, we sequentially insert the remaining 20\% points into the index and randomly remove 20\% points from the existing ones. 

Figure~\ref{fig:update} reports the average insertion and deletion time on two real-world datasets (\texttt{FourSquare} and \texttt{Toronto3d}). 
We can observe that, for the learned indices marked as updatable (i.e., \texttt{IFI}, \texttt{LISA}, and \texttt{RSMI}), they are generally less performant than conventional index structures in terms of update efficiency. 
Moreover, during the data updates, we randomly generate 20 range queries and $k$NN queries\footnote{The query generation process follows the same way as we described in Section~\ref{subsec:data_query_prepare}} at different timestamps to evaluate the query efficiency under a dynamic environment. 
In contrast to the previous findings observed under a \emph{static} environmental condition, our evaluation results indicate that, with regard to the efficiency of range query and $k$NN query processing, learned indices generally incur higher costs compared to non-learned baselines.

The reason is that current updatable learned indices on multi-dimensional data usually employ simple strategies to accommodate dynamic updates like setting a threshold of index retraining, introducing non-negligible cost. 
Moreover, existing techniques for designing updatable 1-D learned index (e.g., ALEX~\cite{DBLP:conf/sigmod/DingMYWDLZCGKLK20} and LIPP~\cite{DBLP:journals/pvldb/WuZCCWX21}) cannot be seamlessly applied to handle the case of multi-dimensional databases. 
This is because, as previously mentioned, a multi-dimensional learned index comprises two primary components: data layouts and learned models. Consequently, dynamic operations on multi-dimensional learned indices necessitate efficient updates to both the underlying data layouts (e.g., grids or partitions) and the learned models (e.g., PGM-Index).

\noindent\underline{\textbf{Takeaways.}} 
Among the \emph{few} learned indices that are capable of handling dynamic operations, their efficiency in terms of insertion, deletion, and dynamic query processing typically falls short when compared to traditional spatial indices, such as the variants of R-tree and $k$d-tree. 

\vspace{-2ex}
\subsection{Deep Dive into Learned Index Performance (Q8)}\label{subsec:exp_cache}
To gain deeper insights into the performance of learned indices, we conduct a comprehensive evaluation of CPU performance counter statistics during the processing of range queries.  
Specifically, Table~\ref{tab:exp_cache} presents the total cache reference count (\texttt{\#CR}), cache miss ratio (\texttt{CMR\%}), total CPU branch count (\texttt{\#BR}), and branch prediction miss ratio (\texttt{BMR\%})\footnote{The hardware performance counters (HPC) are accessed by invoking the standard Linux tool \texttt{perf}.}. 

To better measure the CPU overhead caused by index access, we pick a collection of range queries that are less selective ($Sel\leq 0.1$) to reduce the influence caused by extra overhead of reporting query results. 
In addition, to shed light on the relationship between the query processing efficiency and these CPU performance counter statistics, we mark the cell colors of Table~\ref{tab:exp_cache} from the color map {\color[HTML]{CB7373}{$\blacksquare$}}{\color[HTML]{FFCCC9}{$\blacksquare$}}{\color[HTML]{FFECEC}{$\blacksquare$}}{\color[HTML]{B4FFB4}{$\blacksquare$}}{\color[HTML]{329832}{$\blacksquare$}}.
The color towards the left of the color map (e.g., the dark red color {\color[HTML]{CB7373}{$\blacksquare$}}) indicates the longest query processing time (ranked the three slowest); 
on the other hand, the color towards the right of the color map (e.g., the dark green color {\color[HTML]{329832}{$\blacksquare$}}) indicates the shortest query processing time (ranked the two fastest). 

\begin{figure*}
    \centering
    \includegraphics[width=0.95\textwidth]{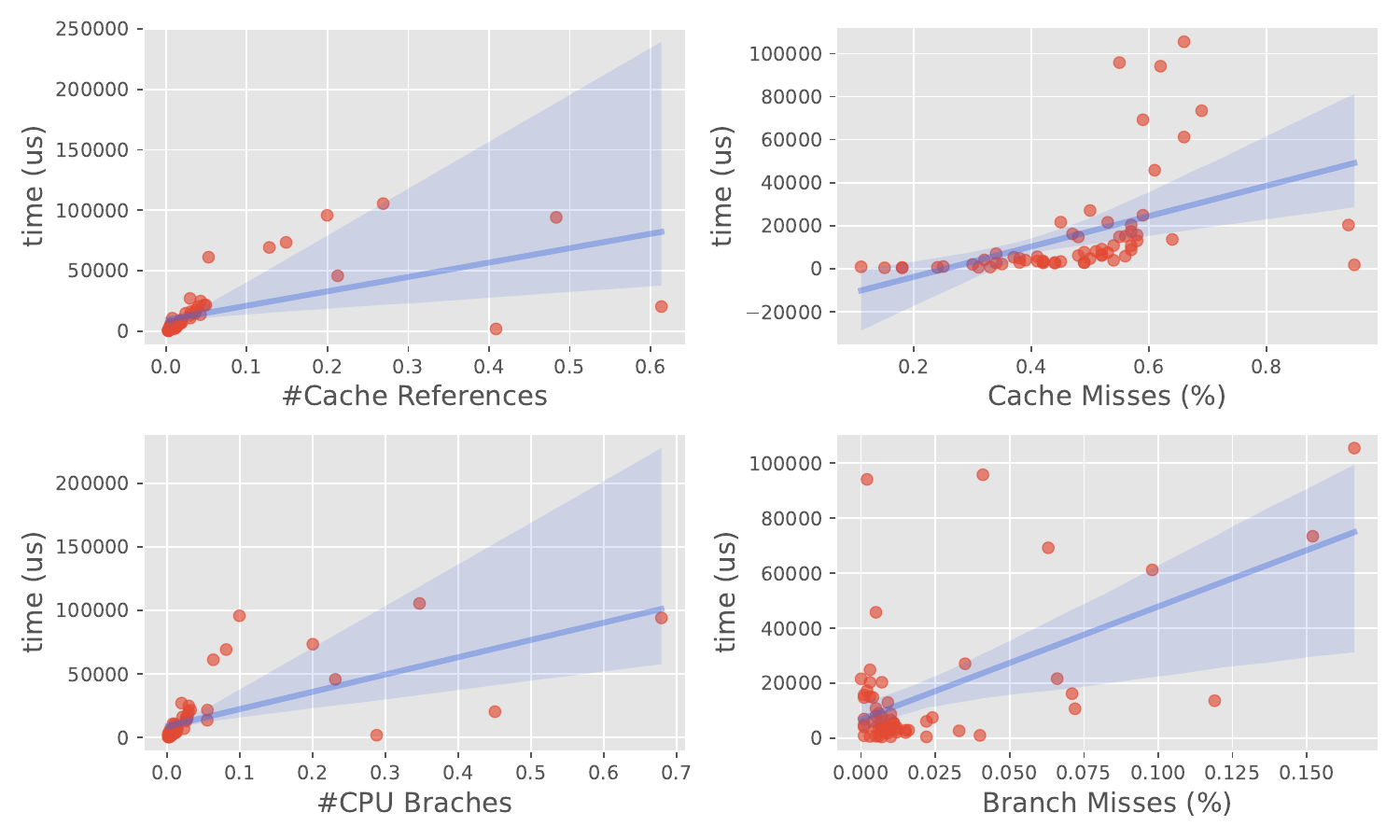}
    \caption{Range query processing time w.r.t.~difference CPU performance counter statistics, i.e., \#Cache References (CR), Cache Miss Rate (CMR), \#CPU Branches (BR), and Branch Miss Rate (BRM). } 
    \label{fig:linear_relationship}
\end{figure*}

Learned indices including \texttt{IFI}, \texttt{LISA} and \texttt{Flood} obtain smaller query processing costs in general, as indicated by their green cell colors in Table~\ref{tab:exp_cache}. 
It could be observed that these learned indices generally obtain smaller values in all of the four CPU performance counter statistics. 
For example, on synthetic dataset \texttt{Normal}, \texttt{Flood} produces 97 million total cache references with a cache miss rate of 33.81\%, where both values are the smallest among all other compared methods (in the same column). 
Meanwhile, \texttt{Flood} produces 253 million CPU branches with a 0.83\% branch prediction miss rate, where both values are the smallest compared with all other methods. 

We then visualize the relationship between range query processing time w.r.t.~4 CPU performance counters in Figure~\ref{fig:linear_relationship}. 
A positive correlation can be individually observed for each CPU performance counter statistics. 
To further quantitatively depict the correlation between the query processing time (i.e., the dependant variable $y$) and the four CPU performance counter statistics (i.e., the independent variables $x_1$: \#Cache References, $x_2$: Cache Miss Rate, $x_3$: \#CPU Branches, and $x_4$: Branch Miss Rate), we perform a linear regression analysis to fit a linear model 
$y=w_0 + w_1 x_1 + w_2 x_2 + w_4 x_4$.  
Note that, we omit feature $x_3$ (\#CPU Branches) as it is highly linearly correlated to $x_1$ (\#Cache References). 
The results are shown in Table~\ref{tab:regression_result} where each independent variable excluding the constant variable is positively correlated to the query processing time with high significance ($P<0.01$). 

\begin{table}[t]
\caption{Regression analysis results. The dependent variable is the range query processing time and independent variables are \texttt{\#CR}, \texttt{CMR\%}, and \texttt{BMR\%}. To avoid the influence of outliers, we perform a robust linear regression with Huber loss~\cite{huber1992robust}. We report the coefficient estimations, standard errors, and significance values based on Z-statistics. }
\label{tab:regression_result}
\begin{tabular}{|c|c|c|c|c|}
\hline
\texttt{Variable} & \texttt{Coef}       & \texttt{Std Err}  & \texttt{Z}      & \texttt{P\textgreater{}|Z|} \\ \hline
\texttt{const}    & -6813.1178 & 2555.357 & -2.666 & 0.008              \\ \hline
\texttt{\#CR}    & 1.804e+05  & 7723.199 & 23.357 & 0.000              \\ \hline
\texttt{CMR\%}    & 1.943e+04  & 5866.650 & 3.312  & 0.001              \\ \hline
\texttt{BMR\%}    & 2.389e+05  & 2.02e+04 & 11.832 & 0.000              \\ \hline
\end{tabular}
\end{table}

As validated by the linear regression analysis, there is a general positive linear correlation between the query processing costs and the four CPU performance counter values. 
However, it does not mean that smaller values in one of the statistics solely (e.g., CPU cache miss rates) necessarily lead to shorter query processing time. 
For example, for grid-based method \texttt{EDG} on the \texttt{Lognormal} dataset, the CPU cache miss rate is 39.31\%, which is smaller than ones of \texttt{Flood}, \texttt{LISA} and \texttt{IFI}. 
But as indicated by the green cell colors, the range query processing time of \texttt{Flood}, \texttt{LISA} and \texttt{IFI} are all less than that of \texttt{EDG}. 
The reason is that the above 3 learned indices have a better capability of pruning unnecessary candidates given a specific query range, leading to less memory footprint (i.e., small total cache references count). 
As an example, the count of total cache references (e.g., 54 million, 55 million, and 55 million for \texttt{IFI}, \texttt{LISA} and \texttt{Flood}, respectively) are all smaller than the 62 million references produced by \texttt{EDG}. 
As a consequence, the overall effect of smaller total cache references results in shorter query processing time.

\begin{table*}[t]
    \centering
    \caption{Summary of evaluation results. The cell colors from left to right {\color[HTML]{CB7373}{$\blacksquare$}}{\color[HTML]{FFCCC9}{$\blacksquare$}}{\color[HTML]{B4FFB4}{$\blacksquare$}}{\color[HTML]{52BE80}{$\blacksquare$}} represent the performance levels of each evaluation item from worst to best.}
    \label{tab:result_summary}
\begin{tabular}{|c||ccc||cccc|}
\hline
\multirow{3}{*}{Index} & \multicolumn{3}{c||}{Index Construction Evaluation}                                                                                                                                                                                         & \multicolumn{4}{c|}{Query Processing Evaluation}                                                                                                                                                                                                                                                     \\ \cline{2-8} 
                       & \multicolumn{1}{c|}{\begin{tabular}[c]{@{}c@{}}Construction\\ Time\end{tabular}}      & \multicolumn{1}{c|}{\begin{tabular}[c]{@{}c@{}}Memory\\ Cost\end{tabular}} & \begin{tabular}[c]{@{}c@{}}Parameter\\ Tunnig\end{tabular} & \multicolumn{1}{c|}{\begin{tabular}[c]{@{}c@{}}Range Query\\ Efficiency\end{tabular}}       & \multicolumn{1}{c|}{\begin{tabular}[c]{@{}c@{}}$k$NN Query\\ Efficiency\end{tabular}}        & \multicolumn{1}{c|}{Scalability} & \begin{tabular}[c]{@{}c@{}}Dynamic\\ Update\end{tabular}  \\ \hline
\texttt{ZMI}           & \multicolumn{1}{c|}{\cellcolor[HTML]{52BE80}\begin{tabular}[c]{@{}c@{}}better than\\ R-tree\end{tabular}}     & \multicolumn{1}{c|}{\cellcolor[HTML]{52BE80}excellent}                                            & \cellcolor[HTML]{52BE80}easy                                                       & \multicolumn{1}{c|}{\cellcolor[HTML]{FFCCC9}\begin{tabular}[c]{@{}c@{}}sometimes better\\ than R-tree\end{tabular}} & \multicolumn{1}{c|}{\cellcolor[HTML]{FFCCC9}\begin{tabular}[c]{@{}c@{}}sometimes better\\ than $k$d-tree\end{tabular}} & \multicolumn{1}{c|}{\cellcolor[HTML]{52BE80}excellent}   & \cellcolor[HTML]{CB7373}not supported                                                      \\ \hline
\texttt{MLI}           & \multicolumn{1}{c|}{\cellcolor[HTML]{FFCCC9}\begin{tabular}[c]{@{}c@{}}comparable\\ to R-tree\end{tabular}}   & \multicolumn{1}{c|}{\cellcolor[HTML]{52BE80}excellent}                                            & \cellcolor[HTML]{52BE80}easy                                                       & \multicolumn{1}{c|}{\cellcolor[HTML]{CB7373}\begin{tabular}[c]{@{}c@{}}no better\\ than R-tree\end{tabular}}        & \multicolumn{1}{c|}{\cellcolor[HTML]{FFCCC9}\begin{tabular}[c]{@{}c@{}}sometimes better\\ than $k$d-tree\end{tabular}} & \multicolumn{1}{c|}{\cellcolor[HTML]{B4FFB4}good}        & \cellcolor[HTML]{CB7373}not supported                                                      \\ \hline
\texttt{IFI}           & \multicolumn{1}{c|}{\cellcolor[HTML]{FFCCC9}\begin{tabular}[c]{@{}c@{}}comparable\\ to R-tree\end{tabular}}    & \multicolumn{1}{c|}{\cellcolor[HTML]{B4FFB4}good}                                                  & N.A.                                                       & \multicolumn{1}{c|}{\cellcolor[HTML]{52BE80}\begin{tabular}[c]{@{}c@{}}robustly better\\ than R-tree\end{tabular}}  & \multicolumn{1}{c|}{\cellcolor[HTML]{CB7373}not supported}                                                                    & \multicolumn{1}{c|}{\cellcolor[HTML]{52BE80}excellent}   & \cellcolor[HTML]{FFCCC9}limited support \\ \hline
\texttt{RSMI}          & \multicolumn{1}{c|}{\cellcolor[HTML]{CB7373}\begin{tabular}[c]{@{}c@{}}much larger\\ than R-tree\end{tabular}} & \multicolumn{1}{c|}{\cellcolor[HTML]{B4FFB4}good}                                                & N.A.                                                       & \multicolumn{1}{c|}{\cellcolor[HTML]{FFCCC9}\begin{tabular}[c]{@{}c@{}}sometimes better\\ than R-tree\end{tabular}} & \multicolumn{1}{c|}{\cellcolor[HTML]{FFCCC9}\begin{tabular}[c]{@{}c@{}}sometimes better\\ than kd-tree\end{tabular}} & \multicolumn{1}{c|}{\cellcolor[HTML]{CB7373}bad}         & \cellcolor[HTML]{FFCCC9}limited support \\ \hline
\texttt{LISA}          & \multicolumn{1}{c|}{\cellcolor[HTML]{FFCCC9}\begin{tabular}[c]{@{}c@{}}comparable\\ to R-tree\end{tabular}}   & \multicolumn{1}{c|}{\cellcolor[HTML]{B4FFB4}good}                                                  & \cellcolor[HTML]{52BE80}easy                                                       & \multicolumn{1}{c|}{\cellcolor[HTML]{52BE80}\begin{tabular}[c]{@{}c@{}}robustly better\\ than R-tree\end{tabular}}  & \multicolumn{1}{c|}{\cellcolor[HTML]{FFCCC9}\begin{tabular}[c]{@{}c@{}}sometimes better\\ than kd-tree\end{tabular}} & \multicolumn{1}{c|}{\cellcolor[HTML]{52BE80}excellent}   & \cellcolor[HTML]{FFCCC9}limited support\\ \hline
\texttt{Flood}         & \multicolumn{1}{c|}{\cellcolor[HTML]{FFCCC9}\begin{tabular}[c]{@{}c@{}}comparable\\ to R-tree\end{tabular}}   & \multicolumn{1}{c|}{\cellcolor[HTML]{B4FFB4}good}                                                  & \cellcolor[HTML]{52BE80}easy                                                       & \multicolumn{1}{c|}{\cellcolor[HTML]{52BE80}\begin{tabular}[c]{@{}c@{}}robustly better\\ than R-tree\end{tabular}}  & \multicolumn{1}{c|}{\cellcolor[HTML]{CB7373}not supported}                                                                    & \multicolumn{1}{c|}{\cellcolor[HTML]{52BE80}excellent}   & \cellcolor[HTML]{CB7373}not supported                                                      \\ \hline
\end{tabular}
\end{table*}

\noindent\underline{\textbf{Takeaways.}} 
There is a general positive correlation between the query processing time and the CPU performance counters, including the counts of cache references, cache miss rates, the number of CPU branches, and the branch miss rate. 
Performant learned indices such as \texttt{Flood}, \texttt{LISA} and \texttt{IFI} have a better capability of pruning unnecessary candidates given a specific range query, resulting in fewer cache references, and eventually leading to faster query processing. 

\subsection{Summary of Results}

Based on the experimental results, we compile a table to summarize our evaluation for each learned index in Table~\ref{tab:result_summary} from two major aspects: index construction and query processing. 

\noindent\underline{\textbf{Index Construction (Q1, Q2, Q6).}}  In terms of the cost of index construction (shown in the Construction Time column in Table~\ref{tab:result_summary}), learned indices using PGMIndex as their internal models are generally as fast as non-learned baselines, while deep learning-based learned indices are much slower than non-learned baselines due to the heavy deep learning runtimes. 

As for the memory footprint of the constructed index (see the Memory Cost column in Table~\ref{tab:result_summary}), multi-dimensional learned indices generally achieve robust and significant space reduction compared to non-learned indices like R-tree. 

In addition, our evaluation results also indicate that much different from the 1-dimensional scenario, 
the error parameter $\epsilon$ for multi-dimensional learned indices has minimal impact on index construction time and and query efficiency. 
Therefore, by a careful trade-off between efficiency and space cost, we suggest setting $\epsilon$ to \textbf{64} to achieve satisfactory performance. 

\noindent\underline{\textbf{Query Processing (Q3, Q4, Q5, Q7).}} 
As for range query processing efficiency (see the corresponding column in Table~\ref{tab:result_summary}), learned indices like \texttt{IFI}, \texttt{Flood}, and \texttt{LISA} can consistently outperform non-learned baselines like R-tree variants for queries with varying selectivity levels. 
The speedup ratio is particularly notable for datasets with uniform distribution and queries with low selectivity. 

The evaluation results of the $k$NN query processing (see the $k$NN Query Efficiency column in Table~\ref{tab:result_summary}) contrast to the conclusion of range queries where there is no learned index that can systematically outperform non-learned baselines like $k$d-tree variants. 
The progressive range search nature hinders efficient $k$NN query processing for learned indices. 

As for the scalability (see the Scalability column in Table~\ref{tab:result_summary}), most methods scale well to large datasets (up to $N=100$M) and moderate dimensions (up to $d=8$), with an exception of \texttt{RMSI}, on which we fail to train the underlying deep learning models when the data size reaches 50M. 

For the capability of handling dynamic operations (see the Dynamic Update column in Table~\ref{tab:result_summary}), the existing learned index structures either cannot support dynamic operations or provide very limited support, which cannot systematically outperform the conventional indices like R-tree and $k$d-tree variants.

\noindent\underline{\textbf{Performance Analysis (Q8).}} 
Our deep-dive analysis into the performance of learned indices in Section~\ref{subsec:exp_cache} reveals that, there is a general positive correlation between the query processing time and hardware performance counter statistics (e.g., cache reference counts and cache miss rates). 
The most performant learned indices such as \texttt{Flood}, \texttt{LISA} and \texttt{IFI} generally have a better capability of pruning unnecessary candidates given a specific range query, resulting in fewer cache references, and a better capability of utilizing cache locality in their data layout design. 
In addition to cache references, methods with a higher number of CPU branches or higher branch miss rates tend to have longer query processing time.

\vspace{-2ex}
\section{Conclusion and Future Studies}\label{sec:conclusion}

\subsection{Conclusion}
Efficient indexing method holds paramount importance for multi-dimensional data management and analytics. 
Inspired by the seminal work of 1-D learned index~\cite{DBLP:conf/sigmod/KraskaBCDP18},  
there is a growing trend of utilizing machine learning models to directly learn the storage layouts for multi-dimensional data. 
However, by thoroughly investigating the experiments reported in existing multi-dimensional learned index studies, we find out that there lacks a comprehensive benchmark to evaluate these indices under a uniform configuration. 
In this study, we first survey the recent multi-dimensional learned indices and deliver a taxonomy based on how learned models are integrated into the data layouts. 
Then, we develop an open-soured benchmark and evaluate 6 representative multi-dimensional learned indices by unifying the index implementation, datasets, query workloads, and evaluation metrics. 

The key experimental findings are threefold. 
\begin{enumerate}[leftmargin=*,label={\bfseries F\arabic*:}]
    \item Compared with traditional indices like R-tree and $k$d-tree variants, multi-dimensional learned indices (especially \texttt{LISA}, \texttt{Flood}, and \texttt{IFI}) can significantly reduce the index size (up to 3 orders of magnitude) while achieving better \textbf{range} query processing efficiency (up to $2.19\times$). 
    \item In terms of $k$NN query processing and dynamic operations, learned indices \textbf{cannot} consistently outperform conventional methods due to the lack of proper query processing algorithms and the intrinsic hardness of updating models and data layouts at the same time. 
    \item We identify the correlations between the performance of learned indices and hardware performance counters, revealing that efficient learned indices are usually cache-efficient. They have stronger candidate pruning power (reflecting on fewer cache reference counts) and better data layouts to maintain spatial locality (reflecting on lower cache miss rates).
\end{enumerate}


\subsection{Future Studies}
Learned index design on multi-dimensional data is a prominent and rapidly growing field in the data management community. 
Our thorough experimental study finds out that, though achieving significant improvements in \textbf{some} cases, there are still multiple technical issues to be addressed for the multi-dimensional learned index before being widely applied to practical systems. 

We identify the following potential research opportunities based on our key experimental findings. 

\noindent\underline{\textbf{Advanced Query Support.}} According to \textbf{F1} and \textbf{F2}, existing multi-dimensional learned indices are capable of accelerating \textbf{range} query processing while significantly reducing the space cost. 
On the other hand, the support to $k$NN queries and more advanced analytical operators like skyline query~\cite{DBLP:conf/icde/BorzsonyiKS01} and spatial joins~\cite{DBLP:conf/sigmod/HjaltasonS98,DBLP:conf/vldb/XiaLOH04} are either limited or missing. 
It would be interesting and challenging to extend existing learned index structures or design new learned data layouts to support these analytical queries efficiently.

\noindent\underline{\textbf{Efficient Updatable Index.}} As we discussed in Section~\ref{subsec:discussion_index} and \textbf{F2}, most of the existing learned indices focus on the \textbf{read-only} workloads. 
Though indices like \texttt{RSMI}~\cite{DBLP:journals/pvldb/QiLJK20} and \texttt{LISA} \cite{DBLP:conf/sigmod/Li0ZY020} support dynamic insertion and deletion via a model-based data layout, it is inevitable to re-train the whole models when the data distribution significantly shifts from the initial one that is used to build the index. 
A fully updatable learned multi-dimensional index requires efforts on novel data layout design and efficient model update strategy. 
In addition, a well-crafted cost model should be established to embed learned indices into DBMS with cost-based optimizers seamlessly. 

\noindent\underline{\textbf{IO-Efficient and Cache-Efficient Learned Index.}} 

\noindent Existing works on learned indices mainly target on \textbf{in-memory} query processing. 
Although \texttt{LISA}~\cite{DBLP:conf/sigmod/Li0ZY020} and \texttt{RSMI}~\cite{DBLP:journals/pvldb/QiLJK20} claim that they can be extended to disk-based indices, unfortunately, they do not adopt any optimization considering the disk access characteristics. 
In addition, although \textbf{F3} claims that some learned index structures are cache-efficient, there are still optimization opportunities to further reduce the cache miss rate by designing new ML-based data layouts that carefully consider the memory hierarchy. 

\noindent\underline{\textbf{Distributed Spatial Analytics.}} To process and analyze web-scale multi-dimensional data, practical solutions usually adopt distributed computation engines like Spark~\cite{spark} to serve the backend, e.g., LocationSpark~\cite{DBLP:journals/pvldb/TangYMOA16}, GeoSpark \cite{DBLP:conf/gis/YuWS15},  Simba~\cite{DBLP:conf/sigmod/XieL0LZG16}. 
As reported in an evaluation study~\cite{DBLP:journals/pvldb/PandeyKNK18}, the index memory overhead of existing systems is generally high. 
Due to the significant space-time trade-off (\textbf{F1}), it would be promising to design new distributed multi-dimensional analytical systems powered by learned indices.


\bibliographystyle{spmpsci}      
\bibliography{main}   

%
%

\end{document}